\documentclass[preprint,12pt]{elsarticle}
\usepackage[top=1in, bottom=1in, left=1in, right=1in]{geometry}
\usepackage{bm}
\usepackage{amsmath}
\usepackage{amssymb}
\usepackage{setspace}
\usepackage{bigints}
\usepackage{array}
\usepackage{multirow}
\usepackage{colortbl}
\usepackage{graphicx}
\usepackage{subfigure}
\usepackage[table,xcdraw]{xcolor}
\usepackage[outdir=./]{epstopdf}
\usepackage{booktabs}
\usepackage{tabu}
\usepackage{bbm}
\usepackage{float}
\usepackage[singlelinecheck=false]{caption}
\usepackage[utf8]{inputenc}
\usepackage[english]{babel}
\usepackage{natbib}
\usepackage{bookmark}
\usepackage{mathrsfs}
\usepackage{color}
\usepackage{mathtools} 
\newcommand{\comment}[1]{}
\usepackage{amssymb}
\usepackage{xfrac}
\usepackage{lscape}
\usepackage{pdflscape}
\usepackage{algorithm}
\usepackage{algpseudocode}
\usepackage{xspace}
\usepackage{booktabs}

\journal{Elsevier}
\begin{document}
\begin{spacing}{1.25}
\begin{frontmatter}

\title{On the quantification and efficient propagation of imprecise probabilities with copula dependence}
\author{ Jiaxin Zhang $^{1, 2}$, Michael Shields $^{2}$ \footnote{Corresponding author.\\ Email: zhangj@ornl.gov (J. Zhang), michael.shields@jhu.edu (M. Shields)} }
\address{$^{1}$Oak Ridge National Laboratory, Oak Ridge, TN 37831}
\address{$^{2}$Department of Civil Engineering, Johns Hopkins University Baltimore, MD 21218}

\begin{abstract}

This paper addresses the problem of quantification and propagation of uncertainties associated with dependence modeling when data for characterizing probability models are limited. Practically, the system inputs are often assumed to be mutually independent or correlated by a multivariate Gaussian distribution. However, this subjective assumption may introduce bias in the response estimate if the real dependence structure deviates from this assumption. In this work, we overcome this limitation by introducing a flexible copula dependence model to capture complex dependencies. A hierarchical Bayesian multimodel approach is proposed to quantify uncertainty in dependence model-form and model parameters that result from small data sets. This approach begins by identifying, through Bayesian multimodel inference, a set of candidate marginal models and their corresponding model probabilities, and then estimating the uncertainty in {\color{black}the} copula-based {\color{black} dependence} structure, which is conditional on the marginals and their parameters. The overall uncertainties integrating marginals and copulas are probabilistically represented by an ensemble of multivariate candidate densities. A novel importance sampling reweighting approach is proposed to efficiently propagate the overall uncertainties through a computational model. Through an example studying the influence of constituent properties on the out-of-plane properties of transversely isotropic E-glass fiber composites, we show that the composite property with copula-based dependence model converges to the true estimate as data set size increases, {\color{black} while an independence or arbitrary Gaussian correlation} assumption leads to a biased estimate. \\

\end{abstract}

\end{frontmatter}
\section{Introduction}

Uncertainty Quantification (UQ) is widely applied to better understand complex stochastic physical and mathematical systems. Typically, computational simulations aim to estimate statistics of the response of a system subject to random inputs. These inputs are commonly modeled by a random vector $\bm{X}$ with their joint probability density {\color{black}$f_{\bm{X}}(\bm{x})$}. The uncertainty associated with the inputs are quantified probabilistically and propagated through a computational model $\mathcal{M}$. The corresponding output $Y = \mathcal{M}(\bm{X})$ is the quantity of interest (QoI), which is uncertain. If the computational model is deterministic, all uncertainties in $Y$ result from the uncertainty in $\bm{X}$. 

Practically, the inputs are often assumed to be mutually independent or to possess a multivariate Gaussian dependence structure because it is simple to model and to fit {\color{black}from} data. Some conventional UQ approaches, for example, importance sampling \cite{melchers2018structural} and polynomial chaos expansions \cite{li1998adaptive}, take advantage of mutually independent inputs. If the inputs are dependent, a number of UQ approaches require to map the model inputs $\bm{X}$ onto an input $\bm{X}^*$ with independent components. When {\color{black}$f_{\bm{X}}(\bm{x})$} has multivariate Gaussian dependence structure, the map corresponds to the Nataf transformation  \cite{nataf1962determination, lebrun2009innovating}. A more general way that maps the input $\bm{X}$ onto $\bm{X}^*$ is the Rosenblatt transformation \cite{rosenblatt1952remarks}, which needs to know the conditional probability distribution functions (pdfs) that are often infeasible in practice. For this reason, the Gaussian dependence assumption is widely applied in the context of UQ. The Gaussian assumption and the associated dependence provides a convenient representation of the input dependencies, but it may introduce a bias in the response estimate if the real dependence structure deviates from this assumption. 

Dependence modeling has recently received widespread attention in the engineering and mathematics communities. This is mainly due to the significant development of copula models \cite{nelsen2007introduction, joe2014dependence, wisadwongsa2018bivariate}, and vine copulas \cite{joe2011dependence, aas2009pair, joe2010tail,nagler2019model, muller2019dependence} in particular.  Copula theory is used to separately model the dependence and the marginal distribution, but it is often limited to low-dimensional problems, typically bivariate or simple copula families, such as Gaussian or Archimedean families \cite{nelsen2007introduction}. Copula-based approaches have been recently used in various dependence modeling studies, for example in reliability and risk analysis \cite{rozsas2017effect, wang2018roles, xu2018failure, he2018failure, wang2018role, pan2019modeling}, sensitivity analysis \cite{wang2018copula, hu2018probability}, and prognostics and health management (PHM) \cite{xi2014copula, xi2019enhanced}. Copulas also have widespread applications in engineering practice, such as ocean and offshore \cite{zhang2015long, masina2015coastal}, wind \cite{warsido2015synthesis}, and earthquake \cite{goda2015multi} engineering.  To overcome the limitation of copula theory in high dimensions, {\color{black}the vine copula} theory was first proposed by Joe \cite{joe1994multivariate, joe1997multivariate} by formulating multivariate copulas as a product of bivariate copulas among pairs of random variables. Bedford and Cooke \cite{bedford2002vines} introduced a graphical model for describing multivariate copulas using pair-copulas, which provides a flexible and easy interpretation. Czado presented a series of productive studies in the context of vine copulas \cite{czado2010pair, czado2013selection} and successfully applied them to financial modeling \cite{dissmann2013selecting, brechmann2012truncated}.  Recently, vine copula approaches have become increasingly attractive in engineering applications \cite{xu2018failure, torre2019general, qiu2019scenario, li2019efficient, niemierko2019d}. 

Conventionally, the dependence structure of multivariate inputs is built probabilistically through a known joint probability measure. Therefore, the first step of copula-based dependence modeling is to identify or assume a reasonable copula or vine copula model for the input variables. However, it may not be straightforward to identify the appropriate copula model when data characterizing the input parameters are sparse. This process may therefore give rise to a form of \emph{epistemic uncertainty} \cite{der2009aleatory} - which is due to a lack of knowledge or data. Epistemic uncertainty plays an essential role in UQ and must be considered, particularly when it arises from a lack of data. 

Many theories have been developed to address the various forms of epistemic uncertainty. It has been argued that epistemic uncertainty needs a different mathematical treatment than \emph{aleatory uncertainty} \cite{ferson1996different} that are naturally stochastic and treated probabilistically.  It remains an open debate as to what that mathematical treatment should be. This desire also has given rise to the field of so-called \emph{imprecise probabilities} wherein epistemic uncertainty contributes a level of ``imprecision" and aleatory uncertainty are quantified by classical probability theory. There are numerous approaches to model this imprecision that include the use of fuzzy sets \cite{zadeh1965fuzzy,dubois2012fundamentals} and measures \cite{wang2013fuzzy}, random sets \cite{berger1994overview, fetz2004propagation, molchanov2005theory, fetz2016imprecise}, intervals and probability boxes \cite{Moore1979, ferson1996different, ferson2002, fetz2004propagation} and Dempster-Shafer theory \cite{Dempster1967, Shafer1976}. Efforts from Walley \cite{walley1991statistical, walley2000towards} have worked to unify these theories under an over-arching theory of imprecise probabilities. An extensive review of many of these imprecise probabilities approaches for engineering applications can be found in \cite{beer2013imprecise}. 

To the author's knowledge, relatively few studies have accounted for the problem of \emph{imprecise dependence modeling} in UQ. Some recent studies focus on the investigations of Sklar's theorem for imprecise copulas using fuzzy theory \cite{montes2015sklar, pelessoni2013imprecise}. Coolen-Maturi et al. \cite{coolen2016predictive} combine nonparametric predictive inference that quantifies the uncertainties through imprecise probability with a parametric copula to model and estimate the dependence structure. Among the most comprehensive studies of UQ with dependence modeling is that conducted by Kurowicka and Cooke \cite{kurowicka2006uncertainty}, who discussed UQ in bivariate as well as high dimensional dependence modeling. More recent works include those of Schefzik et al. \cite{schefzik2013uncertainty}, who propose a general multi-stage procedure called ensemble copula coupling to quantify the uncertainty in complex simulation models, particularly in weather and climate predictions, and Emiliano et al.\cite{torre2019general} who use vine copulas to develop a general data-driven UQ framework for dependence modeling of complex input. 

 In this paper, we investigate copula-based dependence modeling in the context of imprecise probability that specifically results from a lack of data. This is motivated by the difficulty of data collection under complex conditions, for example, long-time cycle and expensive experiments, in engineering practice. When only scarce data is available, it is a challenging task to assign an objective and accurate probability distribution for the random inputs and precisely estimate their dependence relationship. The developed method builds on the previous work of the authors who proposed information-theoretic \cite{zhang2018quantification} and Bayesian \cite{zhang2018effect} multimodel probabilistic methodologies to quantify and efficiently propagate combined aleatory and epistemic uncertainty given small data sets. This work introduces a copula-based dependence modeling, which is flexible enough to capture complex dependence structure. To fully quantify the uncertainty in dependence modeling, we propose a hierarchical Bayesian multimodel approach that allows to first identify a set of candidate marginal models and their associated model probabilities, and then estimate the {\color{black}copula} model-form and model parameter uncertainties, which are {\color{black}conditioned} on the uncertain marginals and their parameters. Using the proposed method, an ensemble of candidate multivariate densities are identified as random inputs that need to be propagated through a complex model to estimate the response of an engineering system. Propagation of these families of densities is particularly difficult because it requires nested Monte Carlo calculations, which are often computationally infeasible even for simple models. This paper proposes a novel efficient importance sampling reweighting algorithm that allows simultaneous propagation of the multiple densities through one Monte Carlo simulation. The proposed method can further achieve an adaptive updating as additional data are collected but without requiring additional computational evaluation. 
  
 This paper is structured as follows. Section 2 provides a brief review of copula-based dependence modeling, particularly bivariate copula theory and vine copula theory. Section 3 presents the uncertainty analysis for copula-based multivariate dependence modeling, including copula uncertainty and marginal uncertainty. An efficient uncertainty propagation with imprecise copula dependence modeling is proposed in Section 4. Section 5 shows an application of the proposed method to the probabilistic prediction of unidirectional composite lamina properties. Some discussions and concluding remarks are given in Section 6. 

 \section{Copula-based modeling of dependence structure}

\subsection{{\color{black}Measures of statistical dependence}}

The most well-known measure of dependence between random variables is the Pearson's correlation coefficient, commonly named simply the correlation coefficient, which measures linear dependence. Considering two random variables $X$ and $Y$ with mean values $\mu_X$ and $\mu_Y$ and standard deviations $\sigma_X$ and $\sigma_Y$, the correlation coefficient $\rho_{X,Y}$ is defined as 
\begin{equation}
\rho_{X,Y} = \frac{cov(X,Y)}{\sigma_X \sigma_Y} = \frac{E[(X-\mu_X)(Y-\mu_Y)]}{\sigma_X \sigma_Y}
\label{eqn:correlation}
\end{equation}
where $E[\cdot]$ is the expectation and $cov$ is the covariance. All correlation coefficient values are bounded in the interval $[-1, 1]$, indicating the degree of linear dependence between two variables. The closer the coefficient is to either 1 or -1, the stronger the correlation between the variables. If the variables are linearly independent, the correlation coefficient is 0.  

Another common measure of dependence is Kendall's $\tau$, or Kendall's rank correlation coefficient, which measures the difference between the concordance and discordance probability and can be used to detect some nonlinear dependence. Let $(X_1, Y_1)$ and $(X_2, Y_2)$ be independent and identically distributed random vectors, then Kendall's tau is defined as
\begin{equation}
\tau_{X,Y} = P[(X_1 - X_2)(Y_1 - Y_2)>0] - P[(X_1 - X_2)(Y_1-Y_2)<0]. 
\end{equation}

Rank correlation can also be expressed using Spearman's $\rho$ (defined as the correlation coefficient -- Eq.\ \eqref{eqn:correlation} -- between the ranks of the variables) and both Kendall's $\tau$ and Spearman's $\rho$ can be shown to be special cases of a generalized rank correlation \cite{kendall1938new}.

However, the information given by a correlation coefficient (Pearson's $\rho$, Kendall's $\tau$, or Spearman's $\rho$) is only enough to define the dependence structure between random variables in special cases, e.g. Gaussian random variables. In general, the complete dependence structure requires knowledge of the full joint distribution. One method to capture the complete dependence structure is to model the joint distribution using a copula. In practice, many data structures exhibit different marginal distributions, nonsymmetric/nonlinear dependencies, and/or tail dependencies between variables. These variables cannot be modeled by a Gaussian or multivariate $t$ distribution. This challenge is overcome by the copula approach, which models the dependencies and marginal distributions separately. 

\subsection{Copula theory}

Consider {\color{black}$F_{\bm{X}}(\bm{x})$} as the $d$-dimensional joint distribution function of the random vector $\bm{X} = (X_1,...,X_d)^T$ with marginal distributions {\color{black}$F_1(x_1),...,F_d(x_d)$}. According to Sklar's theorem \cite{sklar1959fonctions},  there exists a copula $C$ such that for all $\bm{x} = (x_1,...,x_d)^T \in [-\infty, \infty]^d$, 
{\color{black}
\begin{equation}
F_{\bm{X}}(\bm{x}) = C(F_1(x_1),...,F_d(x_d)) \label{eq:sklar}
\end{equation}
}
If {\color{black}$F_1(x_1),...,F_d(x_d)$} are continuous, the copula $C$ is unique. The copula $C$ can be interpreted as the joint distribution function of a $d$-dimensional random vector on $[0,1]^d$ with uniform marginals. 

Sklar's theorem can also be restated with respect to probability densities. The corresponding copula density can be expressed as:
{\color{black}
\begin{equation}
c(F_1(x_1),...,F_d(x_d)) = \frac{\partial C(F_1(x_1),...,F_d(x_d))}{\partial F_1(x_1),..,\partial F_d(x_d)} \label{eq:copula_density}
\end{equation}
}
which implies the joint multivariate pdf can be formulated by 
{\color{black}
\begin{equation}
f_{\bm{X}}(\bm{x}) = c(F_1(x_1),...,F_d(x_d)) \cdot  f_1(x_1) \cdots f_d(x_d)  \label{eq:jpdf_copula_de}
\end{equation}
}
where $f_k(x_k), 1\le k \le d$ are the marginal pdfs. For the bivariate case, Joe \cite {joe1997multivariate} and Nelsen \cite{nelsen2007introduction} provided a rich variety of copula families from the two major classes of \emph{Elliptical} and \emph{Archimedean} copulas. Elliptical copulas are directly derived by inverting Sklar's theorem, shown in Eq.\ (\ref{eq:sklar}). Given a bivariate cumulative distribution function {\color{black}$F_{\bm{X}}(\bm{x})$} with marginals {\color{black}$F_1(x_1)$ and $F_2(x_2)$}, then 
\begin{equation}
C(u_1, u_2) = F(F_{1}^{-1}(u_1), F_2^{-1}(u_2) ) \label{eq:bi_copula}
\end{equation}
is a bivariate copula for $u_1, u_2 \in [0,1]$. One of the most commonly used bivariate elliptical copula is the bivariate Gaussian copula 
\begin{equation}
C(u_1, u_2) = \Phi_{\rho}(\Phi^{-1}(u_1), \Phi^{-1}(u_2) ) 
\end{equation}
where $\Phi_{\rho}$ is the joint cumulative distribution of bivariate standard normal distribution with correlation coefficient $\rho$ and $\Phi^{-1}$ is the inverse standard normal cdf. 

Another common copula is the Student-$t$ copula, whose bivariate density is given by 
\begin{equation}
{\color{black}f_{\bm{X}}(\bm{x})} = \frac{\Gamma(\frac{\nu+2}{2})}{\Gamma(\frac{\nu}{2}) \sqrt{(\pi\nu)^2|\bm{\Sigma}|}} \left( 1+ \frac{(\bm{x}-\bm{\mu})^{\prime}\bm{\Sigma}^{-1}(\bm{x}-\bm{\mu})}{\nu} \right)^{-\frac{\nu+2}{2}} 
\end{equation}
where $\nu$ is the number of degrees of freedom, $\bm{\mu}$ is the mean vector and $\bm{\Sigma}$ is a positive-definite matrix. {\color{black} Since the copula remains invariant under a standardization of the marginal distributions, the copula of a $t(\nu, \bm{\mu}, \bm{\Sigma})$ is identical to that of a $t(\nu, 0, \bm{P})$ distribution where $\bm{P}$ is the correlation matrix implied by the dispersion matrix $\bm{\Sigma}$ \cite{demarta2005t}. Thus, the corresponding Student-t copula is given by
\begin{equation}
C(u_1,u_2) = \int_{-\infty}^{t_{\nu}^{-1}(u_1)} \int_{-\infty}^{t_{\nu}^{-1}(u_2)} \frac{\Gamma(\frac{\nu+2}{2})}{\Gamma(\frac{\nu}{2}) \sqrt{(\pi\nu)^2|\bm{P}|}} \left( 1+ \frac{\bm{x}^{\prime}\bm{P}^{-1}\bm{x}}{\nu} \right)^{-\frac{\nu+2}{2}} d\bm{x}.
\end{equation}
For bivariate case, we simplify the notation to
\begin{equation}
C(u_1, u_2) = t_{\rho, \nu} (t_{\nu}^{-1}(u_1), t_{\nu}^{-1}(u_2))
\end{equation}
where $\rho$ is the off-diagonal element of $\bm{P}$ \cite{demarta2005t}, $t_{\nu}^{-1}$ is defined as the inverse Student-$t$ marginal distribution function with $\nu$ degrees of freedom. Fig.\ \ref{fig:elliptical copula} shows samples from the elliptical copula family with Gaussian and Student-$t$ copulas. Table \ref{tab:elliptical copula table} provides the basic properties of the Gaussian and Student-$t$ copulas. 
\begin{figure}[!ht]    
    \centering
    \includegraphics[height=2in]{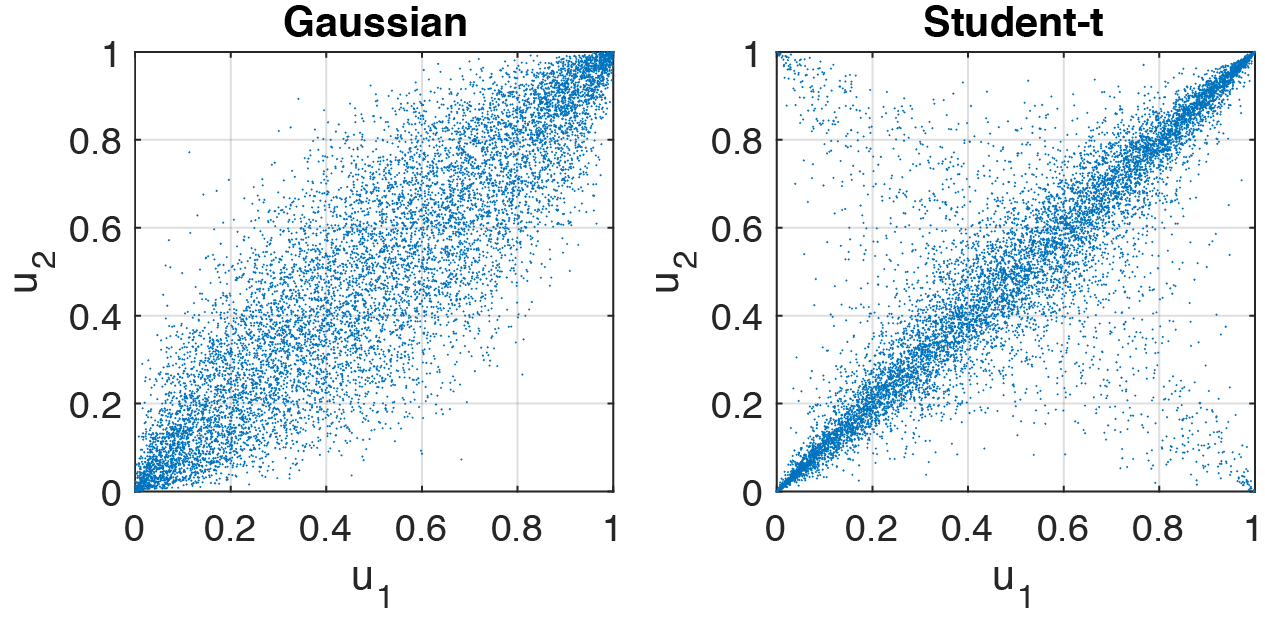}
    \caption{Elliptical copula family. Samples drawn from {\color{black}(left)} Gaussian copula and {\color{black}(right)} Student-$t$ copula}  \label{fig:elliptical copula}
\end{figure}
}

\begin{table}[!ht] \footnotesize
\centering
\caption{Properties and definition of elliptical copula families}
\label{tab:elliptical copula table}
\begin{tabular}{@{}cccc@{}}
\toprule
Elliptical family & Parameter range & Kendall's $\tau$ & Tail dependence \\ \midrule
Gaussian & $\rho \in (-1,1)$ & $\frac{2}{\pi}\arcsin(\rho)$ & 0 \\
Student-$t$ & $\rho \in (-1,1), \nu>2$ & $\frac{2}{\pi}\arcsin(\rho)$ & $2t_{\nu+1}(-\sqrt{\nu+1}\sqrt{\frac{1-\rho}{1+\rho}})$ \\ \bottomrule
\end{tabular}
\end{table}

Another important copula family, Archimedean copulas are defined as
\begin{equation}
C(u_1, u_2) =\psi^{[-1]}(\psi(u_1) + \psi(u_2)) 
\end{equation}
where $\psi$ is the generator function of the copula $C$, which is a continuous strictly decreasing convex function which satisfies $\psi(1) = 0$ and $\psi^{[-1]}$ is defined as 
\begin{equation}
\psi ^{ [-1] }(t)  = \left\{ \begin{matrix} \psi ^{ -1 }(t), \quad 0 \le t \le \psi(0) \\ 0, \quad \psi(0) \le t \le \infty \end{matrix} \right. 
\end{equation}

\begin{figure}[!ht]    
    \centering
    \includegraphics[height=2in]{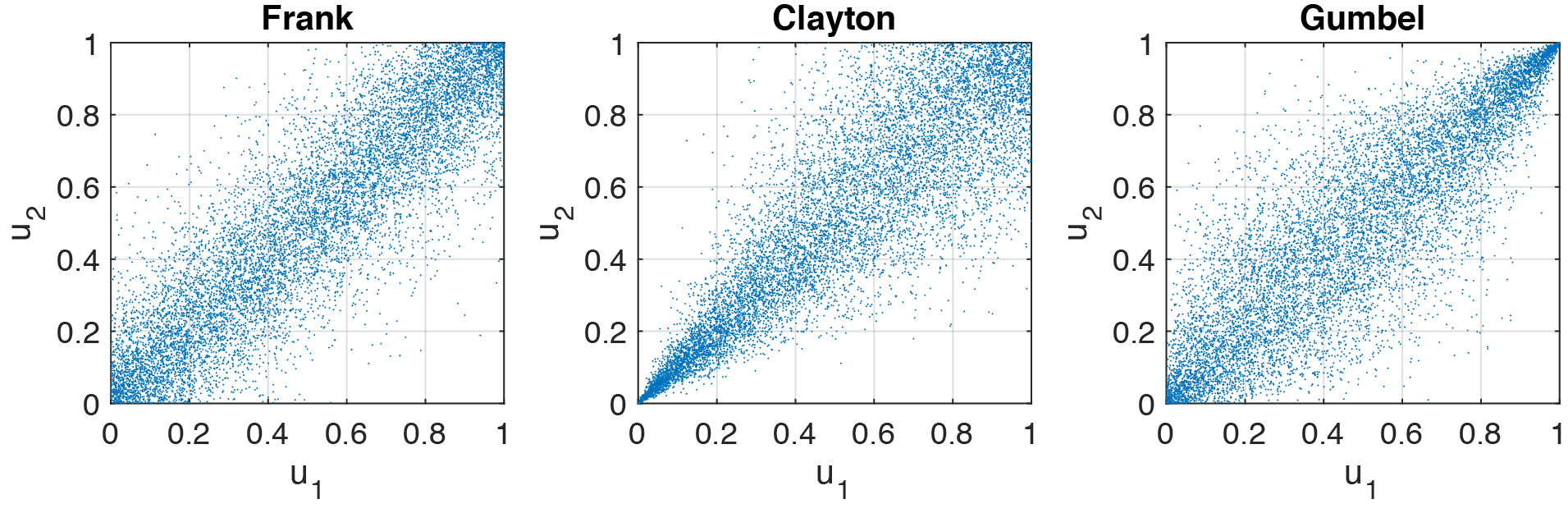}
    \caption{Archimedean copula family. Samples drawn from {\color{black}(left)} Frank copula, {\color{black}(middle)} Clayton copula and {\color{black}(right)} Gumbel copula. }  \label{fig:archimedean copula}
\end{figure}

The most common single parameter Archimedean copulas are the Clayton, Gumbel and Frank \cite{nelsen2007introduction}. Their bivariate copula formulations are shown in Table \ref{tab:Archimedean_1}, with their corresponding properties (generator and Kendall's $\tau$) shown in Table \ref{tab:Archimedean_2} where $D_1(\theta) = \frac{1}{\theta} \int_{0}^{\theta} \frac{t}{e^t-1}dt$ is {\color{black}the} Debye function \cite {joe1997multivariate, nelsen2007introduction}.  Fig.\ \ref{fig:archimedean copula} show examples of samples drawn from these copulas for two random variables $u_1$ and $u_2$. 

\begin{table}[!ht] \footnotesize
\centering
\caption{Definitions of Archimedean copula families}
\label{tab:Archimedean_1}
\begin{tabular}{@{}ccc@{}}
\toprule
Name of Copula & Bivariate copula $C_{\theta}(u_1,u_2)$ & Parameter $\theta$ \\ \midrule
Clayton        & $\left[ \max \left\{ u_1^{-\theta} + u_2^{-\theta} -1, 0 \right\} \right]^{-1/\theta}$               & $\theta \in [-1, \infty) \setminus \left\{0 \right\}$        \\
Frank          & $-\frac{1}{\theta}\log \left[ 1+ \frac{(e^{-\theta u_1}-1)(e^{-\theta u_2}-1)}{e^{-\theta}-1} \right]   $            & $\theta \in \mathbb{R} \setminus \left\{0 \right\} $         \\
Gumbel         & $e^{-\left( (-\log(u_1))^{\theta} + (-\log(u_2))^{\theta} \right)^{1/ \theta}} $               & $\theta \in [1, \infty)$        \\ \bottomrule
\end{tabular}
\end{table}

\begin{table}[!ht] \footnotesize
\centering
\caption{Properties of Archimedean copula families}
\label{tab:Archimedean_2}
\begin{tabular}{@{}ccc@{}}
\toprule
Name of Copula & Generator & Kendall's $\tau$ \\ \midrule
Clayton        & $\frac{1}{\theta}(t^{-\theta}-1)$               & $\frac{\theta}{\theta+2}$        \\
Frank          & $-\log[\frac{e^{-\theta t}-1}{e^{-\theta}-1}]$            & $1-\frac{4}{\theta} + 4 \frac{D_1(\theta)}{\theta} $        \\
Gumbel         & $(-\log t)^{\theta}$               & $1-\frac{1}{\theta}$        \\ \bottomrule
\end{tabular}
\end{table}

\subsection{Vine copulas}
Copula families perform well in the bivariate case, but in arbitrarily high dimension, the choice of adequate copula families is very limited. Elliptical families and Archimedean copulas lack the flexibility to accurately model the dependence structure of high dimensional variables. Simple extensions of these bivariate families offer some improvement, but typically become intricate and introduce additional limitations that, for example, they can not be applied to establish a distribution consistent with arbitrary correlation \cite{brechmann2013cdvine}. 

Vine copulas (also called tree structures) do not suffer from these issues and have been widely used in many fields of application.  Bedford and Cooke \cite{bedford2002vines} introduced a graphical model for describing multivariate copulas using a cascade of bivariate copulas, denoted by \emph{pair-copulas}. This pair-copula construction provides a flexible way to decompose a multivariate probability density into bivariate copulas such that each pair-copula is independent of the others. 

Consider a {\color{black}$d$}-dimensional joint density function {\color{black}$f_{\bm{X}}(x_1, ...,x_d)$} for a random vector {\color{black}$\bm{X} = (X_1,...,X_d)$}. This density can be decomposed based on the law of total probability 
\begin{equation}
{\color{black}f(x_1,...,x_d) = f_n(x_d) \cdot f(x_{d-1}|x_d) \cdot f(x_{d-2} | x_{d-1}, x_d) \cdots f(x_1|x_2,...,x_{d})}.
\end{equation}

From Sklar's theorem, we also know the joint probability density can be formulated as shown in Eq.\ (\ref{eq:jpdf_copula_de}). In the bivariate case, Eq.\ (\ref{eq:jpdf_copula_de}) simplifies to 
\begin{equation}
f(x_1, x_2) = c_{12}(F_1(x_1), F_2(x_2)) \cdot f_1(x_1) \cdot f_2(x_2) 
\end{equation}
where $c_{12}$ is the appropriate \emph{pair-copula density} for the pair of transformed variables $F_1(x_1)$ and $F_2(x_2)$. It is straightforward to write a conditional density 
\begin{equation}
f(x_1| x_2) = c_{12}(F_1(x_1), F_2(x_2)) \cdot f_1(x_1)  \label{eq:2d_copula1}
\end{equation}
in terms of the pair-copula. Similarly, it easily follows for three random variables $X_1$, $X_2$ and $X_3$ as follows
\begin{equation}
f(x_1| x_2, x_3) = c_{12|3}(F(x_1|x_3), F(x_2|x_3)) \cdot f(x_1|x_3) \label{eq:3d_copula1}
\end{equation}
for the appropriate pair-copula $c_{12|3}$ which is used for {\color{black}the} transformed variables $F(x_1 | x_3)$ and $F(x_2 | x_3)$. An alternative decomposition is 
\begin{equation}
f(x_1| x_2, x_3) = c_{13|2}(F(x_1|x_2), F(x_3|x_2)) \cdot f(x_1|x_2) \label{eq:3d_copula2}
\end{equation}
where $c_{13|2}$ differs from the pair-copula in Eq.\ (\ref{eq:3d_copula1}). We can further decompose $f(x_1|x_2)$ in Eq.\ (\ref{eq:3d_copula2}) based on Eq.\ (\ref{eq:2d_copula1})
\begin{equation}
f(x_1| x_2, x_3) = c_{13|2}(F(x_1|x_2), F(x_3|x_2)) \cdot c_{12}(F_1(x_1), F_2(x_2)) \cdot f_1(x_1). 
\end{equation}

By the extension, the conditional marginal can be decomposed into the appropriate pair-copula using the general form given by \cite{aas2009pair, czado2010pair}
\begin{equation}
f(x|\bm{v}) = c_{xv_j|\bm{v}_{-j}}(F({x}|\bm{v}_{-j}), F(v_j|\bm{v}_{-j})) f({x} | \bm{v}_{-j})
\end{equation}
where $v_j$ is an arbitrarily excluded element from vector $\bm{v}$ and $\bm{v}_{-j}$ denotes the vector $\bm{v}$ after excluding $v_j$. Hence,  a multivariate density {\color{black}$f_X(\bm{x})$} can be expressed as a product of bivariate copula density functions with marginal conditional CDFs in the form of {\color{black} $F({x}|\bm{v})$ that can be formulated recursively as follows \cite{joe1997multivariate}
\begin{equation}
F({x}|\bm{v}) = \frac{\partial C_{x,v_j|\bm{v}_{-j}}(F({x}|\bm{v}_{-j}), F(v_j | \bm{v}_{-j}))}{\partial F(v_j | \bm{v}_{-j})} \label{eq:F_conditional}
\end{equation}
where $C_{x,v_j|\bm{v}_{-j}}$ is a bivariate copula distribution function. }

Note that a {\color{black}$d$}-dimensional multivariable density can be factorized into a number of different conditional pair-copulas based on the vine copula construction proposed by Bedford and Cooke \cite{bedford2002vines}. Except regular vine structure (R-vine), there are two special types of regular vines: canonical vine (C-vine) and drawable vine (D-vine).  For the C-vine, each tree has a unique node that is connected to all other nodes, and the corresponding joint pdf {\color{black}$f_X(\bm{x})$} is 
\begin{equation}
{\color{black}f_X(\bm{x}) = \prod_{k=1}^{d}f_k(x_k)\prod_{j=1}^{d-1}\prod_{i=1}^{d-j}c(F(x_j|x_1,...,x_{j-1}), F(x_{j+i}|x_1,...,x_{j-1}))}.
\end{equation}

In contrast,  each tree in a D-vine is a path and the corresponding joint pdf {\color{black}$f_X(\bm{x})$} is 
\begin{equation}
{\color{black}f_X(\bm{x}) = \prod_{k=1}^{d}f_k(x_k)\prod_{j=1}^{d-1}\prod_{i=1}^{d-j}c(F(x_i|x_{i+1},...,x_{i+j-1}), F(x_{i+j}|x_{i+1},...,x_{i+j-1}))}
\end{equation}
where the subscript indices indicate the conditional random variables to be drawn. 

Copula theory and vine copulas are an important tool for modeling the dependence of multivariate densities in either low or high dimension. A following critical question is how to select and estimate all components of a bivariate copula model or tree structure model from limited data. {\color{black} The paper mainly focuses on the bivariate copula model to show how to efficiently quantify the uncertainties associated with copula model selection and the corresponding parameters. The proposed method can be extended to high dimensional problem with dependence given a specified vine copula structure.} The next sections discuss this issue in detail. The next sections discuss this issue in detail.

\section{Statistical inference of copula dependence modeling}

Given a {\color{black}$d$}-dimensional probability density, we can decompose it into products of marginal densities and bivariate copula densities and represent this decomposition with a nested set of trees that fulfill a proximity condition. However, it is often difficult to directly identify a {\color{black}$d$}-dimensional probability density. Instead, more commonly, only data are provided and statistical inference is necessary for model selection and parameter estimation. Small data sets create additional uncertainties which pose a significant challenge to the inference of the copula dependence model. 

Assuming known marginal distributions, copula dependence modeling consists of three principal components: tree structure, copula {\color{black}form} and copula parameters. {\color{black} However, for small data sets, uncertainty uncertainty in the marginals cannot be ignored.} Consequently, the marginal {\color{black}form} and marginal distribution parameters must also be included in the inference process. As a result, the total uncertainty when inferring joint probability model form, $U_{all}$, includes the following five components: 
\begin{equation}
U_{all} =\left\{ U_{t}, U_{cf}, U_{cp}, U_{mf}, U_{mp} \right\}
\end{equation}
where $U_{t}$ is uncertainty in the tree structure, $U_{cf}$ and $U_{cp}$ are the uncertainty in copula families and parameters respectively, and $U_{mf} $ and $ U_{mp}$ represent the uncertainty in marginal distribution families and parameters. To quantify these uncertainties, statistical methods are adopted for model selection and parameter estimation. 

The model uncertainty in tree structure is particularly challenging to address. This is mainly because the possible decomposition of pair-copulas is potentially large, especially in high dimension. Typically, the tree structure is assumed to follow a specified model based on the analyst's knowledge or experience. There are several model selection approaches for specification of tree structures, including optimal C-vine structure selection \cite{czado2012maximum}, Bayesian approaches for D-vine selection \cite{min2011bayesian} and maximum spanning trees for R-vines \cite{dissmann2013selecting}. Here, the tree model selection is not our first priority, so we do not elaborate on these methods. Instead, {\color{black}our emphasis is on how} to efficiently quantify the uncertainties associated with copula {\color{black}form} selection and the corresponding parameters given a specified vine copula structure. 

\subsection{Copula {\color{black}form} selection and parameter estimation}

When a specific vine copula structure is determined, classical statistical approaches, including goodness-of-fit tests \cite{massey1951kolmogorov}, independence test \cite{mcdonald2009handbook} and AIC/BIC \cite{burnham2004multimodel} are capable of handling copula {\color{black}form} selection when data sets are large. When both tree structure and copula {\color{black}form} are known and the data set is large, the copula parameters can be estimated using sequential estimation \cite{aas2009pair, czado2012maximum}, maximum likelihood estimation \cite{stober2012there}, or Bayesian parameter estimation \cite{min2011bayesian, gruber2015sequential}. However, these classical approaches fall short when inferring from small data sets.

Traditionally, statistical inference is applied to select a single ``best" model given a set of candidate models and available data, and the model is the sole model used for probabilistic modeling. Any uncertainty associated with model selection is simply ignored. However, it is often difficult (even impossible) to identify a unique best model without significant (and potentially problematic) assumptions. Consequently, it is necessary to consider model uncertainty and compare the validity of multiple candidate models -- a process referred to as multimodel inference, as introduced by Burnham and Anderson \cite{burnham2004multimodel}. In this study, we generalize the Bayesian multimodel inference developed previously by the authors \cite{zhang2018quantification, zhang2018effect} to include uncertainty in the form and parameters of the copula dependence model. 

Given a data set $\bm{d}$, the model selection problem is to identify the model $M_i$ that``best" fits the data from a collection of $N$ candidate models {\color{black}$\bm{\mathbb{M}}=\{M_j\},\  j=1,\dots, N$}. The notion of best fit varies depending on the selected metric. In the Bayesian setting used here, initial model prior probabilities {\color{black}$\tilde{\pi}_j=p(M_j)$ with $\sum_{j=1}^{N}\tilde{\pi}_j=1$ are assigned to each model $M_j \in \bm{\mathbb{M}}$. According to Bayes' rule, the posterior model probability, given the data $\bm{d}$ can be calculated by
\begin{equation}
{\pi}_j = p(M_j| \bm{d}) = \frac{p(\bm{d}| M_j)p(M_j )}{\sum_{k=1}^{N} p(\bm{d} | M_k)p(M_k )},  \quad j=1,\dots, N \label{eq:Bayes_model}
\end{equation}
having $\sum_{j=1}^{N}{\pi}_j=1$ and where} 
\begin{equation}
p(\bm{d}|M_j)=\int_{\boldsymbol{\theta}_j}p(\bm{d}|\boldsymbol{\theta}_j,M_j)p(\boldsymbol{\theta}_j|M_j)d\boldsymbol{\theta}_j, \quad j=1,\dots,N
\label{eq:Model_evidence}
\end{equation}
is referred as to the marginal likelihood or evidence of model $M_j$.

Commonly, the model $M^* \in \bm{\mathbb{M}}$ with the highest posterior model probability $p(M^*|\bm{d})$ is selected as the single ``best" model. By contrast, Bayesian multimodel inference ranks the candidate models by their posterior model probabilities calculated by Eq. \eqref{eq:Bayes_model} and retains all plausible models with non-negligible probability. Once the plausible models and their associated model probabilities have been identified, model parameter uncertainties are assessed by applying Bayesian parameter estimation. For each model in the set of plausible models, {\color{black}$M_i, \ i= 1,\dots,N_d\ (N_d \le N) $}, we begin by assigning a prior (often a noninformative prior) to the model parameters $\bm{\theta}_i$, denoted ${ p }(\bm{\theta}_i|M_i)$. We then estimate the posterior parameter distribution using Bayes' rule: 
\begin{equation}
{ p }(\bm{\theta}_i|\bm{d},M_i)=\frac { p({ \bm{d}}|\bm{\theta}_i,M_i)p(\bm{\theta}_i|M_i) }{ p(\bm{d}|M_i) } \propto { p({ \bm{d}}|\bm{\theta}_i,M_i)p(\bm{\theta}_i|M_i) }, \quad i=1,\dots, m \label{eq:Bayesian_inference}
\end{equation}
where $p({ \bm{d}}|\bm{\theta}_i,M_i)$ is the likelihood function. The posterior ${ p }(\bm{\theta}_i|\bm{d},M_i)$ is identified implicitly through Markov Chain Monte Carlo (MCMC) without requiring the calculation of model evidence $p(\bm{d}|M_i)$. However, the evidence, as evident from Eq. \eqref{eq:Model_evidence} is critical in Bayesian multimodel inference and needs to be calculated with caution. A detailed discussion of the evidence calculation can be found in \cite{zhang2018effect}.

In the classical setting, a unique set of model parameters $\bm{\theta}_i$ is identified from the posterior samples using, for example, the maximum a posterior (MAP) estimator, 
\begin{equation}
\tilde{\bm{\theta}}_j^{\textup{MAP}}(\bm{d}, M_j)  = \underset {\bm{\theta}_j  }{ \arg \max } \ p(\bm{\theta}_j | \bm{d}, M_j) = \underset {\bm{\theta}_j  }{ \arg \max } \ {p(\bm{d} | \bm{\theta}_j, M_j) p(\bm{\theta}_j | M_j)}.  
\end{equation}

When $p(\bm{\theta}_j | M_j)$ is a noninformative prior, the MAP estimator is equivalent to the maximum likelihood estimate (MLE). Due to a lack of data, the posterior parameter probability will likely possess large variance. Rather than discarding the full uncertainty by selecting a single set of MLE or MAP {\color{black}parameters} or integrating out its variability using Bayesian model averaging \cite{sankararaman2011likelihood}, we retain the full posterior densities for each plausible model. 

\begin{figure}[!ht]    
    \centering
    \includegraphics[height=2in]{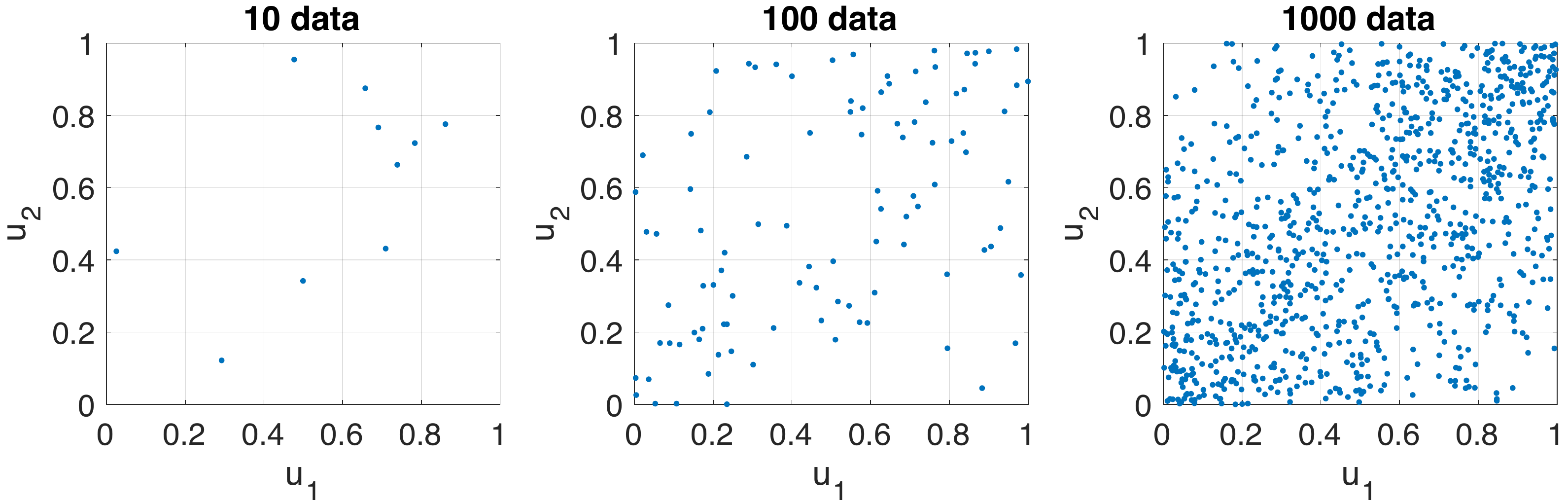}
    \caption{Bivariate correlated data drawn from \texttt{Frank}(3) copula model, showing 10 data, 100 data and 1000 data} \label{fig:frank_bivariate}
\end{figure}

In this work, the Bayesian multimodel inference method is generalized to address copula dependence model selection and parameter estimation. A simple bivariate example is used to illustrate the process and its performance. Consider a bivariate random vector $\bm{u} = [u_1, u_2]$ whose dependence follows the Frank copula model with parameter $\theta=3$ (denoted \texttt{Frank}(3)). Fig.\ \ref{fig:frank_bivariate} shows data sets of varying size drawn from the \texttt{Frank}(3) copula. Notice that, given only 10 data, one cannot decipher a clear dependence relation. Only after 100 data are drawn does the dependence begin to emerge and it finally becomes clear when 1000 points are drawn.

\begin{figure}[!ht]    
    \centering
    \includegraphics[width=3in]{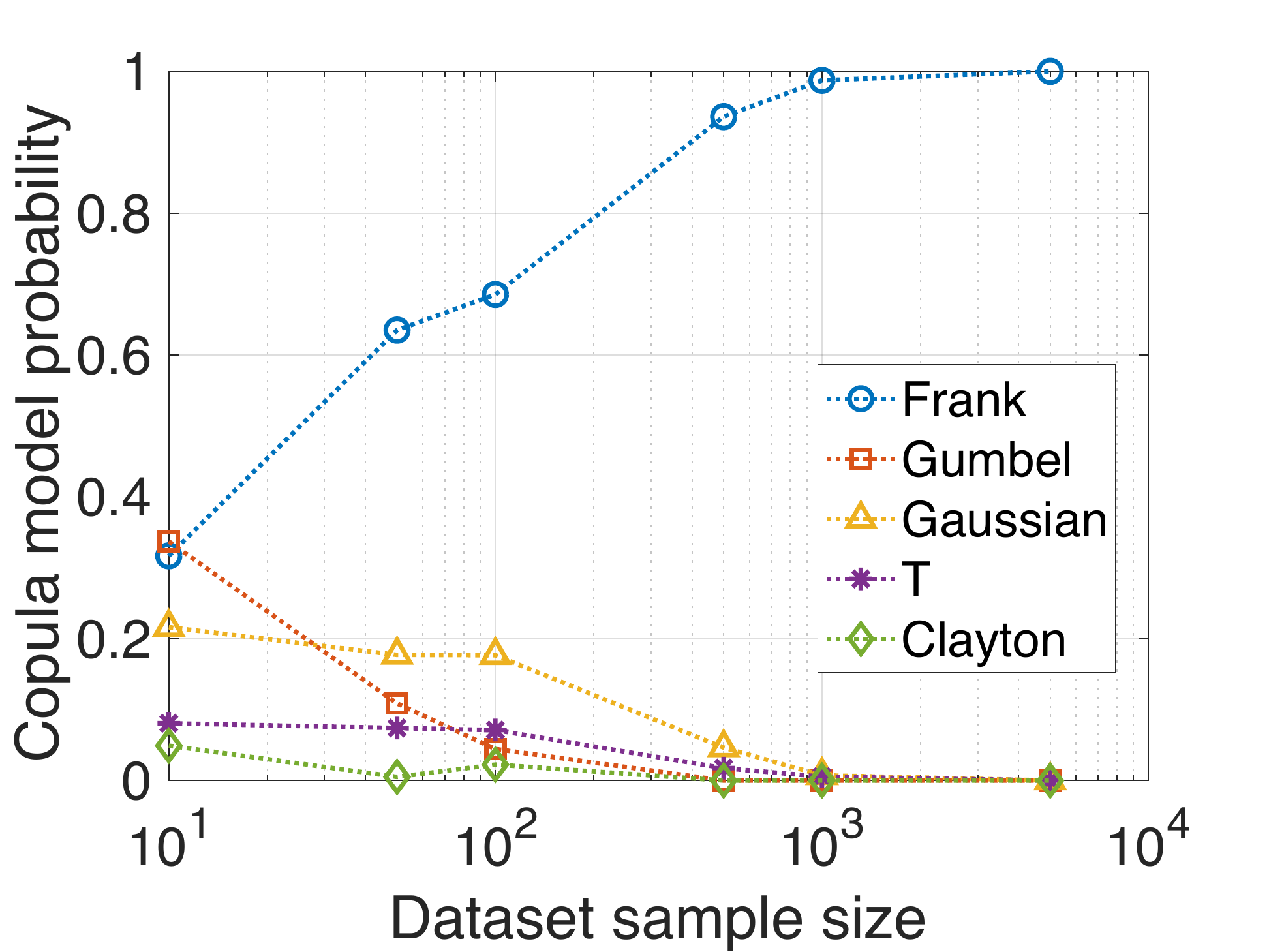}
    \caption{Posterior copula model probability as a function of dataset size} \label{fig:copula_model_probability}
\end{figure}

From these data, Bayesian multimodel inference is first used to quantify the copula {\color{black}form} uncertainty. Five copula models -- the Gaussian, Student-$t$, Clayton, Gumbel and Frank copulas -- are selected as the candidate copula {\color{black}forms}. Without informative prior information, all candidate copula models are assumed to have equal probability. The Monte Carlo method is adopted to compute the evidence from Eq. \eqref{eq:Model_evidence}. Then the posterior copula model probabilities are obtained using Eq. \eqref{eq:Bayes_model}. Fig.\ \ref{fig:copula_model_probability} shows the posterior probabilities for each candidate copula model as a function of dataset size. Notice that the model probability for the Frank copula becomes gradually larger as the data set size increases but the Bayesian multimodel inference does not select the correct Frank copula model conclusively until 1000 correlated data are collected. 

\begin{figure}[!ht]    
    \centering
    \includegraphics[width=3in]{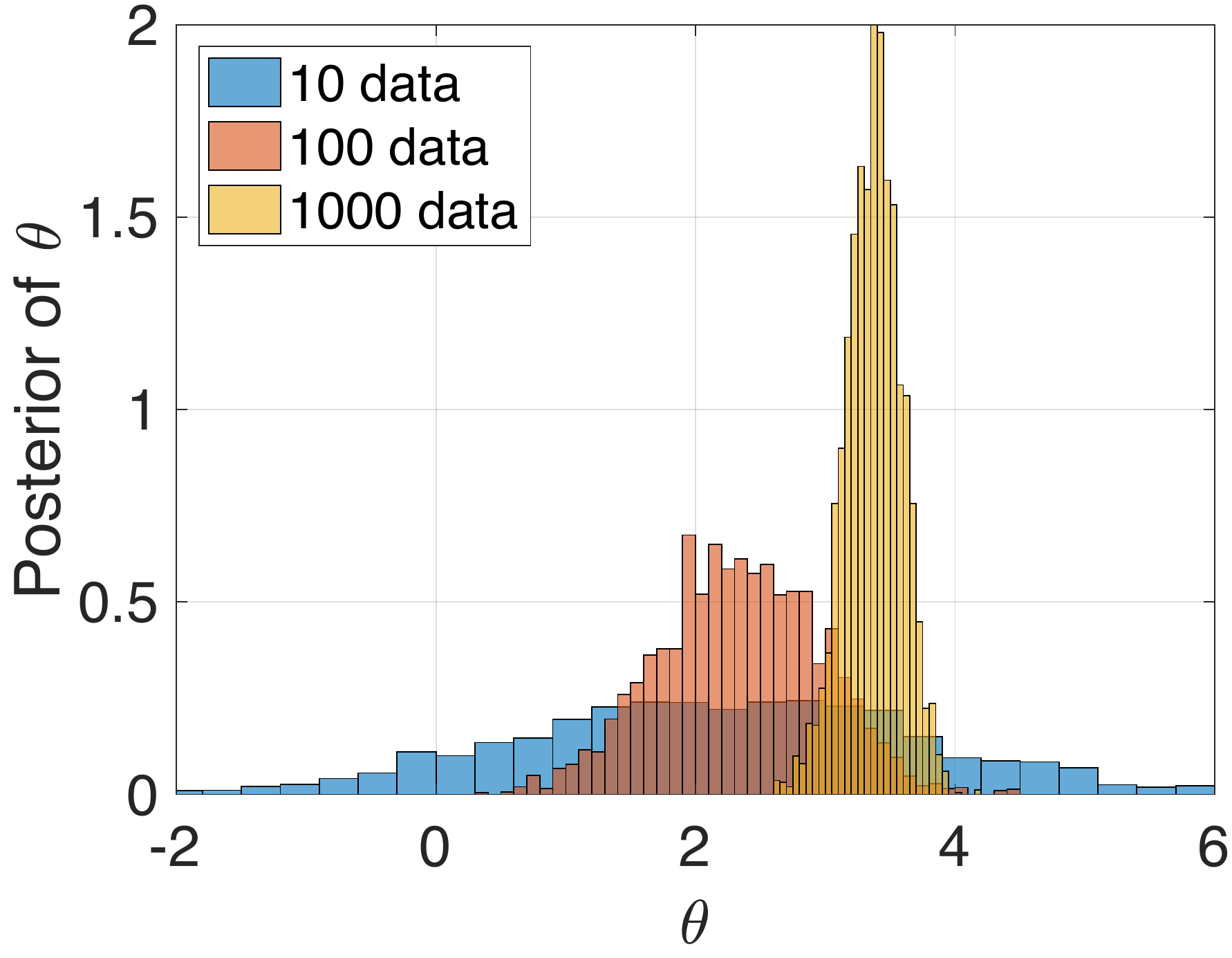}
    \caption{Posterior histogram of the Frank copula model parameter given different data set sizes.} \label{fig:copula_parameter}
\end{figure}

Next, Bayesian inference is employed to estimate the copula parameter for each plausible candidate model. Fig.\ \ref{fig:copula_parameter} shows the posterior probability distribution for the Frank copula parameter $\theta$ for increasing data set size. Note that the posterior variance is large when the data set size is small and the estimate gradually narrows with increasing data set size. Finally, the posterior density with 1000 data converges towards a narrow distribution that includes the true value ($\theta =3$). 

This simple example illustrates the Bayesian multimodel inference process for model selection and parameter estimation of copula dependence modeling. More specifically, it illustrates the fact that inference is inherently imprecise from small data sets. When data sets are small, it is impossible to uniquely identify the copula {\color{black}form} (and the associated copula model parameters) from which the data are drawn. In the following section, we turn our attention to uncertainty in the marginal distributions. 

\subsection{Uncertainty in marginal distributions}
\label{sec:multimodel_UQ}
As observed in authors' previous studies \cite{zhang2018quantification, zhang2018effect, zhang2019efficient}, uncertainty in the marginal distributions play a critical role in uncertainty quantification from small datasets. Consider again for simplicity, the bivariate case where the joint pdf can be expressed as:
\begin{equation}
f_{{\color{black}{X}}}(x_1, x_2) = c_{12}(F_1(x_1, \theta_1), F_2(x_2, \theta_2), \theta_c) \cdot f_1(x_1, \theta_1) \cdot f_2(x_2, \theta_2) \label{eq:bivariate_parameter}
\end{equation}
{\color{black}where $\theta_c$ are the copula parameters.} Given this expression of the joint density, it is clear that the copula model is conditional on the marginals and their parameters, which the previous studies have shown to have very large uncertainties when data sets are small. Consequently, it is necessary to identify copula model probabilities and copula parameter probabilities for each set of inferred candidate marginals. This induces a hierarchy of probabilities that includes both the copula model and {\color{black}the} marginal model. We therefore propose a hierarchical Bayesian multimodel inference method, as illustrated in Fig.\ \ref{fig:c4_hier}. The procedure is summarized {\color{black}for each pair of variables} as follows:
\begin{figure}[!ht]    
    \centering
    \includegraphics[width=7in]{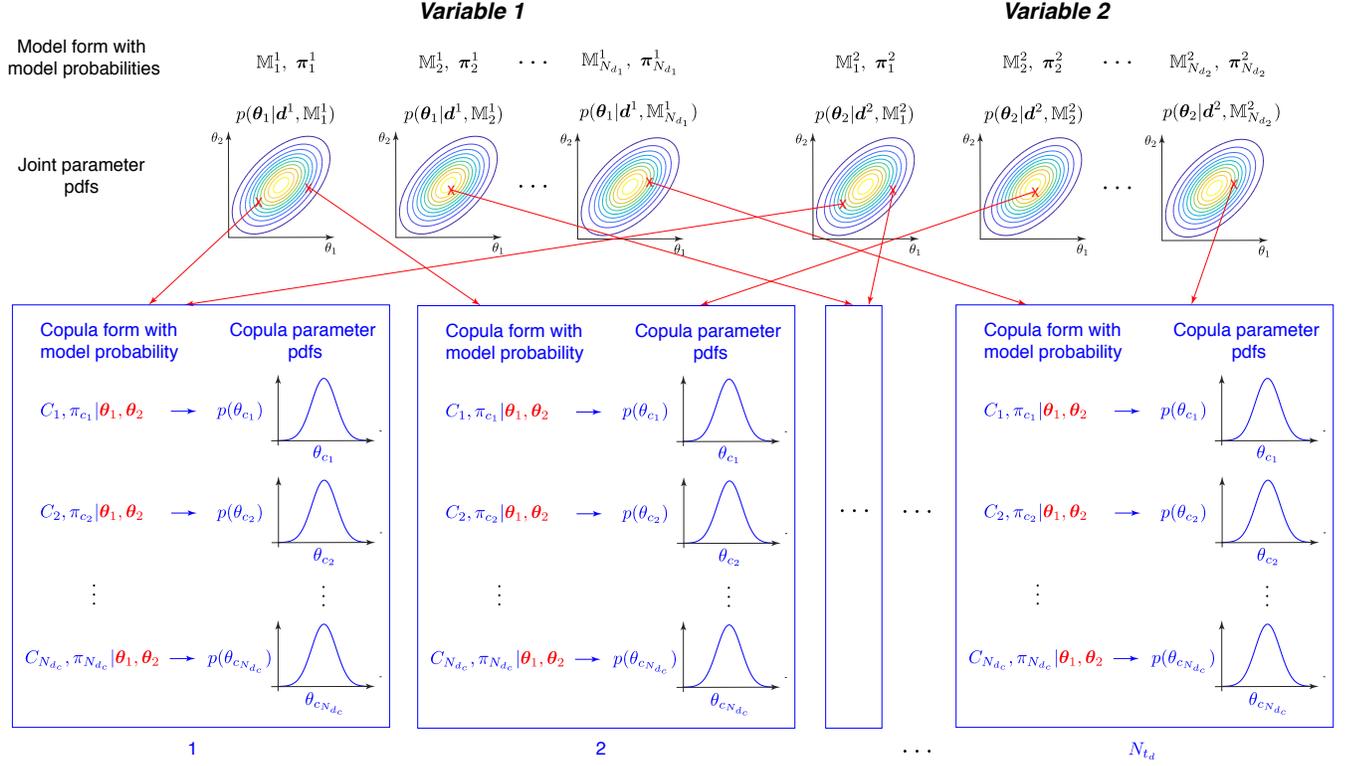}
    \caption{Hierarchy of Bayesian multimodel inference for copulas and marginals} \label{fig:c4_hier}
\end{figure}

\begin{itemize}
\item Step 1: Marginal multimodel inference -- First identify the candidate marginal model set{\color{black}s $\bm{\mathbb{M}}^1=\{M_j^1\}, j=1,\dots,N_{d_1}$ and $\bm{\mathbb{M}}^2=\{M_j^2\}, j=1,\dots,N_{d_2}$} for each variable and compute the marginal model probabilities ${\bm{\pi}}^1=\{\pi_1^1, \pi_2^1, \dots, \pi_{N_{d_1}}^1\}$ and ${\bm{\pi}}^2=\{\pi_1^2, \pi_2^2, \dots, \pi_{N_{d_2}}^2\}$ using Eq. \eqref{eq:Bayes_model}. {\color{black}Notice that this induces a set of $N_{d_1}\times N_{d_2}$ possible marginal pairs.} Then estimate the posterior joint pdf for the marginal parameters for all plausible models, ${ p }(\bm{\theta}_j^1|\bm{d}^1,M_j^1) $, {\color{black}$j = 1,\cdots, N_{d_1}$} and ${ p }(\bm{\theta}_j^2|\bm{d}^2,M_j^2) $, {\color{black}$j = 1,\cdots, N_{d_2}$} using Eq. \eqref{eq:Bayesian_inference}. 
\item Step 2: Define a finite set of marginal distributions -- Theoretically, the above process yields an infinite set of  parameterized probability models. Practically, it is necessary to reduce this to a finite but statistically representative set of {\color{black}$N_{t_d}$ marginal probability model pairs}. This is achieved by randomly selecting a model family for each variable from {\color{black}$\bm{\mathbb{M}}^1$ and $\bm{\mathbb{M}}^2$} with probabilities ${\bm{\pi}}^1$ and ${\bm{\pi}}^2$ respectively, and randomly selecting the parameters of each model from the appropriate posterior joint pdf {\color{black}${ p }(\bm{\theta}_1|\bm{d}^1,M_j^1) $ and ${ p }(\bm{\theta}_2|\bm{d}^2,M_k^2) $}. 
\item Step 3: Copula multimodel inference -- For each pair of marginal distributions {\color{black} $f_1(x_1|\bm{\theta}_1,M_j^1)$ and $f_2(x_2|\bm{\theta}_2,M_k^2)$}, standardize the data using {\color{black}$F_1(\bm{d}^1)$ and $F_2(\bm{d}^2)$}.  Compute the posterior copula model probabilities {\color{black} ${\bm{\pi}}_{c} =\left\{ {\pi}_{c_1},\cdots,{\pi}_{c_{N_{d_c}}}  \right\}$} for each candidate copula model {\color{black}$\left\{C_1, \cdots, C_{N_{d_c}} \right\}$ using Eq. \eqref{eq:Bayes_model} where $N_{d_c}$ is the number of plausible copula models for the specified marginal pair}. Next, estimate the posterior pdf for the copula parameters for each plausible copula model, $p(\bm{\theta}_{c_k} | \bm{d}, C_k)$, {\color{black}$k = 1,\dots,N_{d_c}$} using Eq. \eqref{eq:Bayesian_inference}. As in step 2, a finite set of {\color{black}$N_{t_c}$ ($N_{t_c}$ can be arbitrarily large) copulas (copula models and parameters)} are determined for each marginal pair {\color{black} $\left\{ f_1(x_1|\bm{\theta}_1,M_j^1), f_2(x_2|\bm{\theta}_2,M_k^2) \right\}$}. 
\item Step 4: Identify bivariate joint densities -- Combine the set of marginal densities and copula densities to define the full set of candidate joint densities {\color{black}$f_X(x_1, x_2)$}, as in Eq. \eqref{eq:bivariate_parameter}. This, however, may lead to a prohibitively large number, {\color{black}$N_{t_d} \times N_{t_c}$}, of candidate bivariate densities. In the following section, we discuss a strategy to keep this number tractable. 

\end{itemize} 

The result is a set of {\color{black}$N_{t_d} \times N_{t_c}$} joint distributions that are representative of the uncertainty in marginal model form, marginal parameters, {\color{black}copula} model form, and copula parameters. We now consider how to propagate this set of joint distributions through a computational model. Note that the cost of propagation depends only weakly on {\color{black}$N_{t_d} \times N_{t_c}$}, the number of joint densities in the set. That is, increasing {\color{black}$N_{t_d} \times N_{t_c}$} does not increase the number of model evaluations necessary for uncertainty propagation. Therefore, it is advantageous to make {\color{black}$N_{t_d} \times N_{t_c}$} as large as possible{\color{black}, as undersampling it will result in artificially narrow uncertainty bounds}.


\section{Uncertainty propagation with copula dependence modeling}

In the previous study \cite{zhang2018quantification}, we proposed an efficient algorithm for propagation of the imprecise probabilities characterized by a multimodel set with independent marginals. Here, we extend this algorithm to the propagation of imprecise probabilities with copula dependence modeling. For illustration, and without loss of generality, we derive here the propagation method for bivariate random variables. It's extension to higher-dimensional vectors with copula dependence, particularly vine copulas that rely on a series of bivariate copulas, follows naturally.

\subsection{Importance sampling for bivariate joint probability density}
Consider the performance function $g(\bm{X}_1, \bm{X}_2)$ defining the response quantity of interest for a mathematical or physical system. The aim of uncertainty propagation is to evaluate the expectation $E(g(\bm{X}_1, \bm{X}_2))$ where $(\bm{X}_1, \bm{X}_2) \in \Omega$ is a random vector having bivariate joint probability density $p(\bm{x}_1, \bm{x}_2)$. The classical Monte Carlo estimator is computed as follows:
{\color{black}
\begin{equation}
\mu = E_p[g(\bm{X}_1, \bm{X}_2)] = \int_{\Omega}g(\bm{x}_1, \bm{x}_2)p(\bm{x}_1, \bm{x}_2)d\mathbf{x} \approx \frac{1}{n} \sum_{i=1}^n g(\bm{x}_1^i, \bm{x}_2^i) \label{eq:bivariate_MC}
\end{equation}}
where $E_p[\cdot]$ is the expectation with respect to $p(\cdot)$ and $(\bm{x}_1^i, \bm{x}_2^i)$ are bivariate random samples drawn from $p(\bm{x}_1, \bm{x}_2)$. Importance sampling allows samples to be drawn from an alternate density $q(\bm{x}_1, \bm{x}_2)$ and then reweights the samples to obtain the estimator. The Monte Carlo estimator in Eq.\ (\ref{eq:bivariate_MC}) is modified as:
{\color{black}
\begin{equation}
\begin{aligned}
\mu = E_q\left[ g(\bm{X}_1, \bm{X}_2) \frac{p(\bm{X}_1, \bm{X}_2)}{q(\bm{X}_1, \bm{X}_2)} \right] &= \int_{\Omega}g(\bm{x}_1, \bm{x}_2) \frac{p(\bm{x}_1, \bm{x}_2)} {q(\bm{x}_1, \bm{x}_2)} q(\bm{x}_1, \bm{x}_2) d\mathbf{x} \\
& \approx \frac{1}{n} \sum_{i=1}^n g(\bm{x}_1^i, \bm{x}_2^i) w(\bm{x}_1^i, \bm{x}_2^i) 
\label{eq:bivariate_IS}
\end{aligned}
\end{equation}}
where $E_q[\cdot]$ denotes expectation for $(\bm{X}_1, \bm{X}_2) \sim q(\cdot)$ and the importance weights are defined as:
\begin{equation}
w(\bm{x}_1^i, \bm{x}_2^i) = \frac{p(\bm{x}_1^i, \bm{x}_2^i)}{q(\bm{x}_1^i, \bm{x}_2^i)}. 
\label{eq:weights_bi}
\end{equation}

{\color{black} 
\subsection{Optimal important density for bivariate joint probability density with copula dependence: Derivation}
}

The efficient propagation of multimodel imprecise probabilities is performed by identifying an ``optimal'' importance sampling density, propagating this optimal density, and reweighting the samples according to each distribution in the multimodel set. The optimal sampling density is derived as the distribution that ``best'' matches the multimodel distribution set according to some metric. In the prior work, the authors \cite{zhang2018quantification} derive an explicit analytical optimal importance sampling density given an ensemble of target marginal probability densities that minimizes the total expected mean square difference, {\color{black}{$\mathcal{M}(\mathbb{M}\parallel Q)$, between the model set $\mathbb{M}=\{M_j\}, j=1,\dots,N_d$ and the importance sampling density $Q=q(\bm{x})$ given by:
\begin{equation}
\mathcal{E} = \sum_{j=1}^{{N_d}}E_{\theta}\left[{\mathcal{M}}(M_j  \parallel Q)\right] = {E_{\theta} }\left[ \sum _{ j=1 }^{ {N_d}}\frac{1}{2} \int{  { \left( { p_j(\bm{x} | \bm{\theta} ) } - { q(\bm{x}) }  \right) ^{ 2 }} d\bm{x}} \right],
\label{eq: total_EMSD} 
\end{equation} 
}}

In other words, the following optimization problem is solved:
\begin{equation}
\begin{aligned}
& \underset{q}{\text{minimize}}
& &{\mathcal{L}}(q)=E_{\theta}\left [ \int{{{\mathcal{F}} }(\bm{x}, \bm{\theta}, q(\bm{x}))}d\bm{x} \right] \\
& \text{subject to}
& &{\mathcal{I}}(q) =  \int{q(\bm{x})d\bm{x}}-1=0 \label{eq: opt_EMSD}
\end{aligned}
\end{equation}
where the action functional $\mathcal{F}(\cdot)$ is \textcolor{black}{the total square differences}:
\begin{equation}
{\mathcal{F}(\bm{x}, \bm{\theta}, q(\bm{x}))}={  \frac { 1 }{ 2 } \sum_{j=1}^{{N_d}}{ \left( { p_j(\bm{x} | \bm{\theta}) } - { q(\bm{x}) }  \right) ^{ 2 }}   } \label{eq:MSD_funcitonal} 
\end{equation}
\textcolor{black}{ and $E_{\theta}$ is the expectation with respect to the posterior probability of the model parameters $\bm{\theta}$.} ${\mathcal{I}}(q)$ ensures that $q(\bm{x})$ is a valid pdf. Solving this optimization problem yields a closed-form solution given by the convex mixture model \cite{zhang2018quantification}
\begin{equation}
{q}^*(\bm{x}) ={\color{black}\frac{1}{{N_d}}  \sum_{j=1}^{{{N_d}}}}E_{\theta}\left[{p_j({\bm{x} | \bm{\theta}})}\right] \label{eq: opt_MSD2}
\end{equation}
When the posterior model probabilities are not equal, this solution generalizes as 
\begin{equation}
{q}^*(\bm{x}) = \sum_{j=1}^{{\color{black}{N_d}}}{\pi}_j E_{\theta} \left [ {  p_j(\bm{x}|\bm{\theta})} \right]  \label{eq: opt_MSD3}
\end{equation}
where each term is weighted by the corresponding posterior model probabilities $ {\color{black}{\pi}_j}$ computed by Eq.\eqref{eq:Bayes_model}. The interested reader can find more details in \cite{zhang2018quantification}. 

It is straightforward to generalize this solution from the one-dimensional probability density to multivariate joint probability densities. If the bivariate joint probability density has independent marginals, the optimal sampling density is expressed as:
\begin{equation}
{q}^*(\bm{x}) =\frac{1}{N_{d_1} N_{d_2}}\sum_{i=1}^{N_{d_1}} \sum_{j=1}^{N_{d_2}} E_{\theta}\left[{p_{ij}({\bm{x} | \bm{\theta}})}\right] \label{eq: opt_MSD_2d}
\end{equation}
and the bivariate joint probability density ${p_{ij}({\bm{x} | \bm{\theta}})}$ can be decomposed by marginal distribution $f_1^i (\bm{x}_1|\bm{\theta}_1) $ and $f_2^j(\bm{x}_2|\bm{\theta}_2)$ as follows:
\begin{equation}
{p_{ij}({\bm{x} | \bm{\theta}})}  = f_1^i (\bm{x}_1|\bm{\theta}_1) \cdot f_2^j(\bm{x}_2|\bm{\theta}_2) \label{eq: opt_MSD_2d_1}
\end{equation}
where $N_{d_1}$ and $N_{d_2}$ are the number of candidate probability models for the marginal densities respectively and $N_d= N_{d_1}\cdot N_{d_2}$ is the total number of candidate probability models for the bivariate joint probability density. Thus, the optimal sampling density for independent bivariate joint density can be expanded in terms of the margainals as:
\begin{equation}
\begin{aligned}
{q}^*(\bm{x}) &= \frac{1}{N_{d_1} N_{d_2}}\sum_{i=1}^{N_{d_1}} \sum_{j=1}^{N_{d_2}} E_{\theta}\left[f_1^i (\bm{x}_1|\bm{\theta}_1) f_2^j(\bm{x}_2|\bm{\theta}_2) \right] \\
& = \frac{1}{N_{d_1} N_{d_2}}\sum_{i=1}^{N_{d_1}} \sum_{j=1}^{N_{d_2}} E_{\theta_1}\left[f_1^i (\bm{x}_1|\bm{\theta}_1)  \right] E_{\theta_2}\left[f_2^j(\bm{x}_2|\bm{\theta}_2) \right] \\
&=  \frac{1}{N_{d_1} N_{d_2}}\sum_{i=1}^{N_{d_1}}E_{\theta_1}\left[f_1^i (\bm{x}_1|\bm{\theta}_1)  \right]  \sum_{j=1}^{N_{d_2}} E_{\theta_2}\left[f_2^j(\bm{x}_2|\bm{\theta}_2) \right]
\label{eq: opt_MSD_2d_2}
\end{aligned}
\end{equation}
Again, it is straightforward to show that this solution generalizes for unequal model probabilities as:
\begin{equation}
{\color{black}{q}^*(\bm{x}) =\sum_{i=1}^{N_{d_1}}\pi_i^1 E_{\theta_1}\left[f_1^i (\bm{x}_1|\bm{\theta}_1)  \right]  \sum_{j=1}^{N_{d_2}} \pi_j^2 E_{\theta_2}\left[f_2^j(\bm{x}_2|\bm{\theta}_2) \right]}
\label{eq: opt_MSD_2d_3}
\end{equation}
where {\color{black}$\pi_i^1$} associated with marginal density {\color{black}$f_1^i (\bm{x}_1|\bm{\theta}_1)$} is the posterior model probability for model $M_i$ satisfying {\color{black}$\sum_{i=1}^{N_{d_1}}\pi_i^1 =1$} and {\color{black}$\pi_j^2$} associated with marginal density {\color{black}$f_2^j (\bm{x}_2|\bm{\theta}_2)$} is the posterior model probability for model $M_j$ satisfying {\color{black}$\sum_{j=1}^{N_{d_2}} \pi_j^2=1$} . 
 
If the bivariate joint probability density has copula dependence, with copula density {\color{black}$c_{12}^k(F_1(\bm{x}_1|\bm{\theta}_1), F_2(\bm{x}_2|\bm{\theta}_2) |\bm{\theta}_c)$}, we can express the bivariate joint probability density as:
\begin{equation}
{p_{ij}^k({\bm{x} | \bm{\theta}})}  ={\color{black}c_{12}^k(F_1^i(\bm{x}_1|\bm{\theta}_1), F_2^j(\bm{x}_2|\bm{\theta}_2) |\bm{\theta}_c)} \cdot f_1^i (\bm{x}_1|\bm{\theta}_1) \cdot f_2^j(\bm{x}_2|\bm{\theta}_2) \label{eq: opt_MSD_2d_copula}
\end{equation}
where $k = 1,...,N_{d_c}$ indexes the candidate copula models. Similarly, we can derive the optimal sampling density for dependent bivariate joint probability density with copula dependence as follows. {\color{black}We start by applying the joint density in Eq.\ \eqref{eq: opt_MSD_2d_copula} to the optimal density in Eq.\ \eqref{eq: opt_MSD_2d} where we require an additional summation over all $N_{d_c}$ candidate copula models:
\begin{equation}
{q}_c^*(\bm{x}) = \frac{1}{N_{d_1} N_{d_2} N_{d_c}}\sum_{i=1}^{N_{d_1}} \sum_{j=1}^{N_{d_2}} \sum_{k=1}^{N_{d_c}}  E_{\theta}\left[c_{12}^k(F_1^i(\bm{x}_1|\bm{\theta}_1), F_2^j(\bm{x}_2|\bm{\theta}_2) |\bm{\theta}_c) \cdot f_1^i (\bm{x}_1|\bm{\theta}_1) \cdot f_2^j(\bm{x}_2|\bm{\theta}_2) \right]. 
\label{eq: opt_MSD_2d_copula1}
\end{equation}
Next, let us apply the law of total expectation as:
\begin{equation}
    E[X] = E[E[X|Y]] = \int_Y E[X|Y=y]p(y)dy
\end{equation}
where
\begin{equation}
    X = c_{12}^k(F_1^i(\bm{x}_1|\bm{\theta}_1), F_2^j(\bm{x}_2|\bm{\theta}_2) |\bm{\theta}_c) \cdot f_1^i (\bm{x}_1|\bm{\theta}_1) \cdot f_2^j(\bm{x}_2|\bm{\theta}_2)
\end{equation}
and $Y=y$ is the condition that $\bm{\theta}_1$ and $\bm{\theta}_2$ take specific values, i.e.\
\begin{equation}
    \bm{\theta}_1=\theta_n, \text{ and } \bm{\theta}_2=\theta_m.
\end{equation}
Applying the law of total expectation, the summand in Eq.\ \eqref{eq: opt_MSD_2d_copula1} can be expressed as
\begin{multline}
    \int_{\bm{\theta}_1} \int_{\bm{\theta}_2} E_{\theta}\left[c_{12}^k(F_1^i(\bm{x}_1|\bm{\theta}_1), F_2^j(\bm{x}_2|\bm{\theta}_2) |\bm{\theta}_c, \bm{\theta}_1=\theta_n, \bm{\theta}_2=\theta_m) \cdot f_1^i (\bm{x}_1|\bm{\theta}_1 = \theta_n)  \cdot f_2^j(\bm{x}_2|\bm{\theta}_2 = \theta_m) \right] \cdot \\  p(\bm{\theta_1} = \theta_n, \bm{\theta}_m = \theta_2) d\theta_n d\theta_m
    \label{eqn:exp1-1}.
\end{multline}
Recognizing that the first term is conditioned on $\bm{\theta}_1$ and $\bm{\theta_2}$ taking specific values, the expectation can be written entirely with respect to $\bm{\theta}_c$ and the marginal densities can be taken outside the expectation. We further recognize that $p(\bm{\theta_1} = \theta_n, \bm{\theta}_2 = \theta_m) = { p }(\bm{\theta}_1 = \theta_n|\bm{d},M_i) \cdot { p }(\bm{\theta}_2 = \theta_m|\bm{d},M_j)$ because $\bm{\theta}_1$ and $\bm{\theta}_2$ are independent and inferred from the data for each variable. Hence, Eq.\ \eqref{eqn:exp1-1} becomes:
\begin{multline}
    \int_{\bm{\theta}_1} \int_{\bm{\theta}_2} E_{\theta_c}\left[c_{12}^k(F_1(\bm{x}_1|\bm{\theta}_1), F_2(\bm{x}_2|\bm{\theta}_2) |\bm{\theta}_c, \bm{\theta}_1=\theta_n, \bm{\theta}_2=\theta_m)\right] \cdot f_1^i (\bm{x}_1|\bm{\theta}_1 = \theta_n)  \cdot f_2^j(\bm{x}_2|\bm{\theta}_2 = \theta_m) \cdot \\  { p }(\bm{\theta}_1 = \theta_n|\bm{d},M_i) \cdot { p }(\bm{\theta}_2 = \theta_m|\bm{d},M_j) d\theta_n d\theta_m. 
    \label{eqn:exp2-1}
\end{multline}
Plugging this into Eq.\ \eqref{eq: opt_MSD_2d_copula1} and letting
\begin{equation}
    \hat{c}_{12}^{mn}(F_1^i(\bm{x}_1|\bm{\theta}_1), F_2^j(\bm{x}_2|\bm{\theta}_2)) = \dfrac{1}{N_{d_c}}\sum_{k=1}^{N_{d_c}} E_{\theta_c}\left[c_{12}^k(F_1^i(\bm{x}_1|\bm{\theta}_1), F_2^j(\bm{x}_2|\bm{\theta}_2) |\bm{\theta}_c, \bm{\theta}_1=\theta_n, \bm{\theta}_2=\theta_m)\right]
    \label{eqn:exp3-1}
\end{equation} 
be the expected conditional copula for marginal parameter pair $(\bm{\theta}_1=\theta_n, \bm{\theta}_2=\theta_m)$ gives:
\begin{multline}
{q}_c^*(\bm{x}) = \frac{1}{N_{d_1} N_{d_2}}\sum_{i=1}^{N_{d_1}} \sum_{j=1}^{N_{d_2}}  \int_{\bm{\theta}_1}  \int_{\bm{\theta}_2} \hat{c}_{12}^{mn}(F_1^i(\bm{x}_1|\bm{\theta}_1), F_2^j(\bm{x}_2|\bm{\theta}_2)) \cdot f_1^i (\bm{x}_1|\bm{\theta}_1 = \theta_n)  \cdot f_2^j(\bm{x}_2|\bm{\theta}_2 = \theta_m) \cdot \\  { p }(\bm{\theta}_1 = \theta_n|\bm{d},M_i) \cdot { p }(\bm{\theta}_2 = \theta_m|\bm{d},M_j) d\theta_n d\theta_m. 
\label{eqn:exp4-1}
\end{multline}

Next, recognizing that we likely cannot know ${ p }(\bm{\theta}_1 = \theta_n|\bm{d},M_i)$ and ${ p }(\bm{\theta}_2 = \theta_m|\bm{d},M_j)$ explicitly because we do not have the parameter posterior density in closed form (instead, we have sampled it from MCMC), we will rely on Monte Carlo estimation of the integrals over $\theta_n$, $\theta_m$ with $N_{n}\times N_m\to\infty$ samples such that $\theta_n$ and $\theta_m$ are drawn randomly from the posterior parameter density (i.e.\ from MCMC samples) and allowing us to express the optimal density as
\begin{multline}
{q}_c^*(\bm{x}) = \frac{1}{N_{d_1} N_{d_2} N_n N_m}\sum_{i=1}^{N_{d_1}} \sum_{j=1}^{N_{d_2}}  \sum_{n=1}^{N_n} \sum_{m=1}^{N_m} \hat{c}_{12}^{mn}(F_1^i(\bm{x}_1|\bm{\theta}_1), F_2^j(\bm{x}_2|\bm{\theta}_2)) \cdot f_1^i (\bm{x}_1|\bm{\theta}_1 = \theta_n)  \cdot f_2^j(\bm{x}_2|\bm{\theta}_2 = \theta_m). 
\label{eqn:exp5}
\end{multline}

The optimal sampling density in Eq.\ \eqref{eqn:exp5} can be generalized to account for the posterior model probabilities as follows:
\begin{equation}
{q}_c^*(\bm{x}) = \frac{1}{N_{n}N_{m}}\sum_{i=1}^{N_{d_1}} \sum_{j=1}^{N_{d_2}}  \sum_{n=1}^{N_n} \sum_{m=1}^{N_m} \hat{c}_{12}^{mn}(F_1^i(\bm{x}_1|\bm{\theta}_1), F_2^j(\bm{x}_2|\bm{\theta}_2)) \cdot \pi_i^1 f_1^i (\bm{x}_1|\bm{\theta}_1 = \theta_n)  \cdot \pi_j^2 f_2^j(\bm{x}_2|\bm{\theta}_2 = \theta_m)  
\label{eqn:exp6}
\end{equation}
where the expected conditional copula $\hat{c}_{12}^{mn}(F_1^i(\bm{x}_1|\bm{\theta}_1), F_2^j(\bm{x}_2|\bm{\theta}_2))$ in Eq.\ \eqref{eqn:exp3-1} is replaced by:
\begin{equation}
    \hat{c}_{12}^{mn}(F_1^i(\bm{x}_1|\bm{\theta}_1), F_2^j(\bm{x}_2|\bm{\theta}_2)) = \sum_{k=1}^{N_{d_c}} \pi_c^{k,mn}E_{\theta_c}\left[c_{12}^k(F_1^i(\bm{x}_1|\bm{\theta}_1), F_2^j(\bm{x}_2|\bm{\theta}_2) |\bm{\theta}_c, \bm{\theta}_1=\theta_n, \bm{\theta}_2=\theta_m)\right]
    \label{eqn:exp7}
\end{equation} 
where $\pi_c^{k,mn}$ is the posterior copula model probability conditioned on $\bm{\theta}_1=\theta_n$ and $\bm{\theta}_2=\theta_m$. \\

{\color{black} 
\subsection{Optimal important density for bivariate joint probability density with copula dependence: Implementation}

In the derived form, the optimal sampling density in Eqs.\ \eqref{eqn:exp6} and \eqref{eqn:exp7} is difficult to implement, involving several nested loops. For every pair of marginals $\{f_1^i(\cdot),f_2^j(\cdot)\}$, we need to randomly sample $N_n$ and $N_m$ samples respectively from the parameter densities using MCMC. Then, for each pair of the $N_n\times N_m$ model parameters, we need $N_{\theta_c}$ samples of the copula parameters for each of the $N_{d_c}$ candidate copula models for a total computational complexity of $N_{d_1}\times N_{d_2} \times N_n \times N_m \times N_{d_c} \times N_{\theta_c}$. Here, we propose a Monte Carlo sampling approach to reduce the complexity of this calculation.

This is performed by first populating the marginal sets. That is, we perform the multimodel selection process for the marginal distributions to obtain $\mathbb{M}^1$ and $\mathbb{M}^2$ and the model probabilities $\bm{\pi}^1$ and $\bm{\pi}^2$. Next we, perform Bayesian parameter estimation for each of the marginals in $\mathbb{M}^1$ and $\mathbb{M}^2$, which provides a set of $N_m$ and $N_n$ parameter values following the joint parameter distributions of each model $M_i^1$ and $M_i^2$, respectively. Next, instead of combining all combinations of marginals and parameters ($N_{d_1}\times N_{d_2} \times N_n \times N_m$), we set a feasible value $N_{t_d}$ of total marginal combinations to be considered. Note that while the total number of combinations is likely to be in the millions, e.g.\ $4\times 4\times 1000\times 1000 = 16,000,000$, we generally select $N_{t_d}\approx 1,000$. This set of $N_{t_d}$ probability models is selected by randomly drawing marginals from $\mathbb{M}^1$ and $\mathbb{M}^2$ with probabilities $\bm{\pi}^1$ and $\bm{\pi}^2$ and then randomly drawing their respective parameters from the MCMC samples for each marginal. 

This first simplification reduces the estimator in Eq.\ \eqref{eqn:exp6} to the following form:
\begin{equation}
    q_c^*(\bm{x}) = \dfrac{1}{N_{t_d}}\sum_{l=1}^{N_{t_d}} \hat{c}_{12}^l(F_1^l(\bm{x}_1|\bm{\theta}_1), F_2^l(\bm{x}_2|\bm{\theta}_2))\cdot f_1^l(\bm{x}_1|\bm{\theta}_1=\theta_1^l) \cdot f_2^l(\bm{x}_2|\bm{\theta}_2=\theta_2^l)
    \label{eqn:exp8}
\end{equation}
where $l$ is a single index associated with a pair of marginals randomly selected according to their model probabilities as well as random parameters for each of these marginals selected from their joint posterior pdf.

For each of the $N_{t_d}$ marginal pairs, we perform copula model selection to obtain the copula model probabilities $\pi_c^{l}$ and then, again perform MCMC to obtain samples of the copula parameters following their posterior distribution. To estimate the expected conditional copula, we again reduce the samples from $N_{d_c}\times N_{\theta_c}$, which might be on the order of 10,000, to a smaller number $N_{t_c}$ ($\approx 500$). We estimate Eq.\ \eqref{eqn:exp7} by randomly drawing $N_{t_c}$ copula models according to $\pi_c^{l}$ and randomly drawing the parameter values from the MCMC samples for that model obtained during Bayesian inference. Procedurally, Eq.\ \eqref{eqn:exp7} is re-expressed in the following form for use in Eq.\ \eqref{eqn:exp8}:
\begin{equation}
    \hat{c}_{12}^{l}(F_1^l(\bm{x}_1|\bm{\theta}_1), F_2^l(\bm{x}_2|\bm{\theta}_2)) = \dfrac{1}{N_{t_c}}\sum_{k=1}^{N_{t_c}} c_{12}^k(F_1^l(\bm{x}_1|\bm{\theta}_1), F_2^l(\bm{x}_2|\bm{\theta}_2) |\bm{\theta}_c^k, \bm{\theta}_1=\theta_1^l, \bm{\theta}_2=\theta_2^l)
    \label{eqn:exp9}
\end{equation}
where the superscript $k$ in $c_{12}^k$ denotes that the form of the model for the $k^{th}$ copula is random and follows the model probabilities $\pi_c^l$, while superscript $k$ in $\bm{\theta}_c^k$ denotes that the copula parameters are randomly drawn from the posterior parameter density associated with copula model $c_{12}^k(\cdot)$. 

Eqs.\ \eqref{eqn:exp8} and \eqref{eqn:exp9} are then actually used for optimal sampling density estimation. Overall, this reduces the complexity of the optimal sampling density estimation from $N_{d_1}\times N_{d_2} \times N_n \times N_m \times N_{d_c} \times N_{\theta_c} \sim \mathcal{O}(10^{11}-10^{12})$ to $N_{t_d}\times N_{t_c} \sim \mathcal{O}(10^5-10^6)$, while retaining a statistically representative set of joint probability models from which to estimate the optimal.

We further emphasize here that calculation of the optimal sampling density is generally much less expensive than evaluation of the computational model through which uncertainties are being propagated. Nonetheless, the optimal sampling density must be called for every sample re-weighting, which can lead to additional computational burden. One simple way to alleviate this burden is to compute the optimal joint density once via the approach described above and develop an inexpensive surrogate or lookup table to call it rapidly.

}

The implementation procedure for copula-based optimal sampling density estimation is summarized as Algorithm 1. 
\begin{algorithm} 
{\color{black}
\caption{Copula-based optimal sampling density}
\begin{algorithmic}[1]
\label{algo:copula}
\State Identify the marginal models, $\mathbb{M}^1$ and $\mathbb{M}^2$, and their model probabilities, $\bm{\pi}^1$ and $\bm{\pi}^2$, using Bayesian multimodel inference.
\State Perform Bayesian parameter estimation using MCMC to obtain sample parameters following the posterior parameter density, ${ p }(\bm{\theta}_i|\bm{d},M_i)$ for each marginal model
\State Randomly select a pair of marginals $\left\{ f_1^i(\bm{x}_1 | \bm{\theta}_1 = \theta_n), f_2^i(\bm{x}_2 | \bm{\theta}_2 = \theta_m) \right\}$ by drawing the marginal models with probabilities $\bm{\pi}^1$ and $\bm{\pi}^2$ and randomly drawing the parameters from the MCMC samples of the posterior parameter density. \label{step:2}
\State Identify the candidate copula models and their model probabilities $\pi_c$ for the specific marginal pair using Bayesian multimodel inference.
\State Perform Bayesian parameter estimation using MCMC to obtain sample parameters following the posterior parameter density for each copula model
\State Randomly draw $N_{t_c}$ copula models according to their model probabilities $\pi_c$ and their associated parameters from the MCMC samples for the posterior parameter density.
\State Estimate the expected conditional copula $\hat{c}_{12}^{l}$ according to Eq. \eqref{eqn:exp9} 
\State Determine the expected joint density by multiplying the marginals and copula.  \label{step:5}
\State Repeat Step \ref{step:2} - \ref{step:5} for a large number, $N_{t_d}$, of marginal pairs. 
\State Determine the copula-based optimal sampling density $q_c^*(\bm{x})$ by averaging the $N_{t_d}$ joint densities as shown in Eq.\ \eqref{eqn:exp8}.  
\State (Optional) Create a surrogate optimal sampling density or lookup table to expedite sample re-weighting. 
\end{algorithmic}
}
\end{algorithm}

\subsection{Propagation of imprecise probabilities with copula dependence modeling}
With the constituents outlined in the previous section, the importance sampling reweighting approach for imprecise uncertainty propagation with copula dependence is summarized here and a flowchart is shown in Fig.\ \ref{fig:c4_flowchart}. 
\begin{figure}[!ht]    
    \centering
    \includegraphics[width=4in]{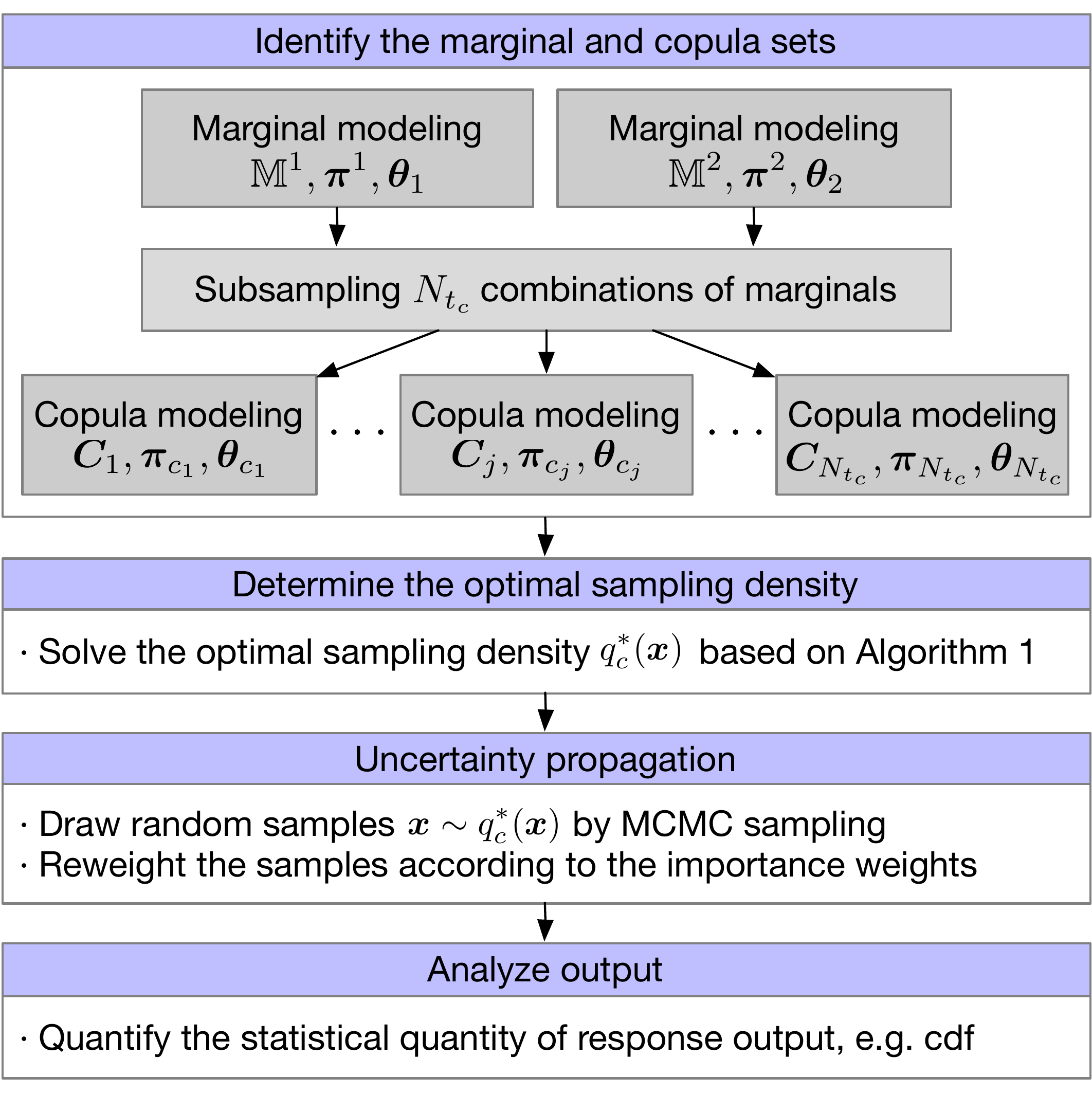}
	\caption{Flowchart for propagation of imprecise probabilities with copula-based dependence modeling} \label{fig:c4_flowchart}
\end{figure}

\begin{itemize}
\item Step 1: Identify the marginal and copula sets -- Given a small data set, the hierarchical Bayesian multimodel inference outlined in Section \ref{sec:multimodel_UQ} is used to identify the candidate sets of margainal distributions and copulas.  We first identify candidate marginal forms and associated model probabilities, and construct combinations of marginals by randomly drawing $N_{t_d}$ marginal pairs. For each pair of marginals, identify copula forms and estimate the copula model probabilities and copula parameters. 
\item Step 2: Determine the optimal sampling density -- Combine all the candidate marginals and associated copulas modeling from Step 1. Solving the optimization problem yields the optimal sampling density $q_c^*(\bm{x})$, shown in Eq. \eqref{eqn:exp6}, which is practically solved as described in Sec. 4.3 (Eqs.\ \eqref{eqn:exp8} and \eqref{eqn:exp9}), i.e.\ according to the Algorithm 1. 
\item Step 3: Uncertainty propagation -- Uncertainty associated with copula-based dependence modeling is propagated using importance sampling with optimal sampling density $q_c^*(\bm{x})$. Samples are drawn from $q_c^*(\bm{x})$ using MCMC sampling and are reweighted for each model according to the importance weights $w(\bm{x}) = p(\bm{x})/q_c^*(\bm{x})$
\item Step 4: Analyze output -- Quantify the distribution of the statistical response quantity of interest. 
\end{itemize} 
}

\section{Application to probabilistic prediction of unidirectional composite lamina properties}

This section applies the proposed methodology to understand the influence of the constituent material properties on the out-of-plane elastic properties (Young's modulus) of a unidirectional composite lamina. 

\subsection{Problem description}

Fiber reinforced composite materials are popular and widely used in many engineering fields because of their attractive properties, for example, high stiffness and strength combined with low weight. In order to evaluate the performance of a composite part, the accurate prediction of its mechanical properties in the layup is important {\color{black}\cite{zhangMAMS2019}}. Several numerical and experimental methods have been proposed to determine the mechanical properties of unidirectional lamina based on the elastic properties of the constituent materials (fibers and matrix)\cite{daniel1994engineering, younes2012comparative}. In this work, the finite element method (FEM) with a representative volume element (RVE) is used to predict the out-of-plane elastic properties of a unidirectional composite lamina given the constituent (fiber and matrix) material properties. 
\begin{figure}[!ht]    
    \centering
    \subfigure[]{\includegraphics[height=1.5in]{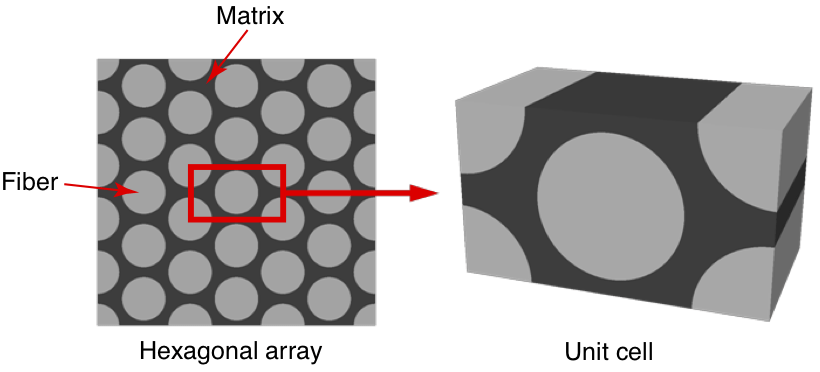}} \quad 
    \subfigure[]{\includegraphics[height=1.5in]{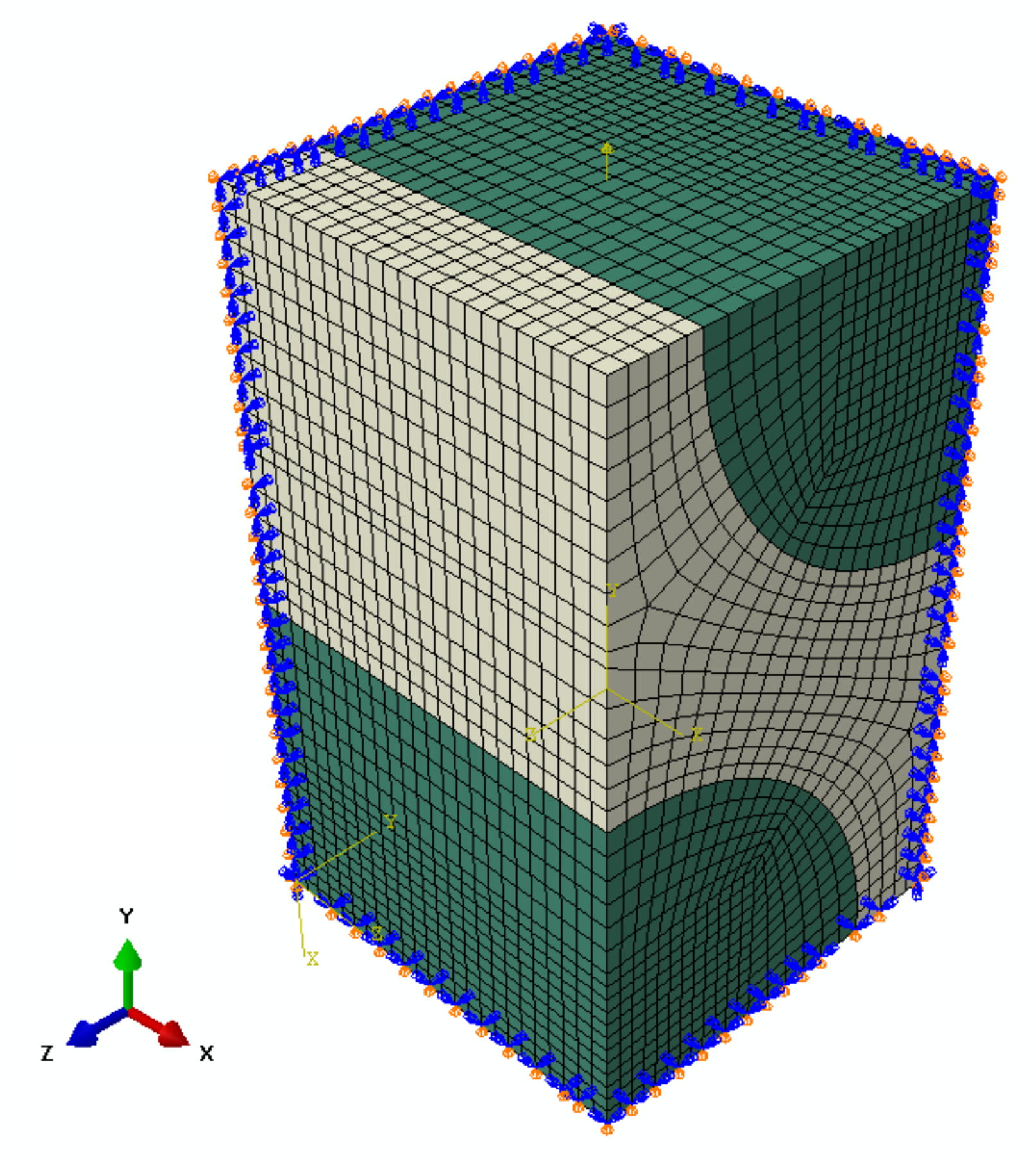}}
    \caption{Unidirectional fiber reinforced composite (a) Hexagonal RVE unit and (b) RVE FEM model}  \label{fig:RVE}
\end{figure}

Typically, unidirectional composites are considered as transversely isotropic materials composed of two phases: a fiber reinforcement phase and a matrix phase, as shown in Fig. \ref{fig:RVE} (a) for a hexagonal packing configuration. Commonly, the reinforced-fiber phase for traditional materials is modeled as isotropic (e.g. glass fibers) or orthotropic (e.g. carbon fiber) and the matrix phase is typically composed of an isotropic material (e.g. epoxy). The overall mechanical properties of transversely isotropic unidirectional fiber reinforced lamina with a hexagonal packing geometry are determined by five independent engineering constants which are given by the following compliance matrix: 
\begin{equation}
C = \left[ \begin{matrix}  1/E_{11} & -\nu_{12}/E_{11}  & -\nu_{12}/E_{11}  & 0 & 0 & 0\\
-\nu_{12}/E_{11}  & 1/E_{22}  & -\nu_{23}/E_{22} & 0 & 0 & 0 \\
-\nu_{12}/E_{11}  & -\nu_{23}/E_{22}  &1/E_{22} &0 &0 &0 \\
0 & 0 & 0 & 1/G_{23} & 0 & 0 \\
0 & 0 & 0 & 0& 1/G_{12} & 0 \\
0 & 0 & 0 & 0& 0 & 1/G_{12} \\
 \end{matrix} \right] 
 \label{eq: C}
\end{equation}
where $E_{11}$ and $E_{22}$ are the longitudinal and transverse Young's moduli respectively, $G_{12}$ and $G_{23}$ are the longitudinal and transverse shear moduli, $\nu_{12}$ is the major Poisson's ratio and $\nu_{23}$ is the minor Poisson's ratio. The transverse shear modulus is determined from the minor Poisson's ratio $\nu_{23}$ and elastic modulus $E_{22}$ as \cite{hashin1983analysis}:
\begin{equation}
G_{23} = \frac{E_{22}}{2(1+\nu_{23})} 
\end{equation}

Experimental determination of the in-plane lamina properties are typically straightforward and generally provide accurate values for these properties. However, the out-of-plane lamina properties are difficult to obtain experimentally \cite{king1992micromechanics, gipple1994measurement, soden2004lamina}, and consequently numerical prediction becomes an attractive alternative to predict these lamina properties. In this example, we focus on the determination of the elastic modulus $E_{22}$ which is an independent out-of-plane lamina property. 

The overall mechanical properties in Eq. \eqref{eq: C} depend on the constituent properties (fibers and matrix). Table \ref{tab:c4_mean_cov} shows the four independent constituent material properties and the fiber volume fraction, which are needed to define the lamina properties for the isotropic resin and fiber materials. 

\begin{table}[!ht]  \footnotesize
\centering
\caption{Constituent material properties of E-Glass fiber/LY556 Polyester Resin composites}
\label{tab:c4_mean_cov}
\begin{tabular}{@{}cccc@{}}
\toprule
Material property & Physical meaning & Mean value & Coefficient of variation \\ \midrule
$V_f$                & Fiber volume fraction                & 0.6          & 0.05 \\
$E_m$                & Matrix's Young's modules                & 3.375          & 0.05 \\
$\nu_m$                & Matrix Poisson's ratio                & 0.35          & 0.05 \\
$E_{1f}$               & Fiber Young's modules along 1 direction                & 73.01          &0.05 \\
$\nu_{12f}$              & Fiber Poisson's ratio along 1-2 direction                 & 0.228          & 0.05 \\ \bottomrule
\end{tabular}
\end{table}

In this work, we study a common composite lamina fabricated from E-glass fibers and LY556 polyester resin matrix. The finite element method is employed to construct a three-dimensional RVE with two symmetry planes in the $x-y$ and $x-z$ directions and periodic boundary conditions, as shown in Fig. \ref{fig:RVE} (b). The model has a total 22750 nodes and 20448 C3D8R solid elements and is solved using the commercial solver Abaqus. 


\subsection{Identification of probabilistic input model}
From engineering experience, the five inputs in Table \ref{tab:c4_mean_cov} may be correlated or dependent and thus one task is to identify the dependence relationship among these five random variables from data. Commonly, the matrix properties $E_m$ and $\nu_m$ are considered to be dependent and the fiber properties $E_{1f}$ and $\nu_{12f}$ are dependent. However, fiber and matrix properties are independent of one another and the fiber volume fraction is often assumed independent of constituent properties.  Therefore, the five probability inputs are composed of two bivariate dependent models and one independent variable: $\left\{E_m, \nu_m \right\}$, $\left\{E_{1f}, \nu_{12f} \right\}$ and $\left\{ V_f \right\}$. 
\begin{figure}[!ht]    
    \centering
    \includegraphics[width=5in]{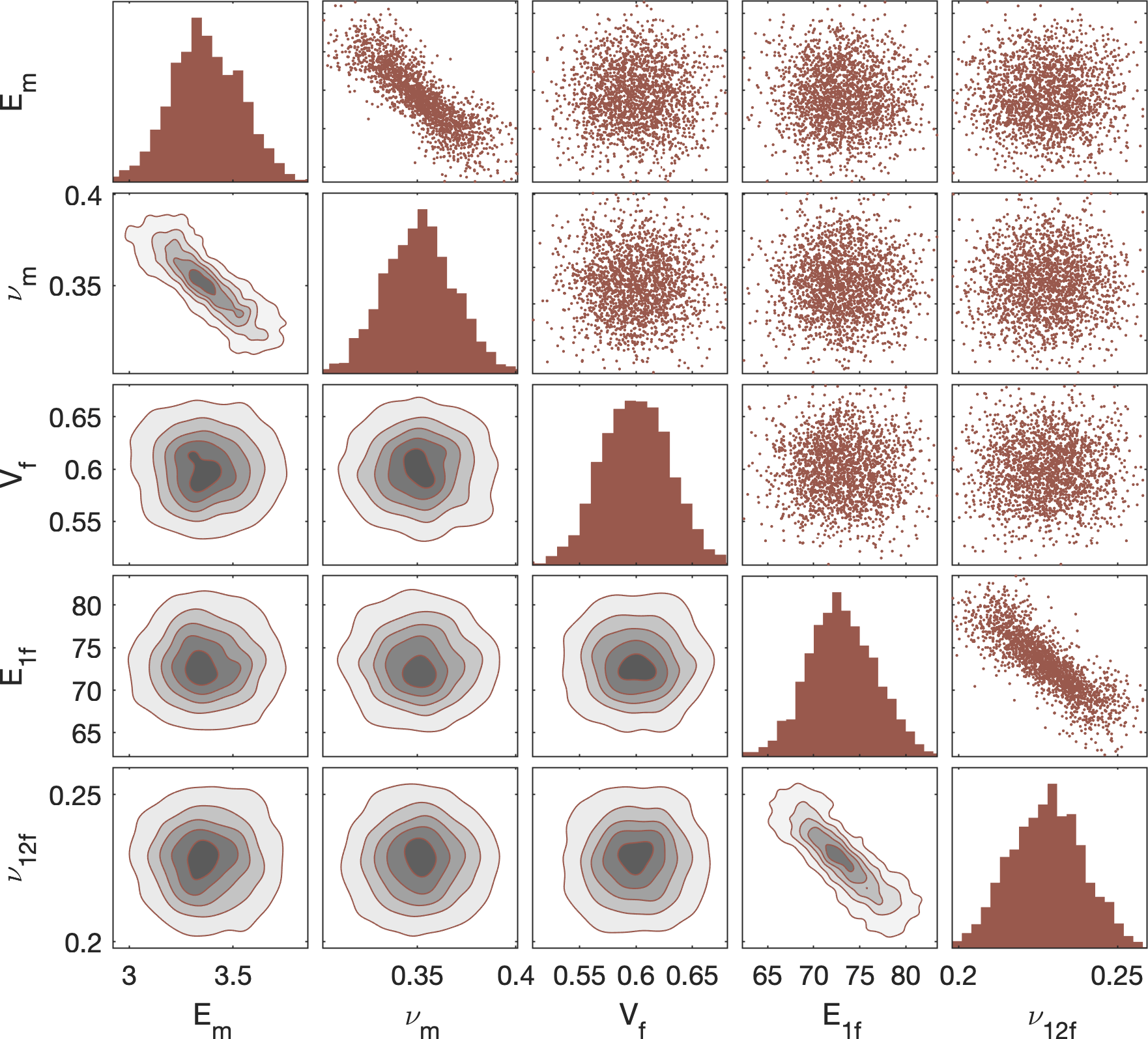}
    \caption{Dependent probabilistic input model} \label{fig:de_input}
\end{figure}

Although this type of composite materials has been used extensively in many engineering applications, statistical data for its constituent properties are very limited. Typically, only nominal design values are provided without adequate guidance on their variability.  The nominal values in Table \ref{tab:c4_mean_cov} were compiled from the literature for each constituent property and candidate probability distributions were identified for each property. The interested readers can find an extensive list of references for the relevant data and literature in the authors' recent work \cite{zhangigsa2019}. 

\begin{figure}[!ht]    
    \centering
    \subfigure[]{\includegraphics[height=2in]{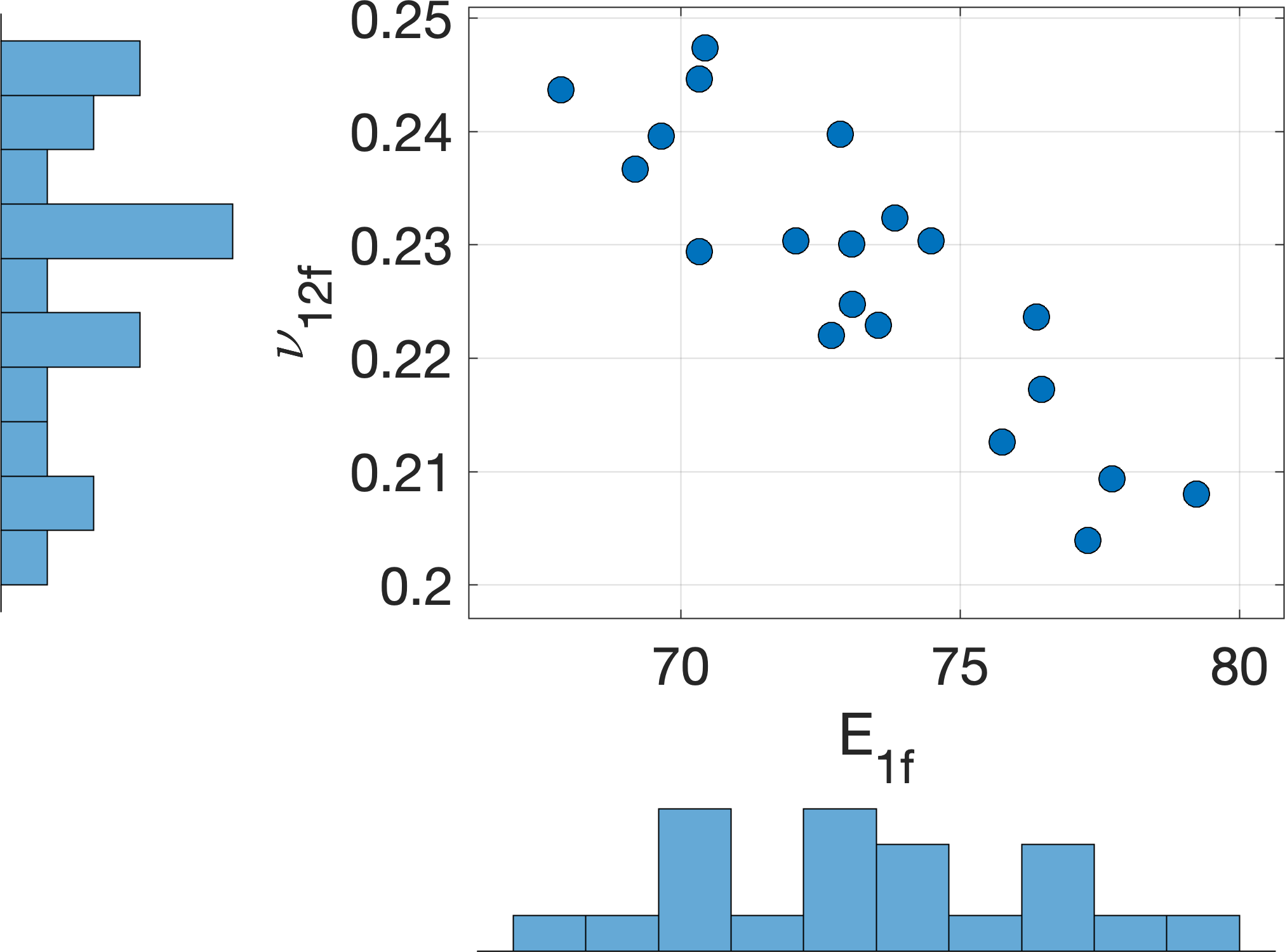}} \quad
     \subfigure[]{\includegraphics[height=2in]{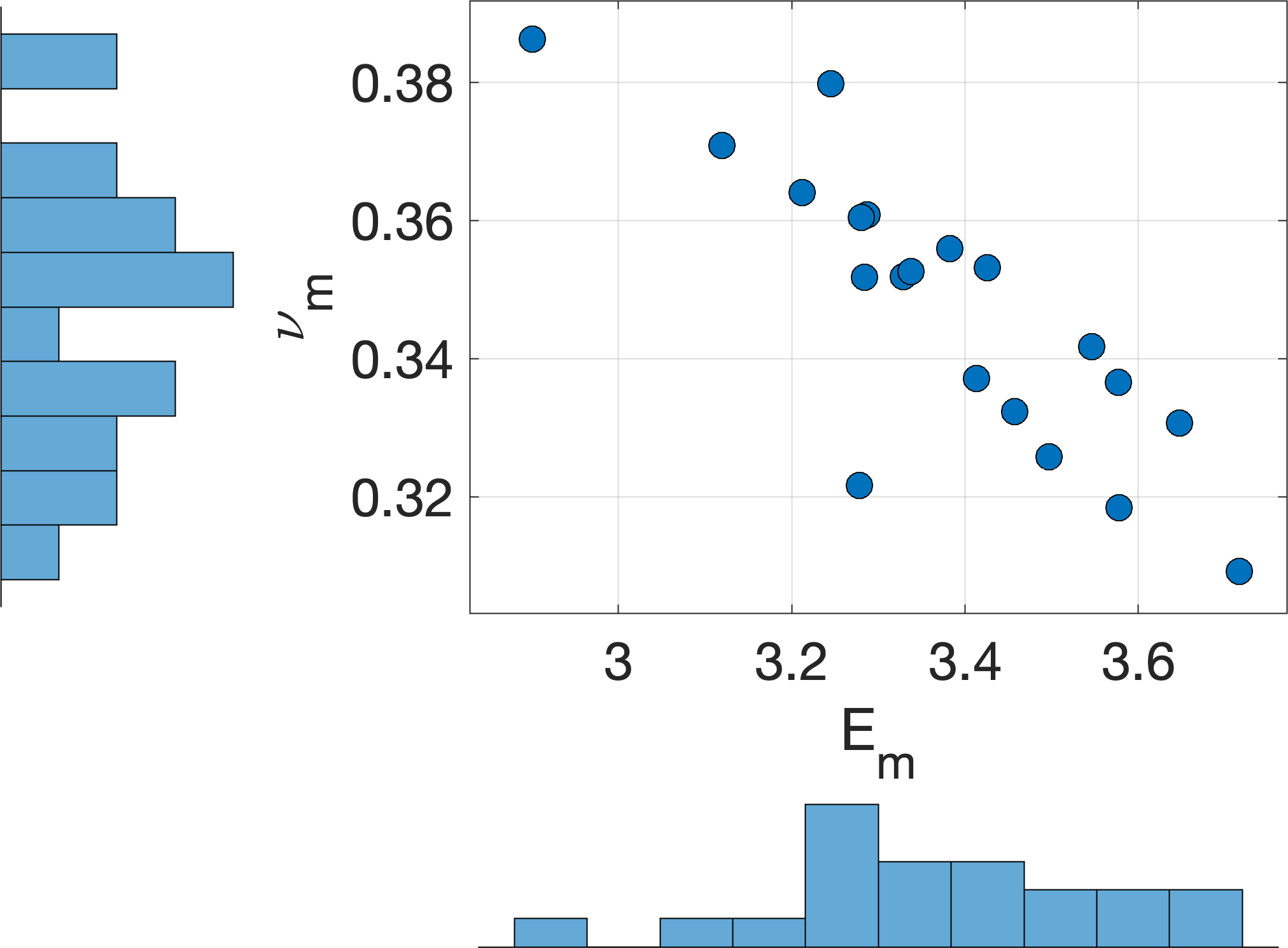}} 
    \caption{20 randomly generated constituent material properties that serve as the initial dataset (a) fiber property and (b) matrix property}  \label{fig:c4_20data}
\end{figure}

Due to a lack of statistical data for characterizing the constituent material properties, it is difficult to assign accurate and objective probabilistic models for the properties, specifically the dependence model for the constituent properties. For reference purposes, we assume normal distributions with nominal mean value in Table \ref{tab:c4_mean_cov} and 5\% coefficient of variation (COV) as the ``true" marginal distributions for each fiber and matrix property. The matrix properties $\left\{E_m, \nu_m \right\}$ and the fiber properties $\left\{E_{1f}, \nu_{12f} \right\}$ are assumed to be strongly correlated with a ``true" Frank copula with parameter $\theta=-10$. Fig. \ref{fig:de_input} shows the ``true" probabilistic input model, which includes the marginal histogram and dependence relationship between each of these input variables. It can be observed that $\left\{E_m, \nu_m \right\}$ and $\left\{E_{1f}, \nu_{12f} \right\}$ have a strong dependence that follows the true Frank(-10) copula model. We assume this probabilistic model to be the truth and generate 20 random data, as shown in Fig. \ref{fig:c4_20data} for the joint matrix and fiber properties. These serve as the initial data from which uncertainty needs to be quantified and propagated. Clearly, a single bivariate dependence model cannot be precisely identified from these data -- although it is clear that the properties are dependent. 

\subsection{Probabilistic prediction of composite properties}
\label{sec:composite3}
\begin{figure}[!ht]    
    \centering
    \subfigure[]{\includegraphics[height=2in]{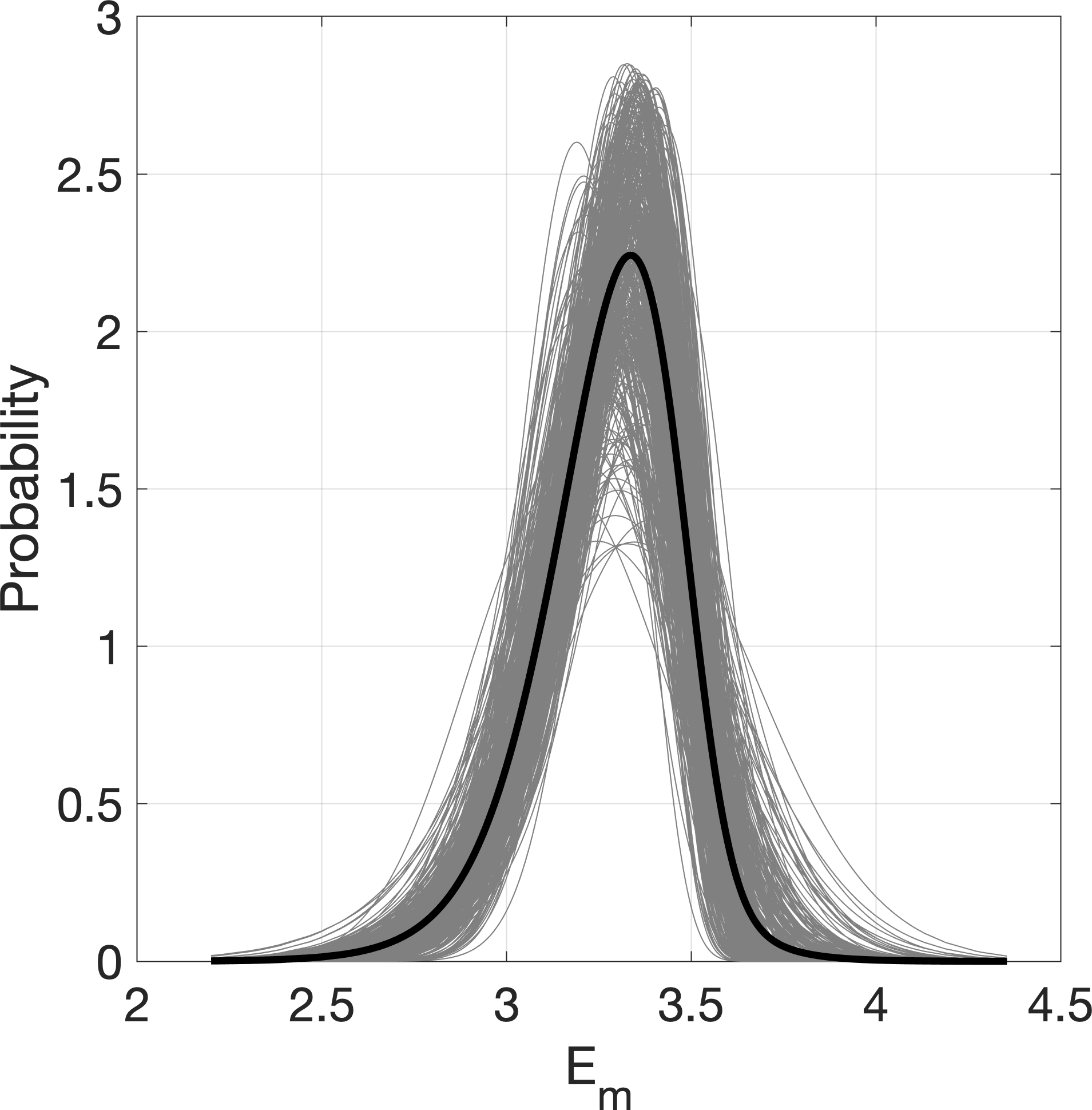}} \quad
    \subfigure[]{\includegraphics[height=2in]{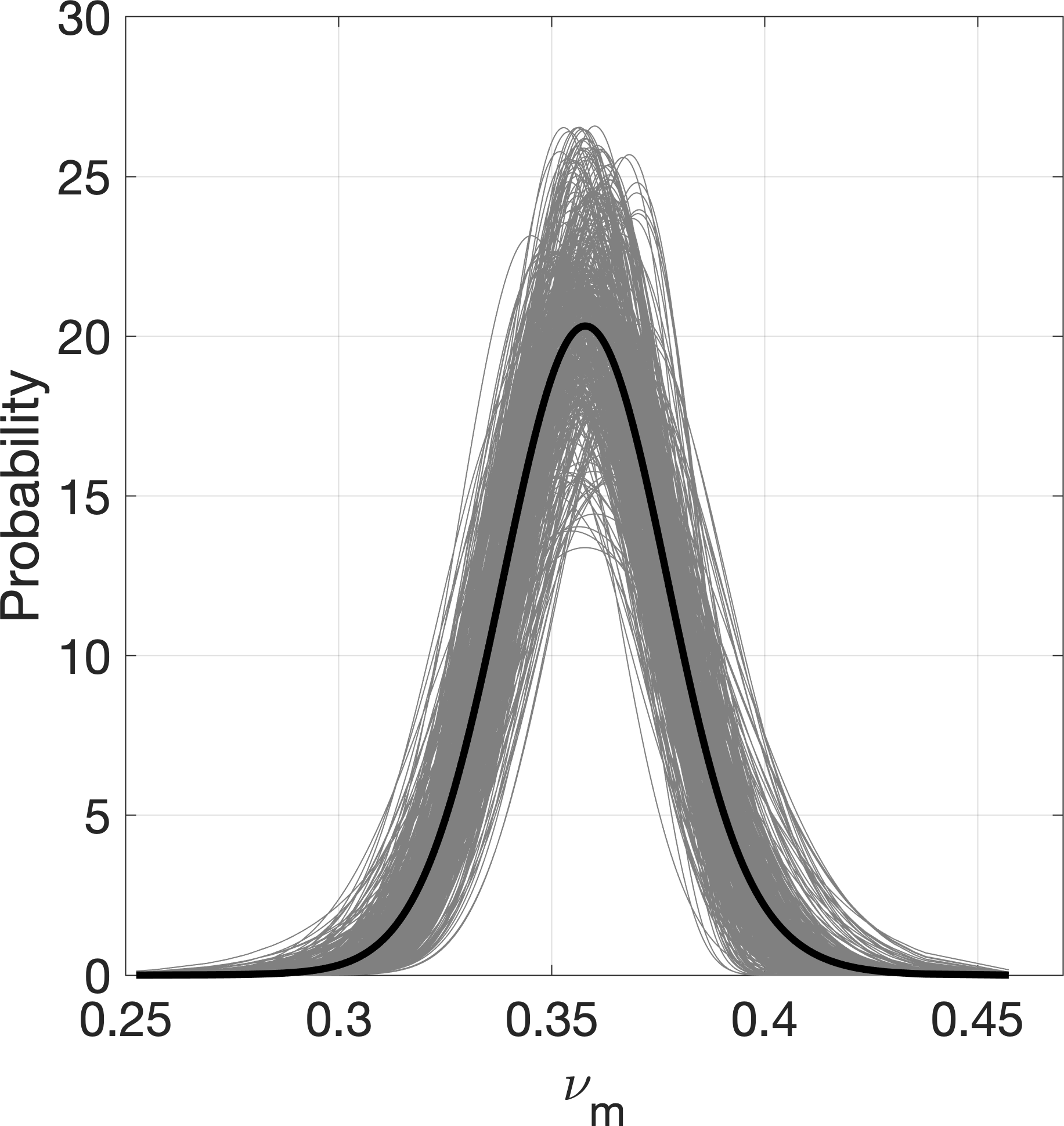}}\quad
    \subfigure[]{\includegraphics[height=2in]{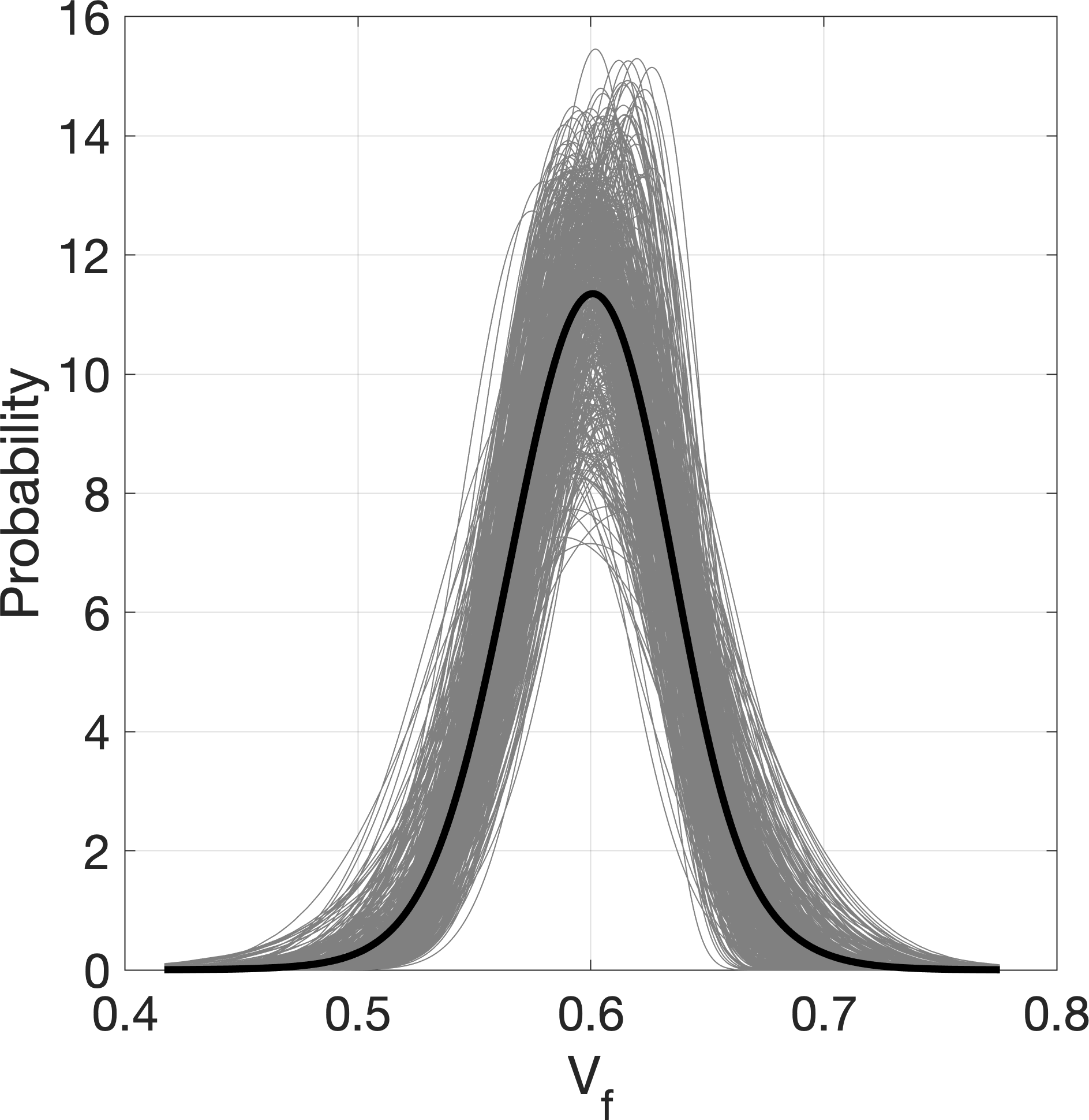}}\quad
    \subfigure[]{\includegraphics[height=2in]{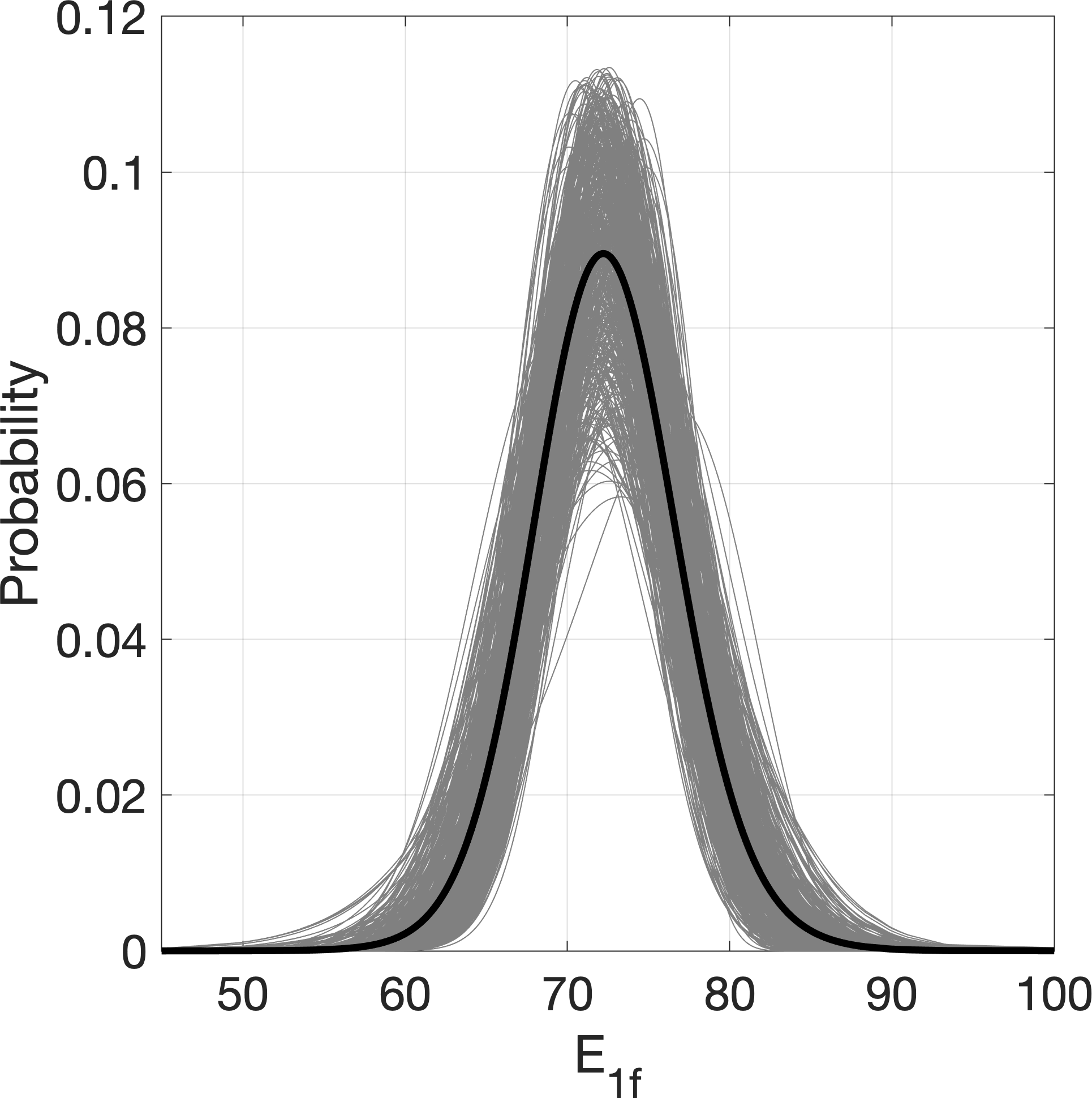}}\quad
    \subfigure[]{\includegraphics[height=2in]{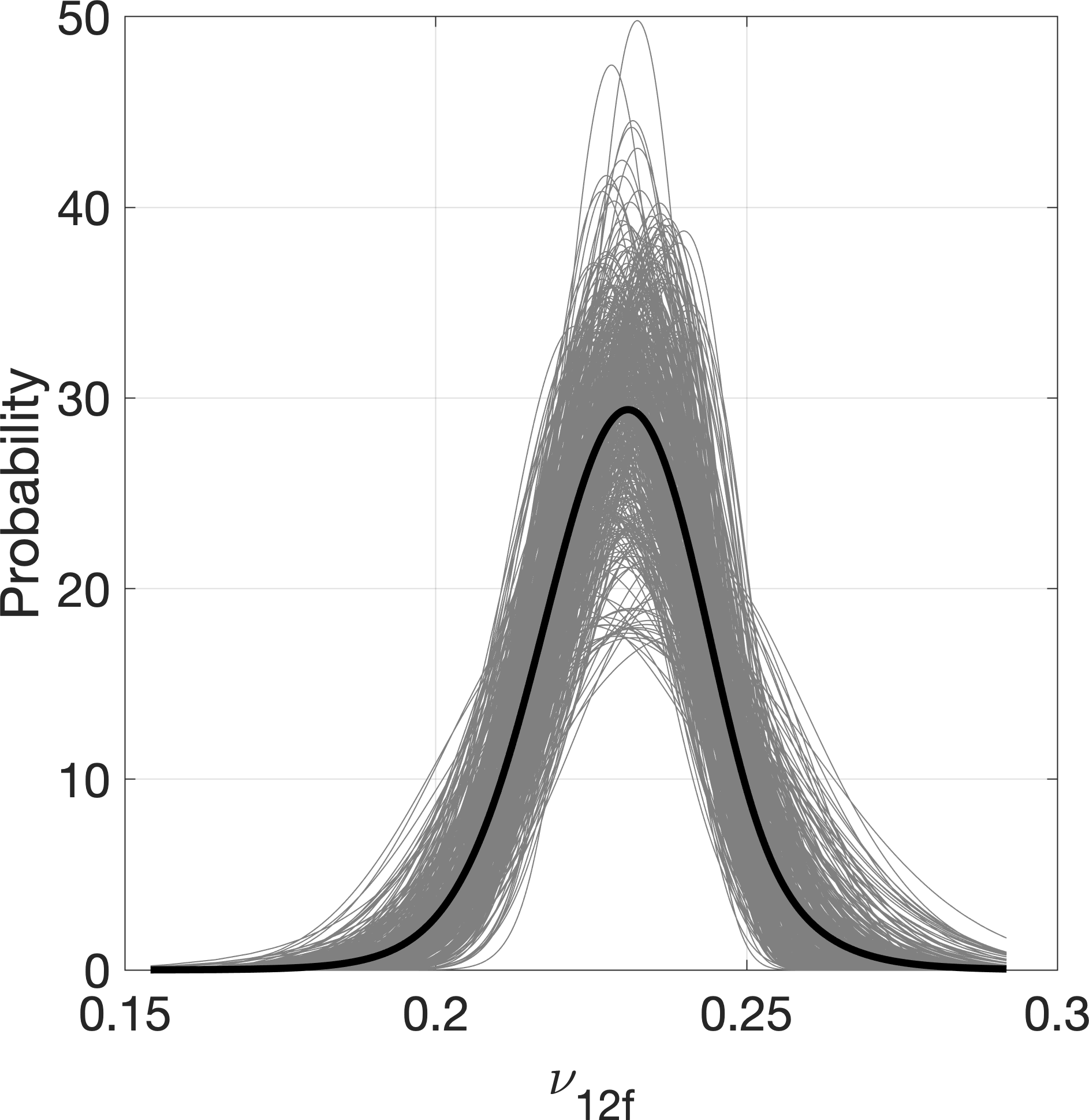}}
    \caption{Multiple candidate probability densities for marginals (a) $E_m$, (b) $\nu_m$, (c) $V_f$, (d) $E_{1f}$ and (e) $\nu_{12f}$}  \label{fig:c4_marginal1}
\end{figure}

The multimodel inference approach proposed herein is applied to this problem, given the limited data characterizing the constituent material properties and their clear dependencies. We first identify a set of candidate marginal probability models, which include the Gaussian, Gamma, Lognormal and Weibull distributions.  The Bayesian multimodel approach in Eq. \eqref{eq:Bayes_model} is used to estimate the posterior model probabilities and the corresponding model parameter uncertainties are estimated by Bayesian inference using MCMC sampling. Combining these model-form and model parameter uncertainties, we therefore obtain an ensemble of plausible probability densities for the five input variables shown in Fig. \ref{fig:c4_marginal1}. 

In this example, we identify 500 candidate densities for each marginal such that the total number of combinations of these marginal distributions is $500^5 = 3.125^{13}$, which is computationally prohibitive. Instead, a representative 1000 marginal pairs are compiled by Latin hypercube sampling. To evaluate the elastic modulus $E_{22}$, 5,000 random samples are drawn from the optimal sampling density, shown in the thick black thick curves in Fig. \ref{fig:c4_marginal1}, for each material property and computational model evaluations are performed using FEM. {\color{black}Hence, the computational advantage of the approach lies in the vastly reduced number of model evaluations needed to propagate the full model set. In this case, we need only 5,000 simulations where conventional multi-loop Monte Carlo approaches require on the order of $5,000^3$ simulations to cover the full set of copulas, marginals, and marginal parameters. For the composite model used herein, the 5,000 simulations take approximately 28 cpu-hours to complete, making the conventional strategy infeasible.}

\begin{figure}[!ht]    
    \centering
    \subfigure[]{\includegraphics[height=2in]{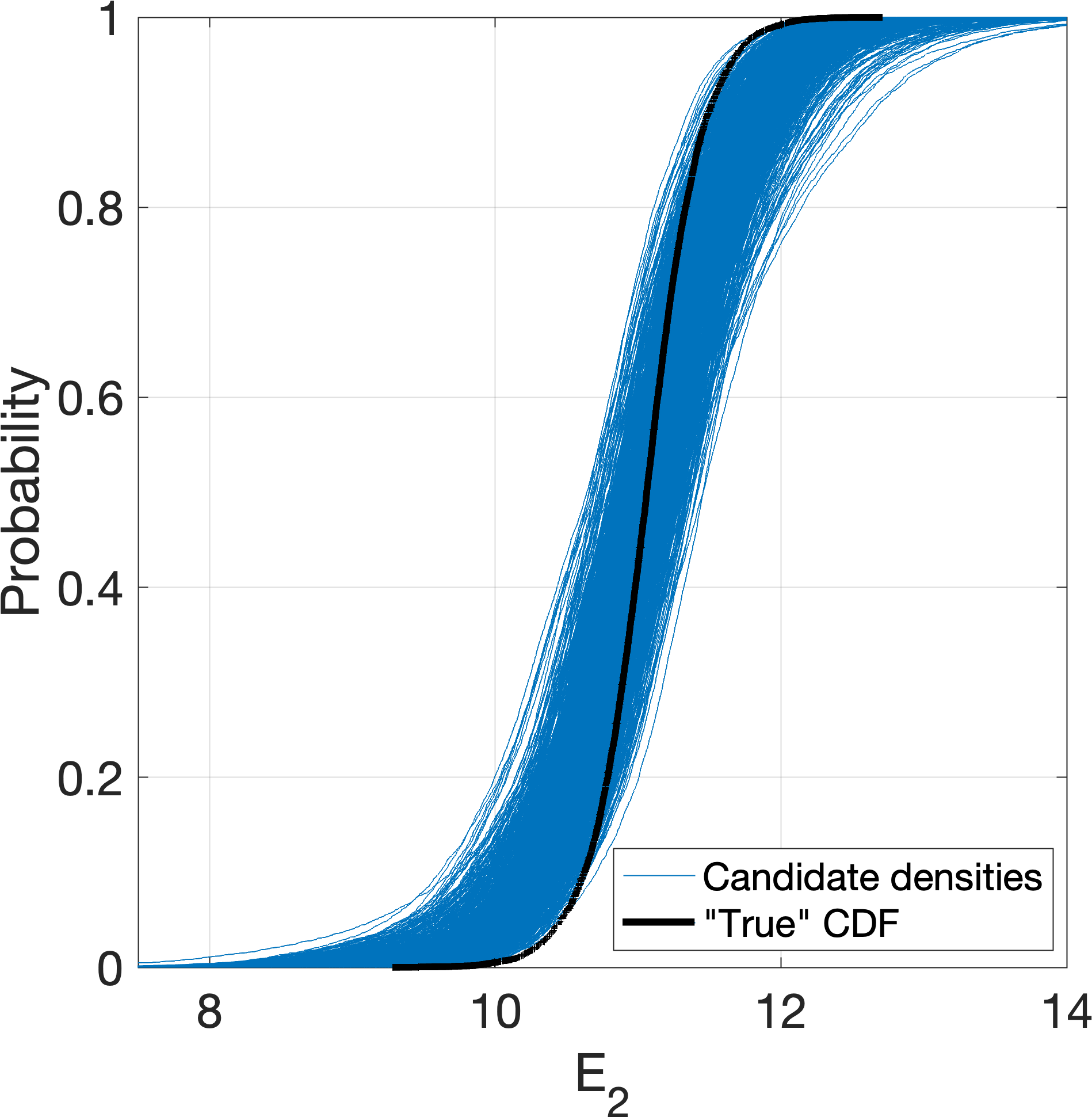}} \quad
    \subfigure[]{\includegraphics[height=2in]{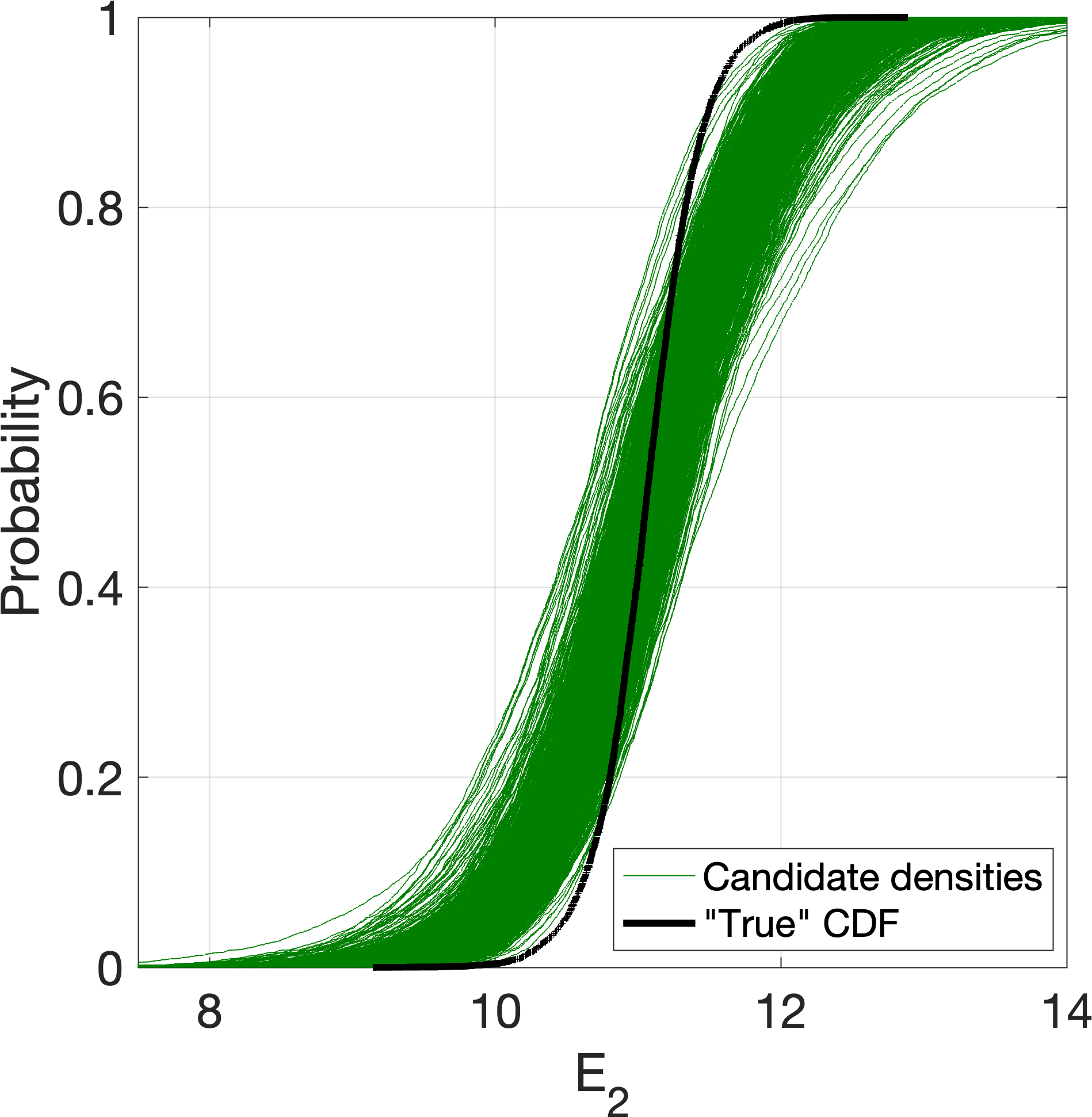}} \quad 
    \caption{Collection of candidate empirical CDFs for Young's modulus $E_{22}$ given the initial 20 data, assuming (a) independent marginal distributions and (b) Gaussian correlation}  \label{fig:c4_marginal2}
\end{figure}

If the multivariate input is assumed independent, we can easily achieve the probabilistic prediction of overall material property $E_{22}$ by multiplying each marginal. {\color{black} Fig. \ref{fig:c4_marginal2} shows the cloud of candidate empirical CDFs for $E_{22}$ based on multimodel inference from the 20 data assuming the marginals are independent and Gaussian correlated with $\rho=0.8$. The ``true" CDF in Fig. \ref{fig:c4_marginal2} (with variable dependence) is shown in black. Note that the collection of CDFs compiled under the independence assumption (blue) and Gaussian correlation (green) as well as true estimate with dependence (black) seem to overlap -- suggesting that perhaps the independence assumption is sufficient to bound the elastic properties. However, as we show next, this result underestimates the uncertainty in $E_{22}$. }

\begin{figure}[!ht]    
    \centering
    \subfigure[]{\includegraphics[height=2in]{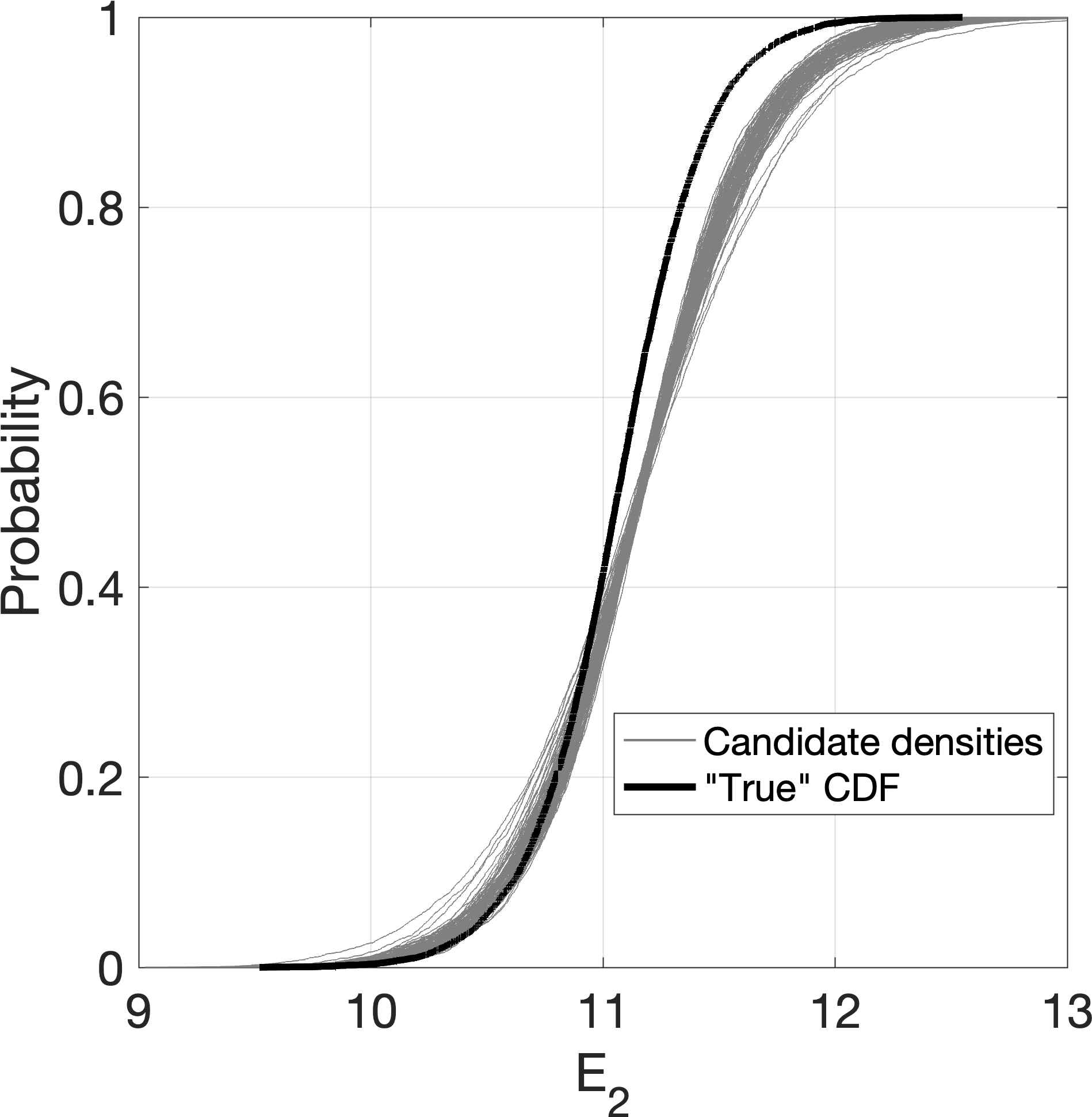}} \quad 
    \subfigure[]{\includegraphics[height=2in]{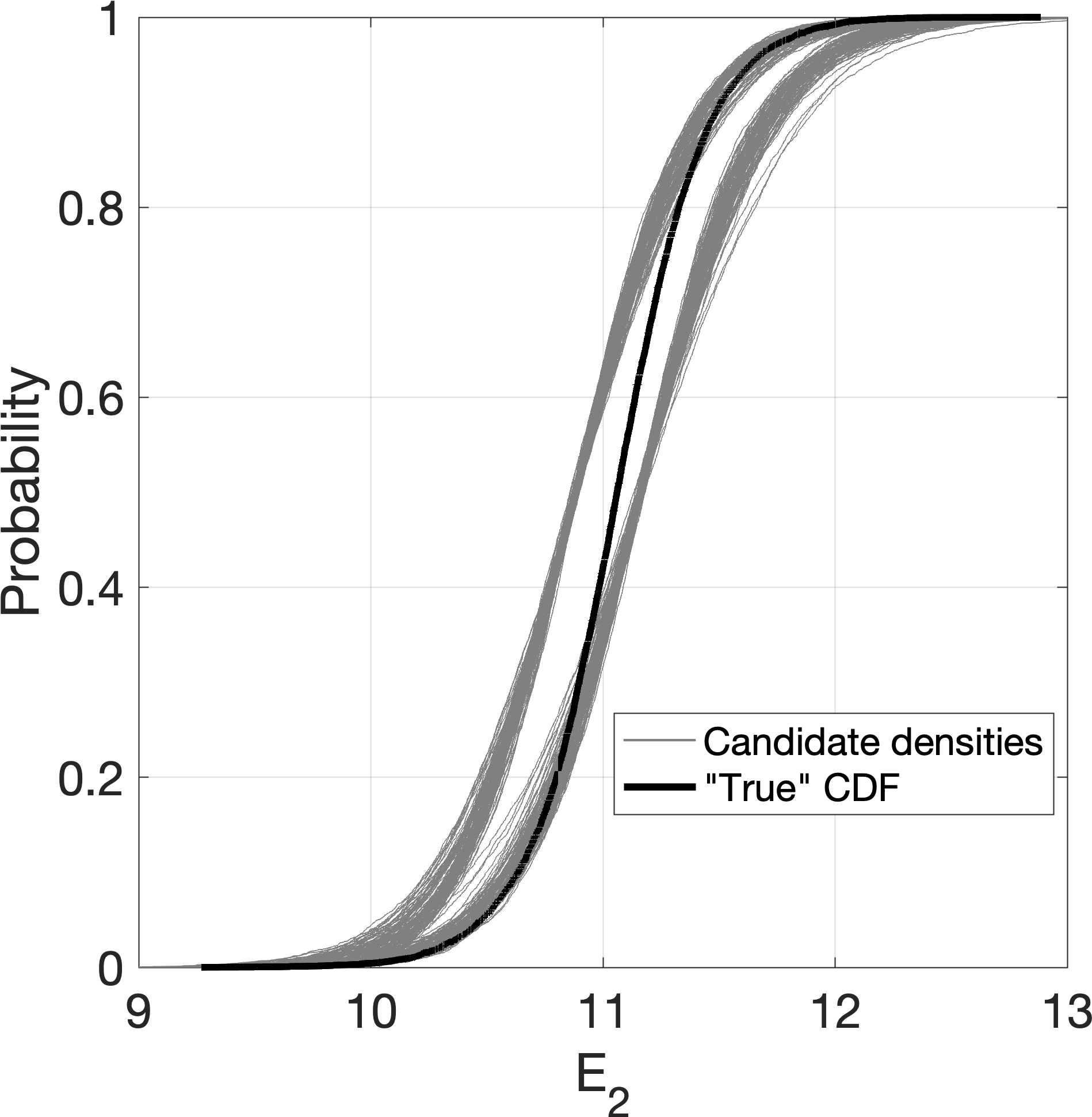}} \quad 
    \subfigure[]{\includegraphics[height=2in]{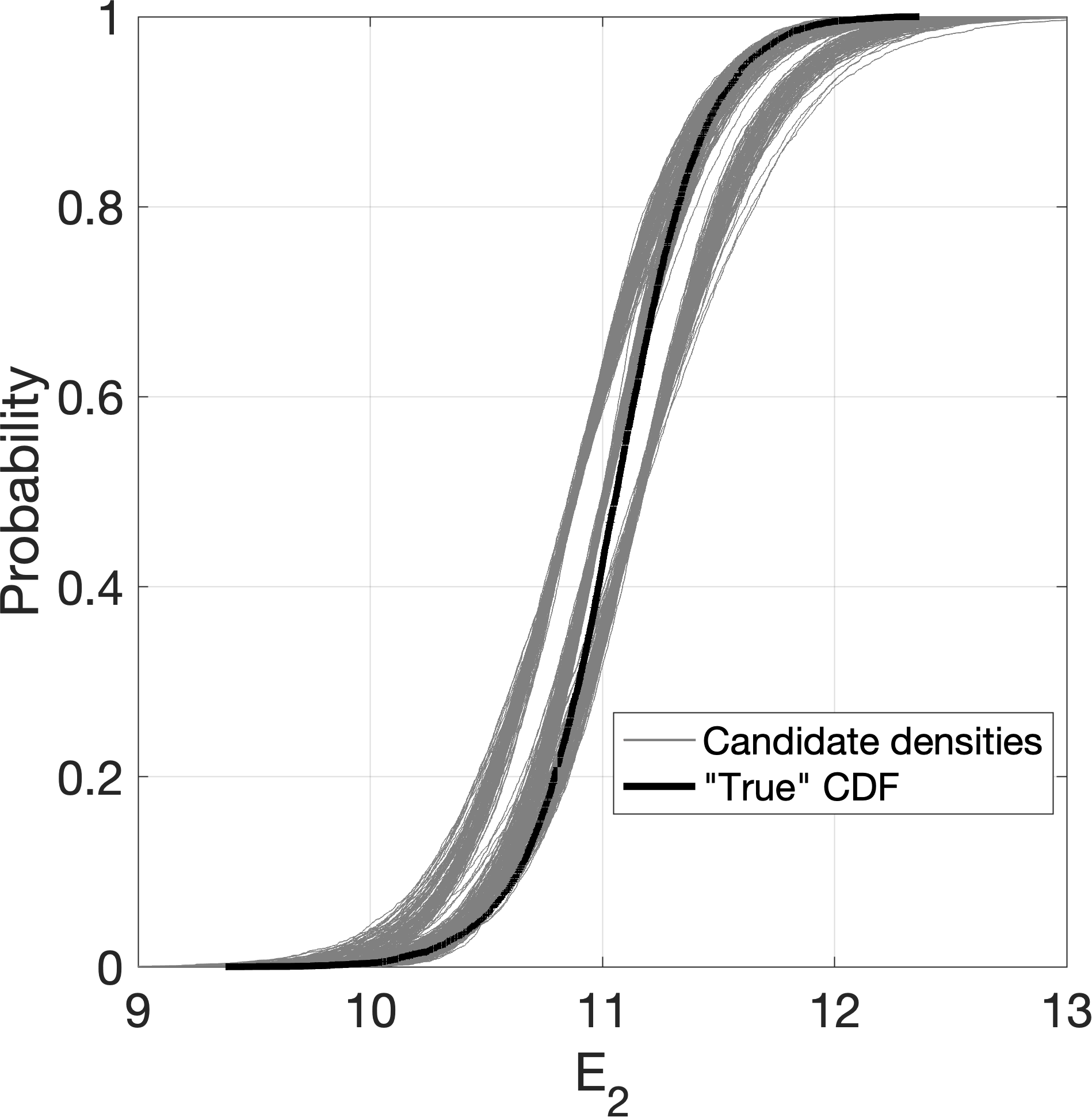}}
    \caption{Collection of candidate empirical CDFs for Young's modulus $E_{22}$ with only copula uncertainty given (a) one pair of marginals, (b) two pairs of marginals and (c) three pairs of marginals}  \label{fig:c4_copula_1mar}
\end{figure}

To account for variable dependence, for each pair of marginals we must identify a set of candidate copulas. For this we perform the hierarchical Bayesian multimodel selection for the Gaussian, Clayton, Frank and Gumbel copulas. We first compute the posterior copula model probabilities and then compute the associated joint parameter densities. For each pair of marginals, we then construct an ensemble of copula model sets by randomly selecting the copula models and copula parameters. Finally, the optimal sampling density in Eq.\ \eqref{eqn:exp8} is determined and employed for propagation of the multiple candidate densities with copula dependence. Fig.\ \ref{fig:c4_copula_1mar} shows three examples illustrating the influence of copula dependence uncertainty for specific marginal density pairs. Notice that, when the marginals are assumed to be independent a single cdf for $E_{22}$ is generated. However, with uncertainty in the copula dependence, there are several candidate pdfs for each pair of marginal densities. In other words, the uncertainty associated with the spread in the sets of cdfs in Figure \ref{fig:c4_copula_1mar} is ignored if we assume independent marginals. 

\begin{figure}[!ht]    
    \centering
{\includegraphics[height=2in]{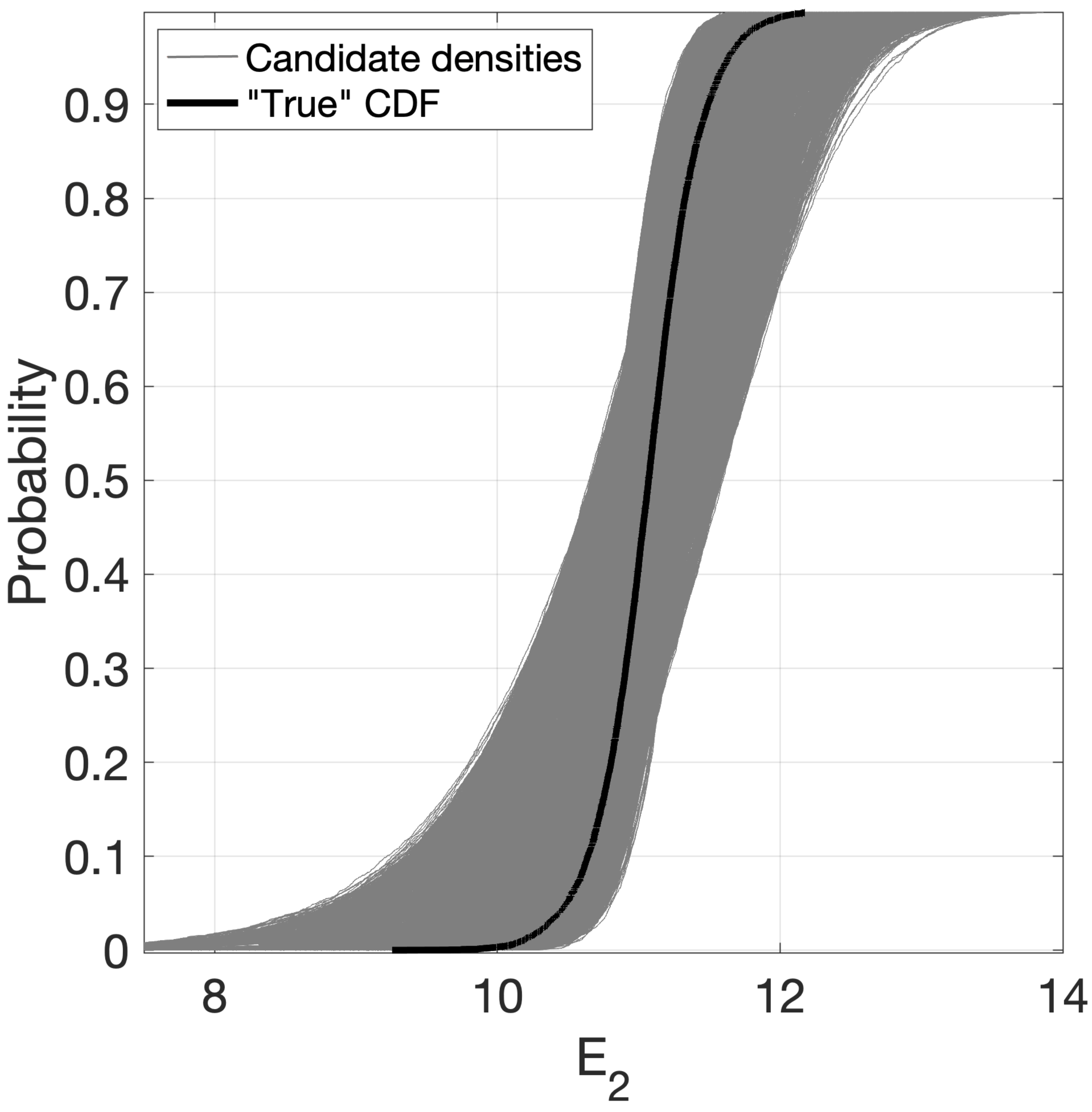}} 
    \caption{Total collection of candidate empirical CDFs for Young's modulus $E_{22}$ with uncertainty in dependence modeling given 20 data.}  \label{fig:c4_copula_total_20}
\end{figure}

{\color{black}When we combine the uncertainties from the copula model and marginal model together in Figure \ref{fig:c4_copula_total_20}, we see that the overall uncertainty is considerably wider than it was when assuming the marginals to be independent or Gaussian correlated (Fig.\ \ref{fig:c4_marginal2}). That is, the candidate densities with dependence modeling show a much wider band than the densities with independent or Gaussian correlated assumption. }


\subsection{Influence of dataset size}

In this section, we investigate the convergence of the composite material properties as a function of dataset size. As discussed in the previous section, small datasets led to large uncertainties including the copula model and marginal model in the composite material properties. This raises a critical question: ``How much data is necessary to gain adequate confidence in the probabilistic prediction of composite material properties?"

\begin{figure}[!ht]    
    \centering
    \subfigure[]{\includegraphics[height=1.5in]{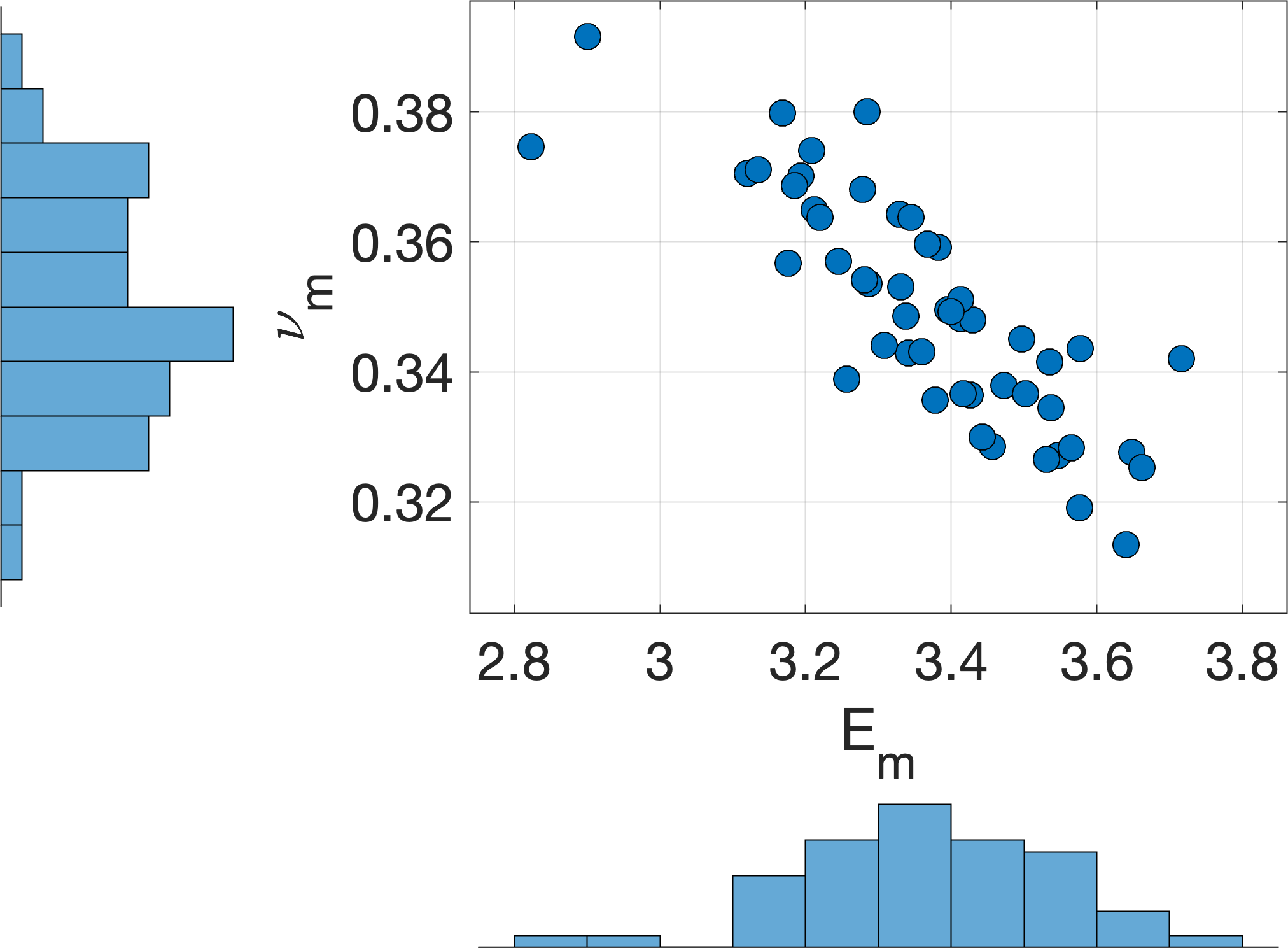}} 
    \subfigure[]{\includegraphics[height=1.5in]{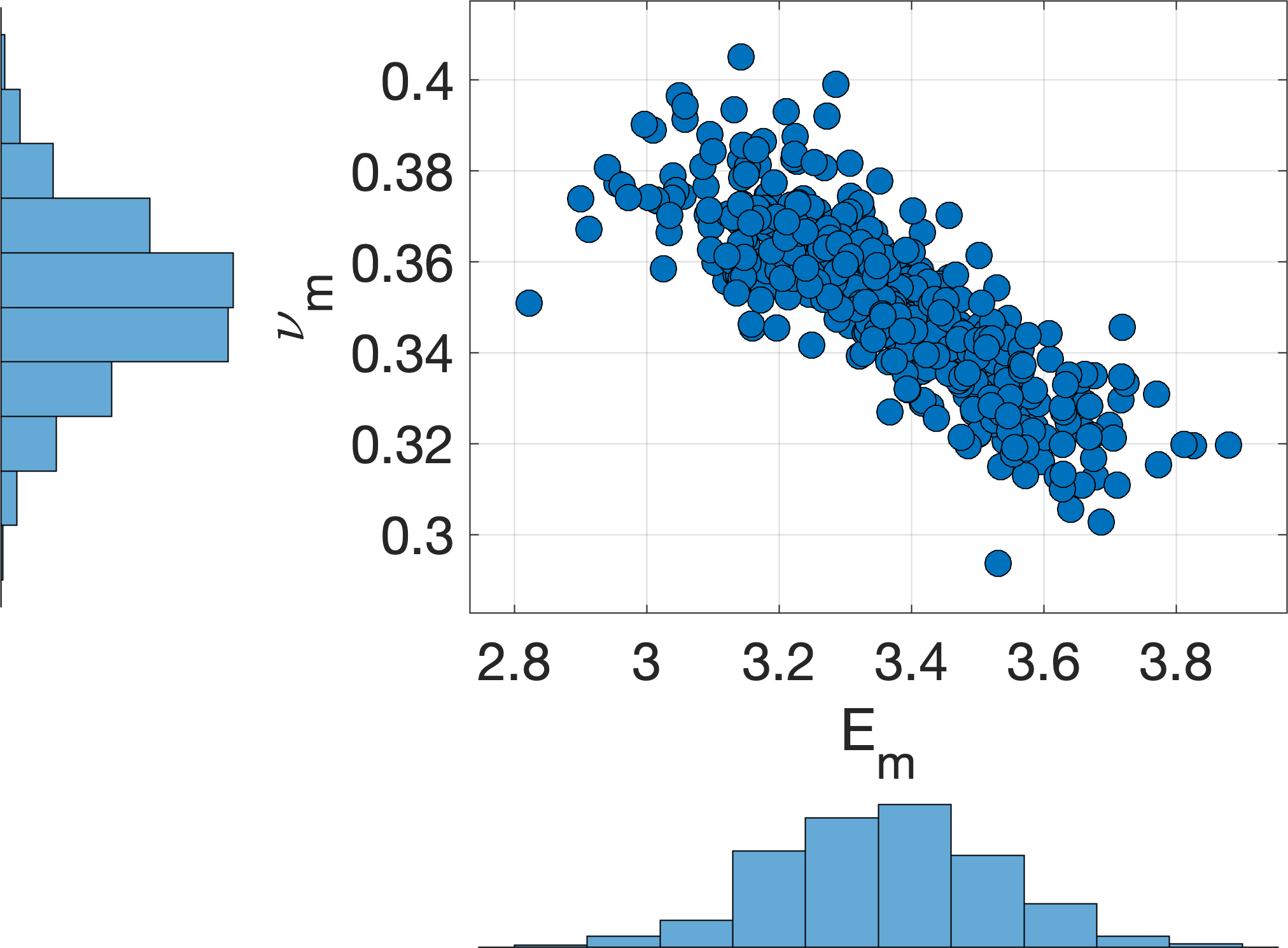}}
    \subfigure[]{\includegraphics[height=1.5in]{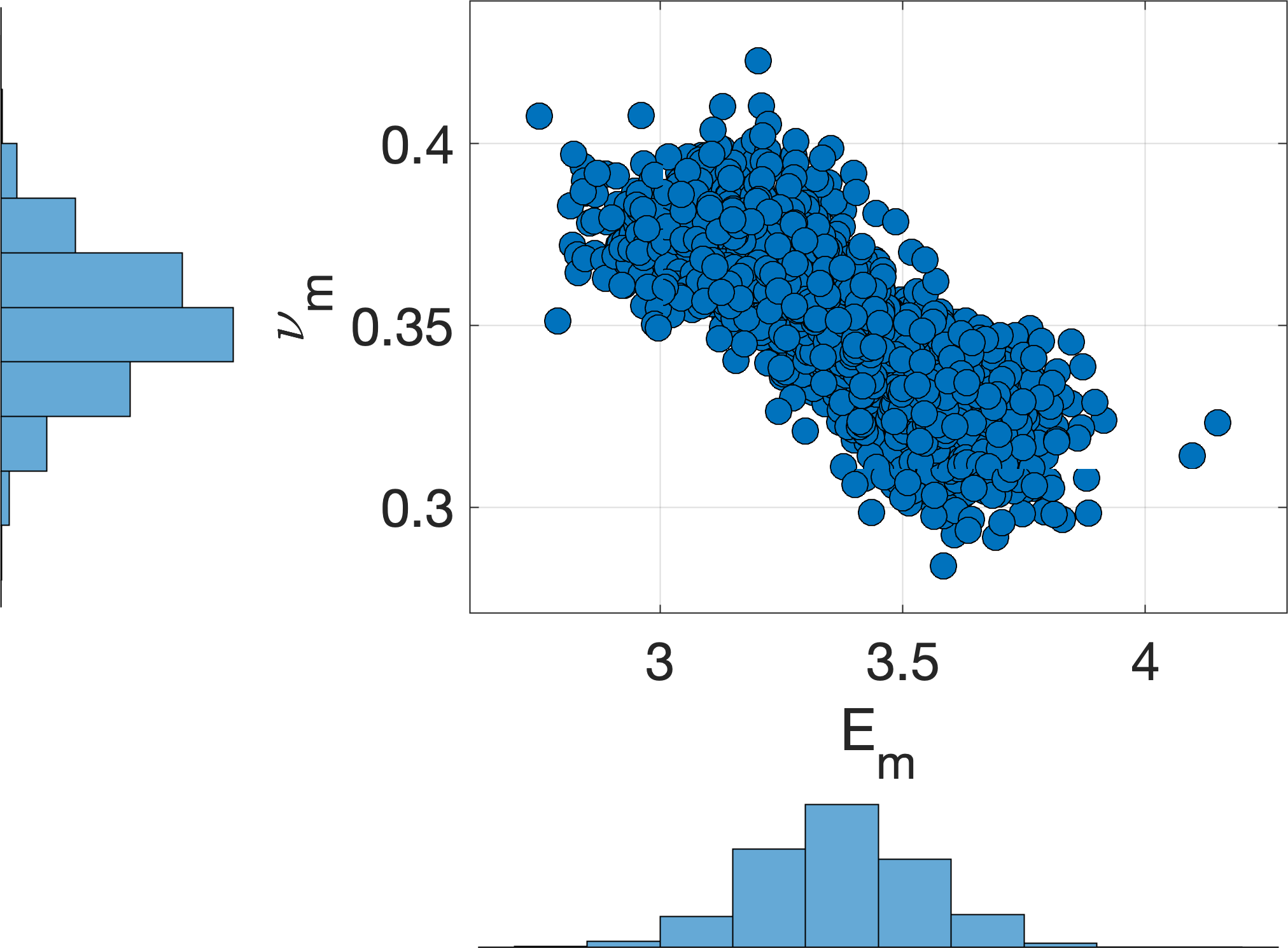}} \\
    \subfigure[]{\includegraphics[height=1.5in]{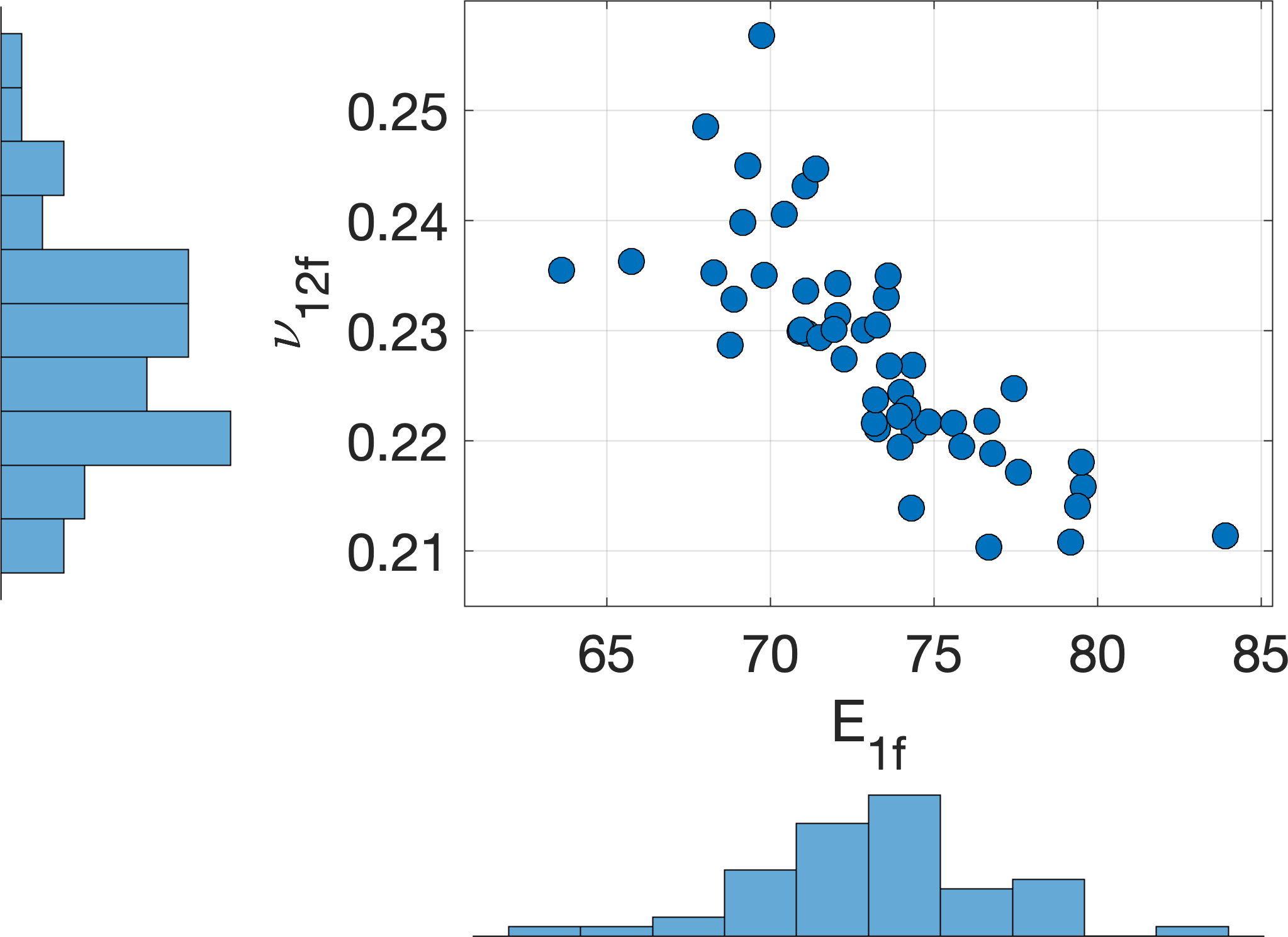}}
    \subfigure[]{\includegraphics[height=1.5in]{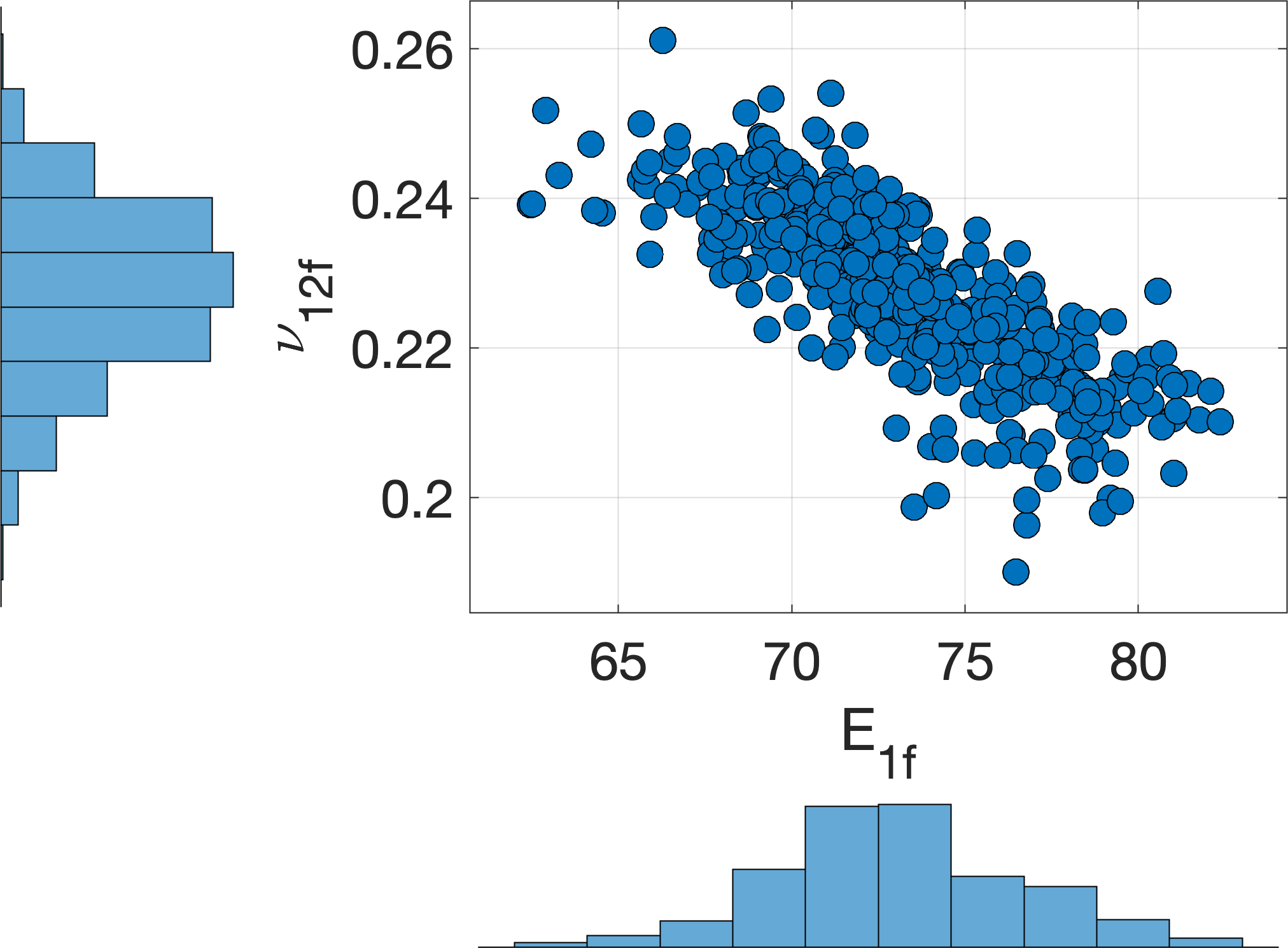}} 
    \subfigure[]{\includegraphics[height=1.5in]{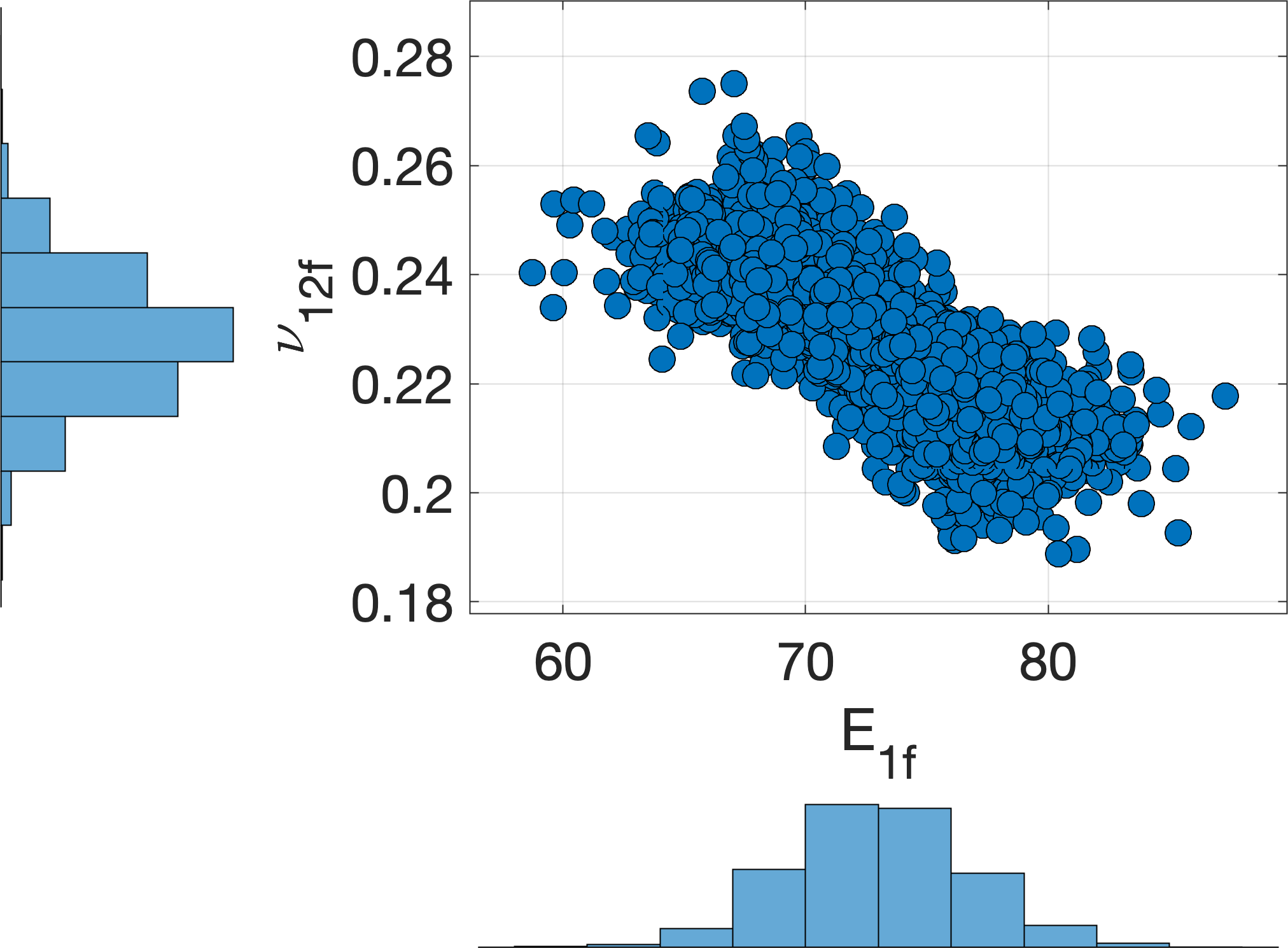}}
    \caption{Increasing data set size for dependent matrix and fiber properties: (a,d) 50 data, (b,e) 500 data and (c,f) 5000 data}  \label{fig:c4_more_data}
\end{figure}

Here, additional data are generated from the true joint probability density. We begin with the initial 20 data  and increase to 50 data, 500 data and 5000 data, as shown in Fig.\ \ref{fig:c4_more_data}. As the data set size increases, we more clearly see the true dependence emerge. Both the normal marginals become increasingly pronounced and the nature of the underlying copula dependence becomes clear.

\begin{figure}[!ht]   
    \centering
    \subfigure[]{\includegraphics[height=2in]{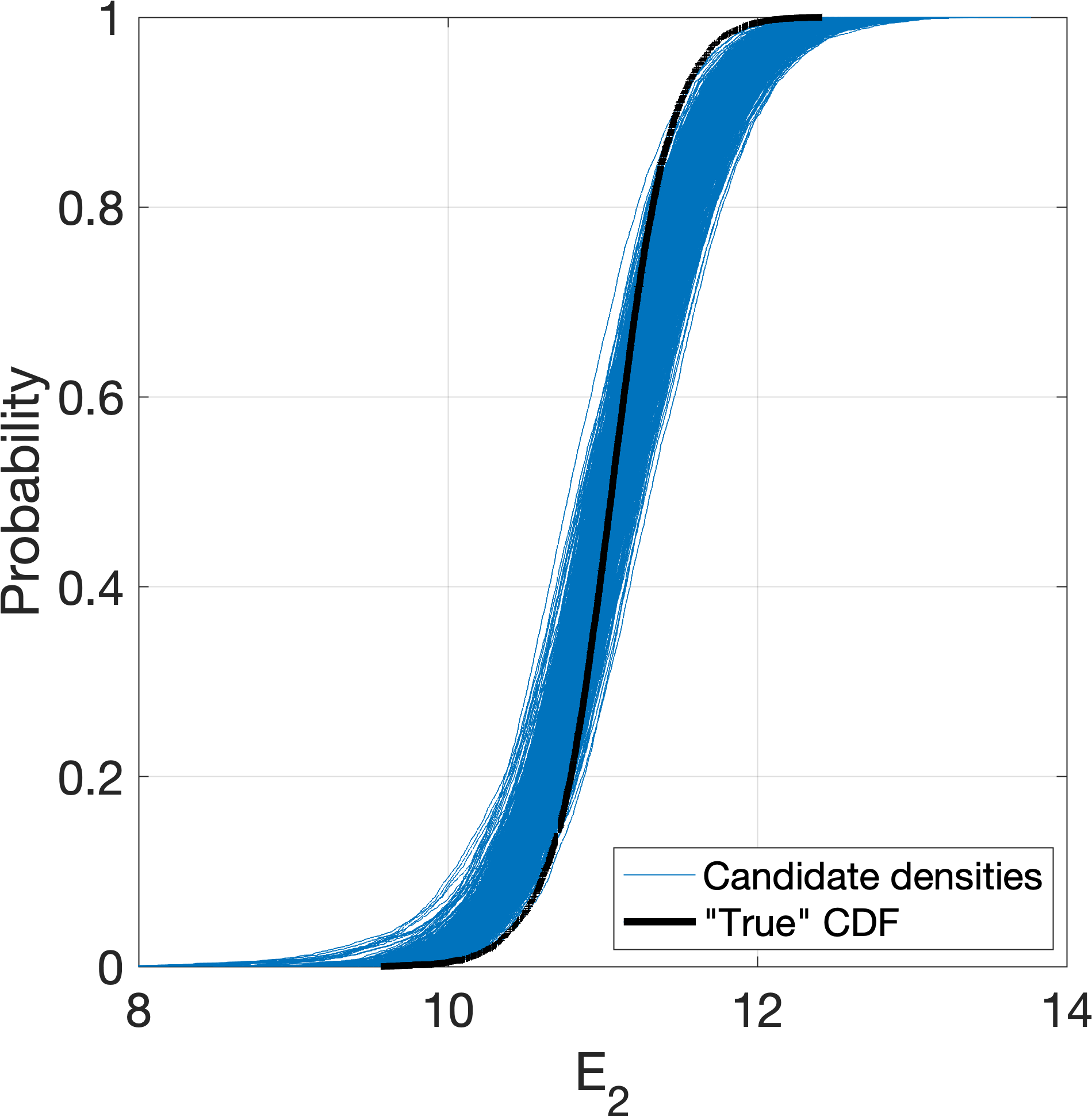}} \quad
    \subfigure[]{\includegraphics[height=2in]{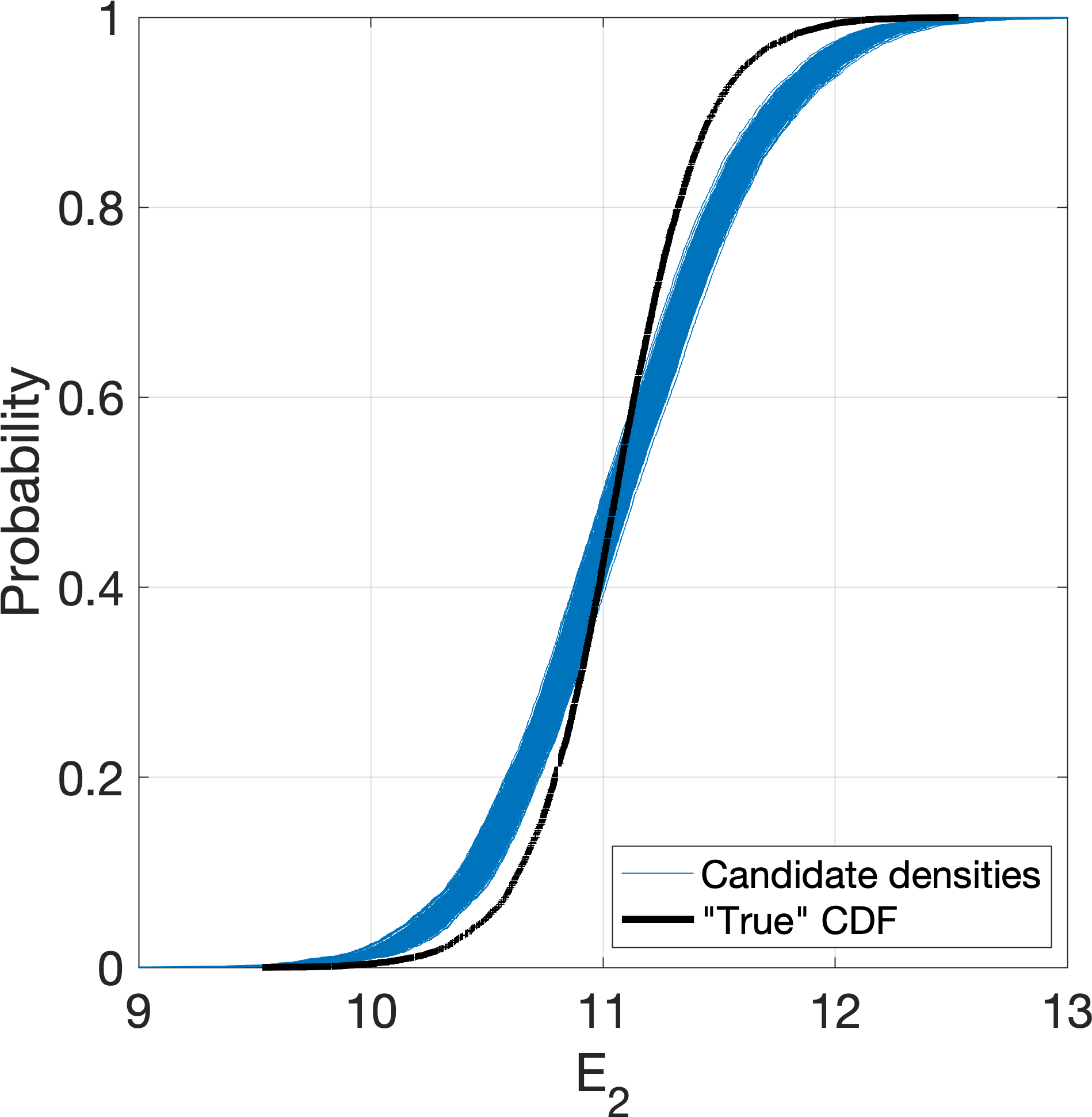}} \quad
    \subfigure[]{\includegraphics[height=2in]{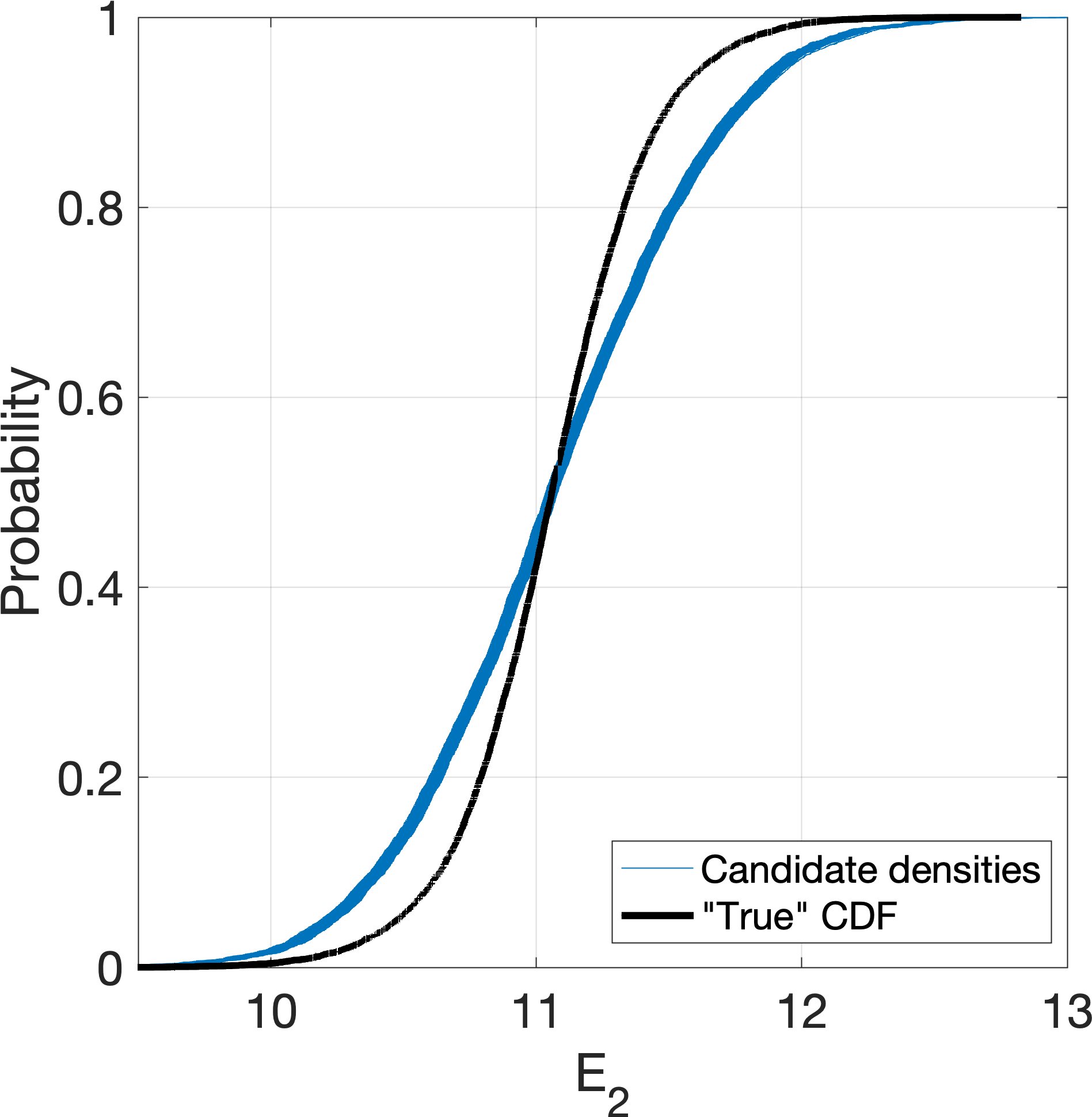}} \\
        \subfigure[]{\includegraphics[height=2in]{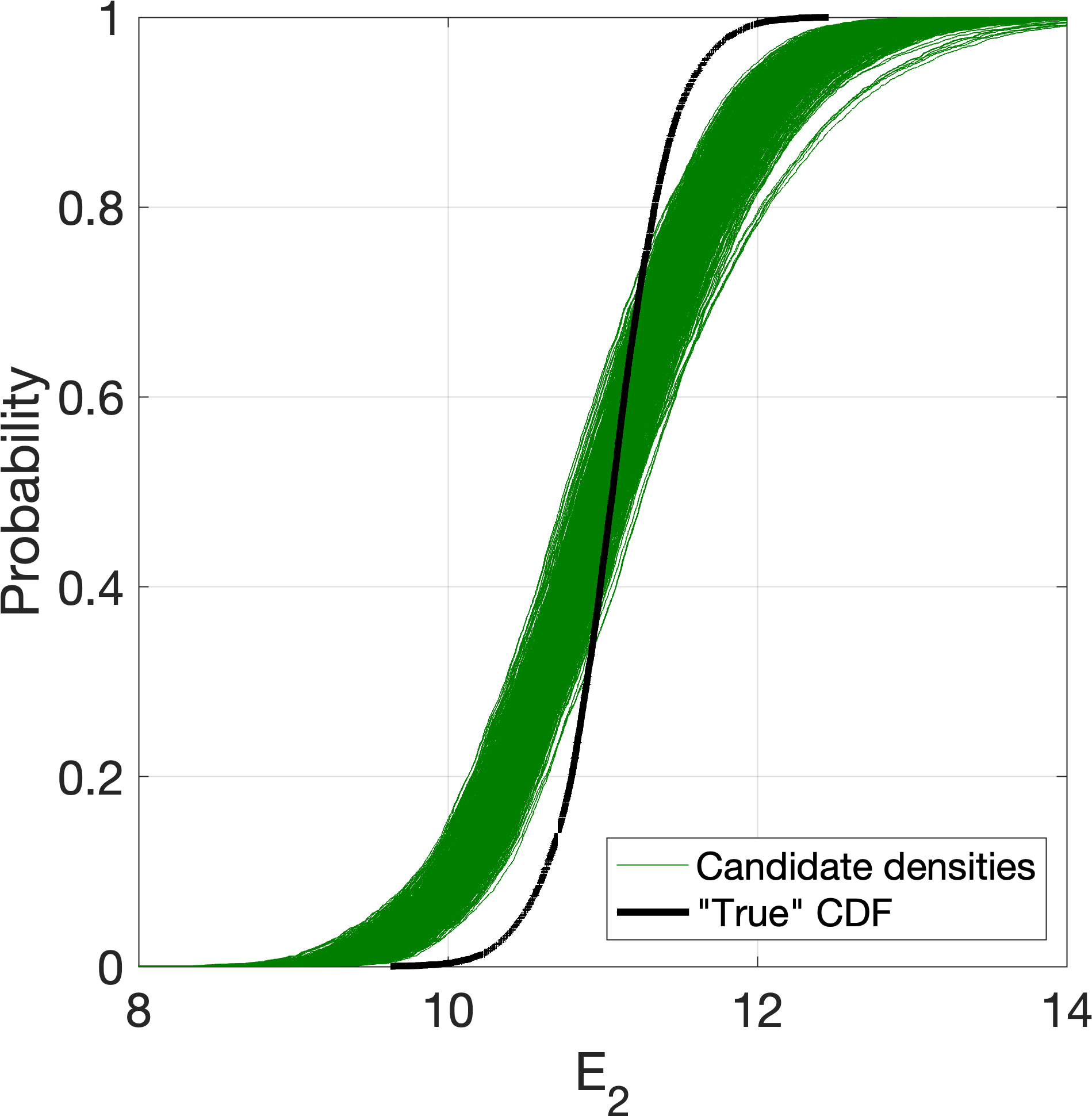}}  \quad
     \subfigure[]{\includegraphics[height=2in]{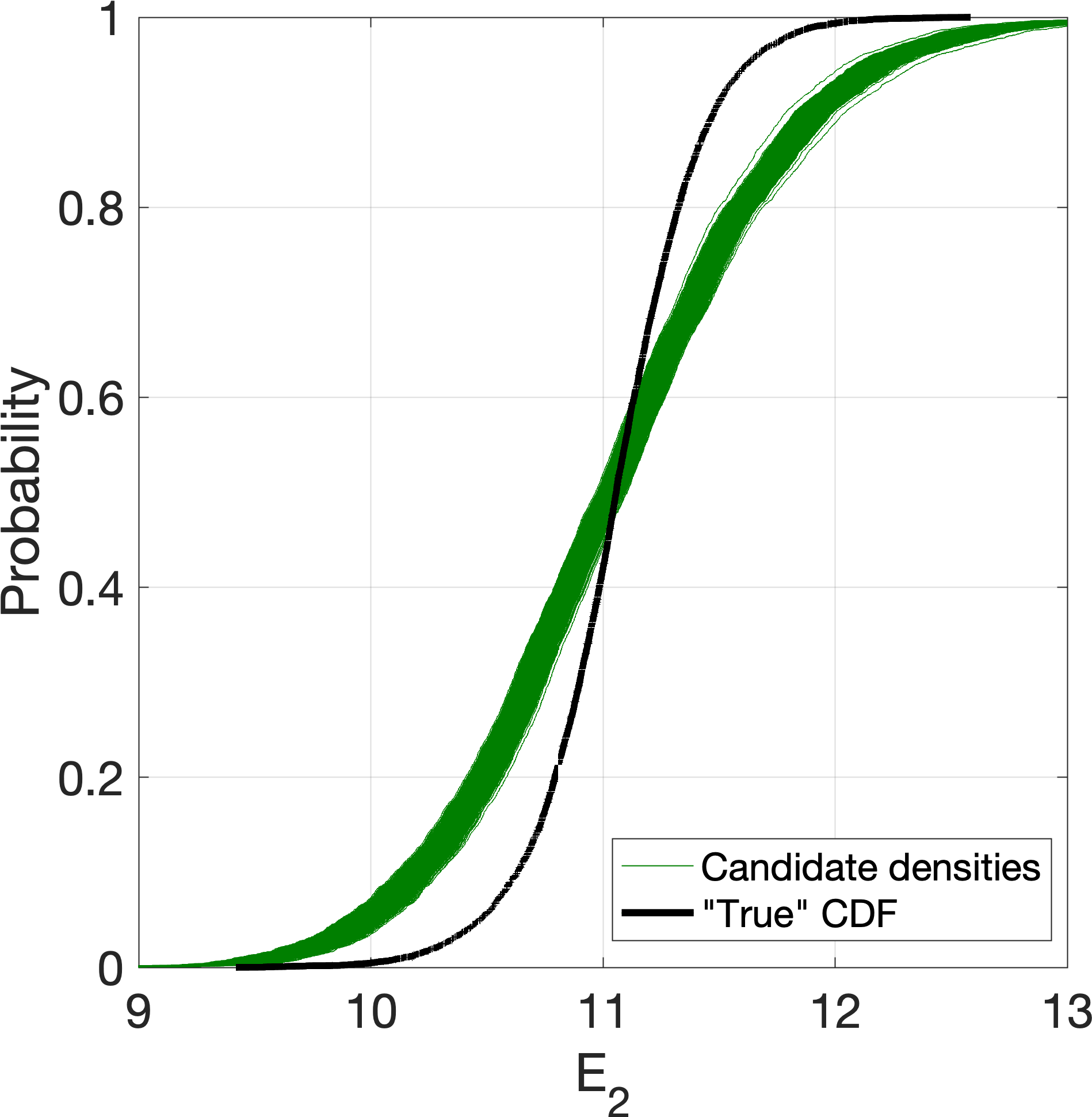}} \quad 
    \subfigure[]{\includegraphics[height=2in]{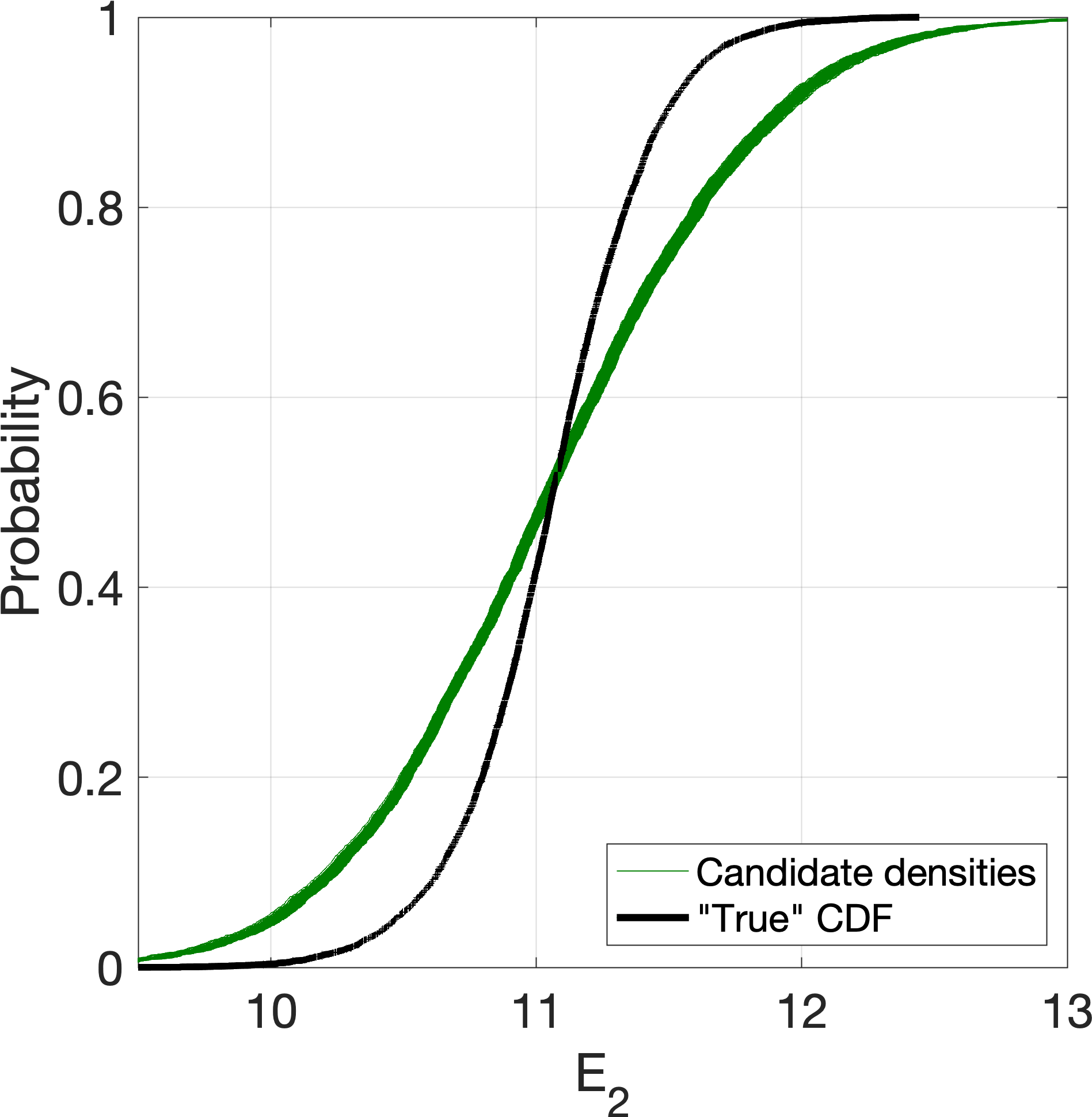}} \\
    \subfigure[]{\includegraphics[height=2in]{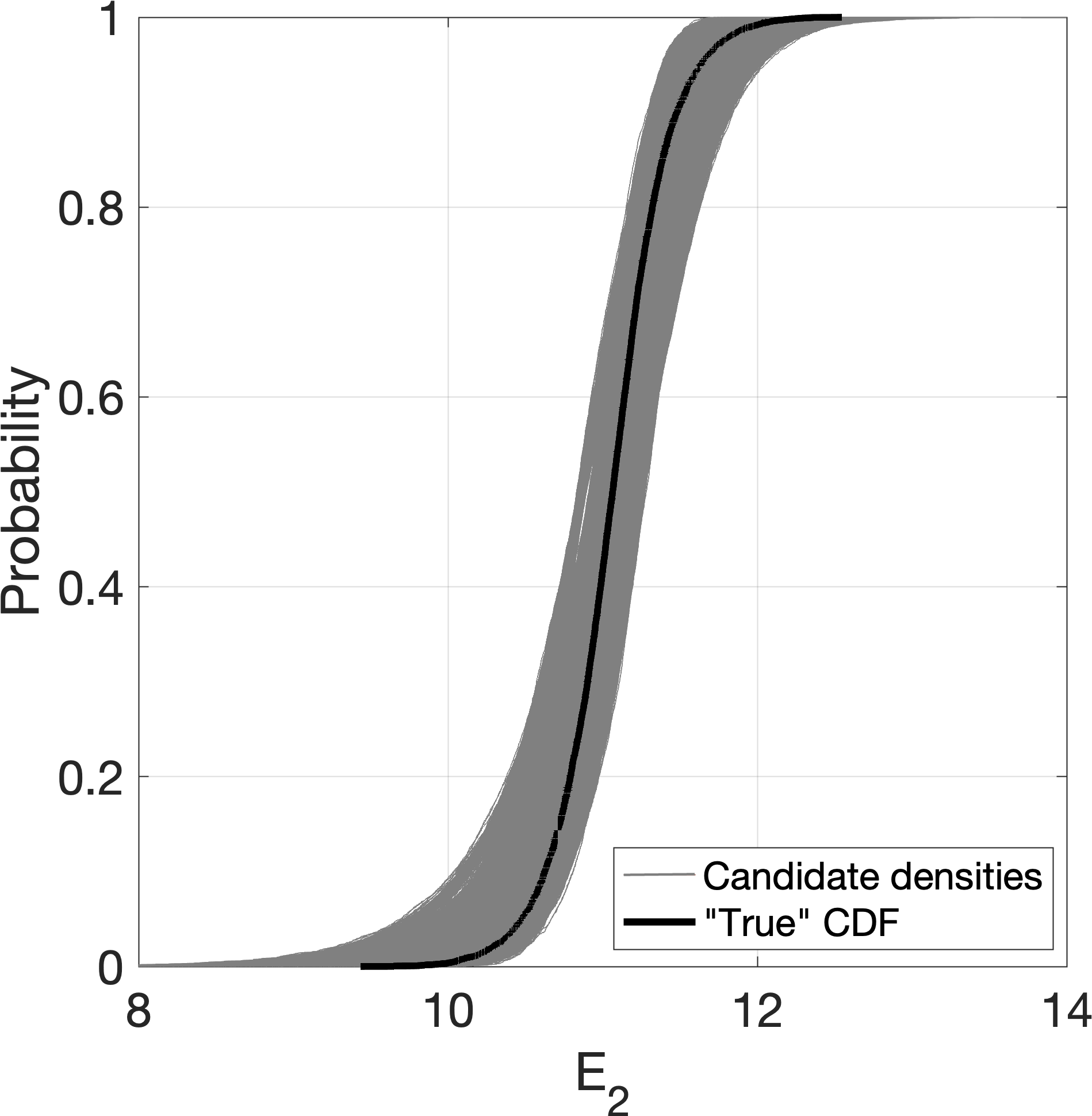}}  \quad
     \subfigure[]{\includegraphics[height=2in]{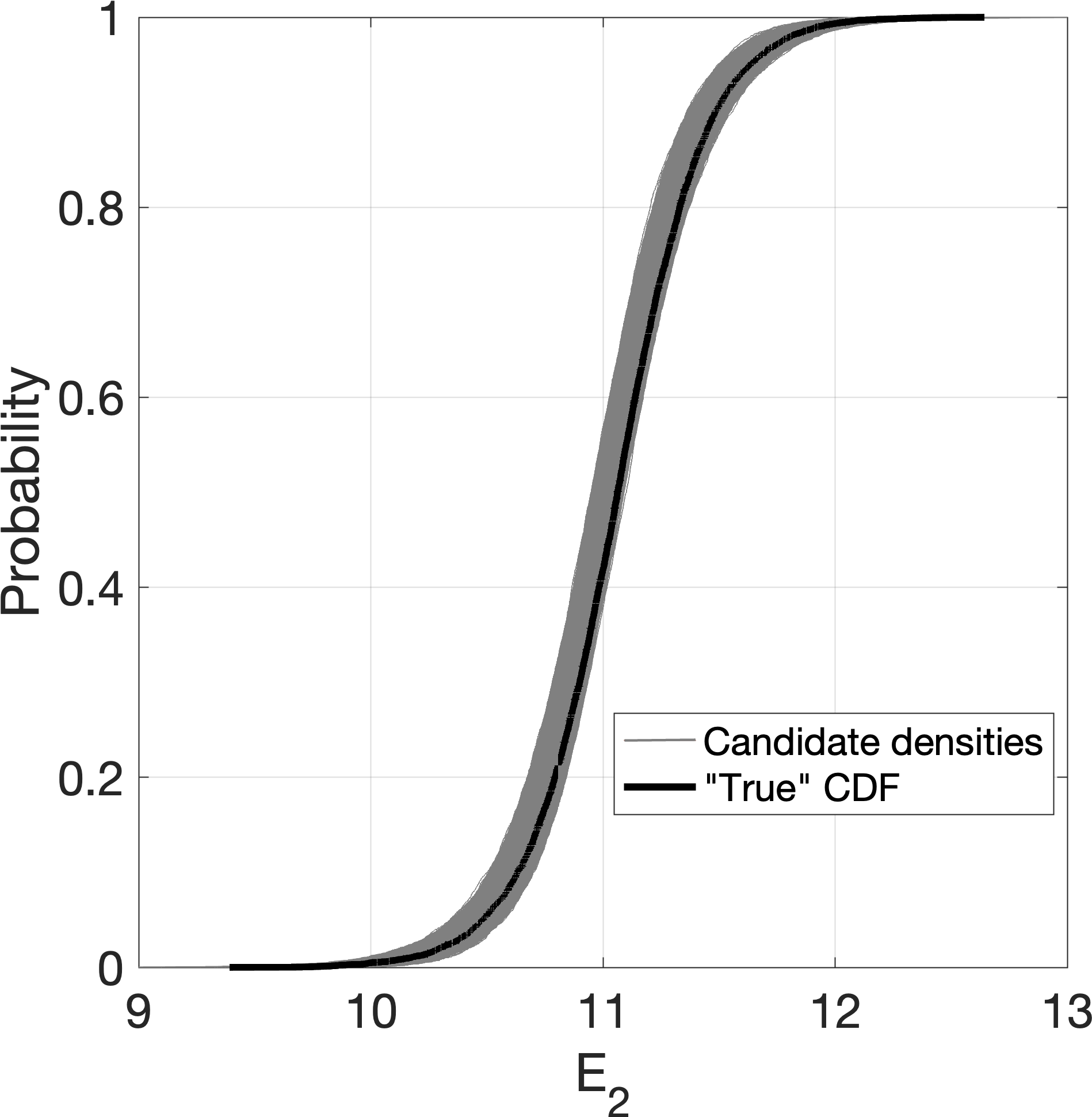}} \quad 
    \subfigure[]{\includegraphics[height=2in]{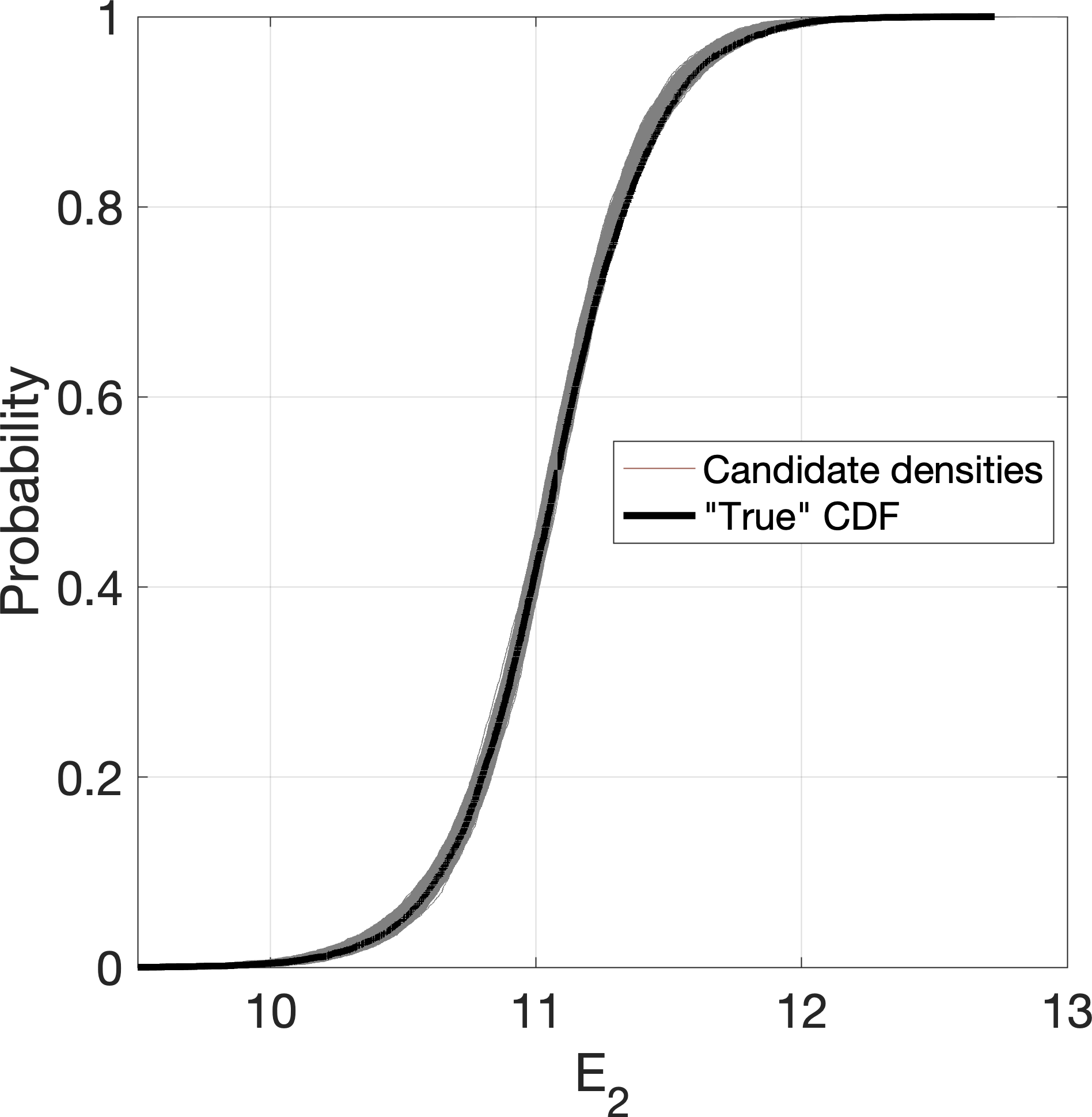}} 
    \caption{Uncertain CDFs for transverse elastic modulus $E_{22}$ with increasing data set size under the assumption of independent marginals (a-c), Gaussian correlation (d-f) and accounting for copula dependence (g-i): (a,d,g) 50 data, (b,e,h) 500 data and (c,f,i) 5000 data.}  \label{fig:c4_more_data_cdf}
\end{figure}

{\color{black}
Fig. \ref{fig:c4_more_data_cdf} shows the results of the multimodel uncertainty propagation to estimate the cdf of the transverse modulus $E_{22}$ for increasing data set size. The figure shows the convergence of the approach under assumptions of independent marginals (Figure \ref{fig:c4_more_data_cdf}a-c), Gaussian correlation (Figure \ref{fig:c4_more_data_cdf}d-f) and with dependence included (Figure \ref{fig:c4_more_data_cdf}g-i). The true cdf (with the known joint probability densities) is shown for reference. As expected, in all three cases the band of cdfs narrow as additional data are collected -- i.e.\ uncertainty in the prediction of $E_{22}$ is reduced. However, we notice that under the assumption of independent marginals and Gaussian correlation, the band of cdfs do not converge to the true cdf. Instead, there is a bias introduced by the assumption of independent and Gaussian correlated marginals . Only when we account for the variable dependence in the multimodel UQ approach are we able to converge to the true cdf of the modulus. This is an important conclusion because it shows that, altough uncertainty bands generated under the incorrect assumption of independence \textbf{may} initially bound the true probability distribution, they (i) are likely to underestimate the uncertainty in the estimated distribution as shown in Section \ref{sec:composite3}, and (ii) provide biased bounds on the true probability distribution that will not converge as the data set size increases.}

\section{Conclusion}

In this work, we propose a hierarchical multimodel approach to investigates the effect of uncertainties associated with small data sets for quantifying and propagating probabilistic model inputs with dependencies. The joint CDF of the probabilistic model inputs is composed of marginal distributions and copulas, which are modeled separately. The proposed approach is set in a hierarchical Bayesian multimodel inference framework, where the model-form and model parameter uncertainties associated with marginals are first quantified, and uncertainties associated with the copula are conditioned on specified marginal pairs. This results in an ensemble of joint probability densities that represent the imprecise probabilities in the assignment of probability model inputs with {\color{black}statistical dependence}. A novel importance sampling reweighting algorithm is derived to efficiently propagate the imprecise probabilities through a mathematical or physical model, which is often computationally intensive. The proposed approach therefore estimates the uncertainty in the quantity of interest given multiple candidate model input distributions at a low computational cost when compared with the typical nested Monte Carlo simulations. 


The methodology is demonstrated on an engineering application which aims to understand the influence of constituent properties on the overall out-of-plane properties of a transversely isotropic E-Glass fiber/LY556 Polyester Resign composites. A strong correlation between the constituent properties (fibers and matrix) is assumed and described using a Frank copula model. The results show that the assumption of independent {\color{black} and arbitrary Gaussian correlated} marginals in the imprecise UQ modeling both underestimates the uncertainty in predictions of the modulus and yields biased statistical estimates. When copula-based dependence is integrated into the multimodel UQ framework, the model achieves more realistic bounds on the uncertainty and more accurate probabilistic predictions.


\section{Acknowledgement}
The work presented herein has been supported by the Office of Naval Research under Award Number N00014-16-1-2582 with Dr. Paul Hess as the program officer. {\color{black} The work of J. Zhang was supported by the U.S. Department of Energy, Office of Science, Office of Advanced Scientific Computing Research, Applied Mathematics program under contract ERKJ352; and by the Artificial Intelligence Initiative at the Oak Ridge National Laboratory (ORNL). ORNL is operated by UT-Battelle, LLC., for the U.S. Department of Energy under Contract DEAC05-00OR22725.} The authors are grateful to Prof.\ Stephanie Termaath for providing models and support for materials applications. 

\bibliographystyle{elsarticle-num}
\bibliography{paper}

\end{spacing}
\end{document}